%% file: Main.tex
\pdfoutput=1
\documentclass[11pt,a4paper]{article}

\usepackage[margin=1in]{geometry}
\usepackage{amsfonts}
\usepackage{amssymb}
\usepackage{amsmath}
\usepackage{graphicx}
\usepackage{authblk}

\usepackage{xcolor}
\usepackage[caption=false]{subfig}
\usepackage{capt-of}
\usepackage{amsmath}
\usepackage[export]{adjustbox}
\usepackage{multirow}
\usepackage{cite}
\usepackage{url}
\usepackage[hidelinks]{hyperref}
\usepackage{xspace}

\hypersetup{
  pdftitle={Supervised and LLM-Augmented Cardinality Estimation for Similarity Queries on Knowledge Graphs},
  pdfauthor={Georgios Chatzigeorgakidis, Dimitrios Skoutas, Alkis Simitsis},
  pdfkeywords={cardinality estimation, knowledge graphs, similarity queries, machine learning, query optimization}
}

\definecolor{brass}{rgb}{0.71, 0.65, 0.26}
\newcommand\mnote[1]{\textcolor{black}{#1}}

\newcommand{\eat}[1]{}

\newcommand{\SGSIM}{\mbox{SG$_{sim}$}\xspace}
\newcommand{\SGSPLIT}{\mbox{SG$_{split}$}\xspace}

\newcommand{\baseline}{\mbox{LMKG}\xspace}
\newcommand{\MODELa}{\mbox{SG$_{sim}$-MLP}\xspace}
\newcommand{\MODELb}{\mbox{SCE-KG}\xspace}

\everydisplay\expandafter{\the\everydisplay\footnotesize}

\begin{document}

\title{Supervised and LLM-Augmented Cardinality Estimation for Similarity Queries on Knowledge Graphs}

\author{Georgios Chatzigeorgakidis}
\author{Dimitrios Skoutas}
\author{Alkis Simitsis}
\affil{Athena Research Center, Greece\\
\texttt{gchatzi@athenarc.gr}, \texttt{dskoutas@athenarc.gr}, \texttt{alkis@athenarc.gr}}

\date{}

\AddToHookNext{shipout/foreground}{%
  \setlength{\unitlength}{1pt}%
  \put(0,-66){\makebox[\paperwidth]{\parbox[b]{\textwidth}{\centering\small\color{blue}%
    ``This is the authors' version of the work. It is posted here for your
    personal use. Not for redistribution. The definitive version has been
    presented at Int.\ J. of Semantic Computing 2026.''}}}}

\maketitle

\begin{abstract}
    \input{Abstract}

\end{abstract}

\input{Introduction}
\input{Background}
\input{Problem}
\input{Methods}
\input{Evaluation}
\input{Conclusions}

\section*{Acknowledgements}
This work was partially supported by the EU projects STELAR (101070122),  WiseFood (101181895), and DataGEMS (101188416).

\bibliographystyle{abbrv}
\bibliography{References}

\end{document}

%% file: Abstract.tex
Over the past decades, query cardinality estimation on relational databases has been proven crucial, enabling the result size estimation of all sub-plans of each query. This allowed query optimizers to design efficient query plans, towards minimizing the query response time. Lately, knowledge graphs are being preferred over traditional databases in various applications, due to their ability to represent complex relationships and semantic meaning in a domain-specific context.
In this work, we present a novel approach for supervised similarity-aware cardinality estimation on knowledge graphs that incorporates similarities into the model training process, thus treating them as first-class citizens. To this end, we present \MODELb, a model that employs a novel encoding scheme for cardinality estimation, named \SGSPLIT.
We further extend our approach by leveraging a Large Language Model (LLM) to generate training queries that resemble realistic user workloads, and show that combining LLM-generated and random walk samples consistently improves estimation accuracy on test workloads generated by two independent LLMs. Our experimental evaluation reveals that our approach outperforms the state-of-the-art techniques in terms of prediction error, achieving 40x and 37x improvement for star and chain queries, respectively (95th percentile), whereas its online overhead accounts for less than 0.05\% of the average query execution time and less than 1\% of the average query planning time, showing a sub-millisecond latency in all cases.

%% file: Introduction.tex
\section{Introduction}
\label{sec:intro}

A Knowledge Graph (KG) is a network that semantically describes entities in a domain of interest and the relationships between them. Over the last years, the interest in KGs and their adoption has been increasing, both in academia and in industry~\cite{DBLP:journals/csur/HoganBCdMGKGNNN21}. There exist both open KGs, which are published online, e.g., Wikidata, as well as enterprise KGs, which are internal to a company, e.g., describing products and customers. Moreover, depending on their contents and focus, KGs can be encyclopedic, providing general knowledge about the world, commonsense, describing daily-life concepts, or domain-specific, representing specialized knowledge in a certain domain. KGs have several applications, including advanced search and recommendations~\cite{DBLP:journals/tkde/GuoZQZXXH22}, while also having a promising synergistic role with Large Language Models, which they can complement and enhance by offering factual, up-to-date knowledge and interpretability~\cite{DBLP:journals/corr/abs-2306-08302}.

The two most commonly used data models to represent KGs are RDF and Property Graphs~\cite{DBLP:journals/cacm/SakrBVIAAAABBDV21}. The former, which is a recommendation by the World Wide Web Consortium (W3C), represents facts as triples in the form
$<$$subject,~predicate,~object$$>$,
using Internationalized Resource Identifiers (IRIs) to refer to entities and relationships. The latter offers more flexibility, allowing each entity or relationship to be associated with a label and a set of key-value pairs. The SPARQL query language, which is also a W3C recommendation, is used to query RDF graphs, whereas Cypher, Gremlin and G-CORE are languages for querying property graphs~\cite{DBLP:journals/csur/AnglesABHRV17}.

Storing, indexing, and querying KGs, especially RDF graphs, has received a lot of attention~\cite{DBLP:journals/vldb/SagiLPH22}. Yet, cardinality estimation, which is a crucial component for query optimization~\cite{DBLP:journals/vldb/LeisRGMBKN18}, is still underexplored for KGs. Cardinality estimation refers to predicting the result size of a given query. It is a difficult and long-standing problem that has been primarily studied in the context of relational databases, with most recent approaches moving from techniques such as histograms and samplings to learned cardinality estimation based on deep learning~\cite{kim2022learned,Wan+21}.

However, unlike relational databases, KGs do not have a fixed, predefined, and binding schema. Although in many aspects this is an advantage offering greater flexibility, it makes querying a KG, including cardinality estimation, more challenging. 
Recent work has shown that state-of-the-art techniques for relational databases do not perform as well for KGs~\cite{davitkova2021lmkg}. To fill this gap, \baseline~\cite{davitkova2021lmkg} has proposed a deep-learning approach for cardinality estimation in RDF graphs. The main novelty of the work lies in designing an encoding, termed SG encoding, that represents the queries as subgraph patterns. However, \baseline only supports queries without similarity predicates and, as verified experimentally, fails to estimate query cardinalities in such scenarios accurately.

Learned cardinality estimation has also been investigated for similarity queries~\cite{JiGN21,Yao+21}. In this type of query, given a query point $p$, a distance function, and a threshold $r$, the goal is to estimate the number of data points that have distance to $p$ no more than $r$. Similarity search is a fundamental problem in information retrieval, data integration, and recommendation, all of which are also typical application areas of KGs~\cite{DBLP:journals/tkde/GuoZQZXXH22}. 
Motivated by this, in this paper, we introduce and address the novel problem of learned cardinality estimation for KG queries, including similarity predicates.

This paper extends our previous research~\cite{supervised2026} by (a)~elaborating on the theoretical foundations of MLPs, (b)~introducing a clustering-based segmentation scheme, (c)~proposing an LLM-augmented strategy for training data generation, and (d)~expanding our experimental analysis to include four new studies. In particular, we present \MODELb, a method that employs a fine-grained MLP architecture, which, in synergy with the novel \SGSPLIT encoding, treats node similarities as first-class citizens. \MODELb shows excellent performance and effectiveness compared to the state of the art, and especially for queries containing many triples.

The main contributions of our work are as follows:

\begin{itemize}
    \item We introduce the novel problem of learned cardinality estimation for KG queries, including similarity predicates. 
    \item \mnote{We propose \textit{\SGSPLIT}, an encoding for incorporating the similarity predicates in learned models.}
    \item We present \MODELb, a novel method that treats similarities as first-class citizens.
    \item We improve estimation accuracy via a clustering-based segmentation scheme that operates on similarity thresholds and partitions the training set into smaller subsets.
    \item We present the results of an extensive experimental evaluation on three real-world and one synthetic datasets.
\end{itemize}

The rest of the paper is structured as follows.
Section~\ref{sec:related} discusses related work.
Section~\ref{sec:problem} formulates the similarity-aware cardinality estimation problem on KG.
Section~\ref{sec:similarity} describes the \MODELb method.
Section~\ref{sec:exp} presents our experimental evaluation, and Section~\ref{sec:conclusions} concludes the paper.

%% file: Background.tex
\section{Related Work}
\label{sec:related}

The work most related to ours can be classified into the following areas: (a) cardinality estimation in relational databases, (b) learning approaches to cardinality estimation in relational databases, (c) similarity-aware cardinality estimation, and (d) cardinality estimation on knowledge graphs.

\paragraph{Cardinality estimation in relational databases}

Query optimizers depend on accurate cardinality and selectivity estimates, especially for intermediate results, to produce reasonable plans. Toward this direction, many approaches have been proposed employing a variety of techniques, e.g.,
histograms \cite{BrCG01, DeGR01, GKTD05, MuDe88, PoGI99},
entropy \cite{MHK+07, ReSu12},
samples \cite{HNSS96, EsNa06}
and index-based sampling \cite{LRG+17},
histograms and samples \cite{MuMK18},
probabilistic models \cite{Cher16, GeTK01, TzDJ11},
randomized hashing and data sketching \cite{CaBS19},
exponential back-off~\cite{Sack14}, dynamic sampling~\cite{Bayl21}, and so on.
These approaches work with assumptions (e.g., attribute value independence -- AVI,
uniformity, data independence) and when these assumptions cannot be met, the techniques usually fall back to an educated guess.
However, in practice the estimates are routinely wrong by orders of magnitude, causing slow queries and unpredictable performance.

Recently, a significant amount of works have studied various learning techniques for cardinality estimation in the context of relational databases~\cite{Han21,TsSi22}.
\cite{WHT+19}~builds local models, each focusing on one part of the DB schema.
\cite{OBGK18}~evaluates deep learning techniques with respect to trade-offs among model size, training time, and prediction accuracy.
\textit{CRN}~\cite{HayS20}~uses deep learning to estimate cardinalities using the containment rates among queries.
\textit{DeepDB}~\cite{Hil+20}~proposes a data-driven approach that supports workload and data changes, fast retraining, and reasonable accuracy for unseen queries.
\textit{FLAT}~\cite{Zhu+21} builds on top of DeepDB~\cite{Hil+20}, considering the level of dependency between attributes.
\textit{Naru}~\cite{YLK+19} and \textit{NeuroCard}~\cite{Yan+20b} use autoregressive models to learn the joint distribution of relations, with the latter focusing on learning the full outer join of all tables.
\textit{UAE}~\cite{WuCo21} uses data and queries to accurately estimate cardinalities, even when data or workload shifts.
\cite{Yao+21,JiGN21} learn estimating the cardinality for similarity search.

Supervised methods work better in static environments, where they avoid re-training.
Several data-driven approaches adopt incremental training, thus becoming more dynamic. 
Most approaches use \textit{q-error} as the de facto loss function or quality metric. 
\textit{Flow-Loss}~\cite{Neg+21} presents a novel loss function that considers the quality of the estimated cardinalities in terms of query runtime. 
\cite{Wan+21}~argues that the proposed solutions are not yet production-ready due to issues like high training/inference time, black-box decisions, and difficulty in recovering after data updates.

\paragraph{Similarity-aware cardinality estimation} 

The above approaches deal with cardinality estimation for queries without similarity predicates. Recently, a new line of work has emerged that focuses on estimating the cardinality for similarity queries. In \cite{vu2021learned}, the authors introduce a machine learning-based query optimizer for spatial joins that considers dataset skewness and algorithm complexity. The optimizer consists of three models, the first of which performs cardinality estimation. Sun et al.~\cite{JiGN21}, employ Deep Neural Networks (DNNs) for learned cardinality estimation on similarity queries. To overcome the inherent complexity of DNNs, the authors derive a query and data segmentation technique before training a separate model for each segment. In~\cite{Yao+21}, Wang et al. attempt to leverage the monotonicity of the cardinality with respect to the similarity threshold by quantizing the similarity thresholds to a pre-defined number of bins before training. Then, they forward them to an encoder-decoder-based DNN. The decoder consists of several smaller decoder modules, one for each separate threshold bin. Their output is then aggregated to provide the estimated cardinality.

\paragraph{Cardinality estimation on knowledge graphs}

Although the problem of learned cardinality estimation in KGs is critical in many applications, the research community has only recently considered it.
To the best of our knowledge, \baseline~\cite{davitkova2021lmkg} is the only research approach to the learned cardinality estimation problem in KGs. 
\baseline addresses SPARQL queries, especially star and chain queries, over RDF graphs. To avoid introducing a high overhead, the neural network employed by \baseline is a multilayer perceptron (MLP). To achieve better performance, \baseline constructs different models per query type and/or size. LMKG feeds queries to the model using a novel encoding that represents each query as a subgraph pattern: each query triple is first converted to numerical form by encoding its terms with either one-hot or binary encoding, after which the full query is encoded via subgraph (SG) encoding, consisting of two binary vectors, $X$ of size $n$ and $E$ of size $e$, and a matrix $A \in \{0,1\}^{n*n*e}$, where $n$ is the number of nodes within the subgraph and $e$ the number of predicates within the subgraph; together, $A$, $X$, and $E$ capture the query’s structure, with $A$ indicating node connections and predicates, $X$ representing node features, and $E$ representing predicate features (see Figure~\ref{fig:encodings}(b)). However, \baseline only supports equality predicates. Hence, it cannot effectively estimate the cardinality of queries that include similarity predicates, which is the focus of our paper.

%% file: Problem.tex
\section{Problem Formulation}
\label{sec:problem}

\subsection{The cardinality estimation problem on KG}

In this work, we focus on RDF graphs and SPARQL queries. This is without loss of generality because a translation between property graphs and directed edge-labeled graphs is possible without loss of information~\cite{DBLP:conf/amw/AnglesTT19}.

Accordingly, a KG $G$ consists of a finite set of triples $\textbf{t}$, where each triple $t = (s, p, o)$ comprises a subject $s$, a predicate $p$ and an object $o$. Subjects and predicates are URIs, whereas an object may be a URI or a literal.

A SPARQL query $q$ consists of triples forming a graph pattern, where each triple $t$ can have an arbitrary number of variables (unbound terms). Moreover, we assume a similarity function $sim$ which assigns a similarity score  to any pair of nodes in the graph. For instance, this may refer to Jaccard similarity or edit similarity for string literals, or cosine similarity computed on node embeddings. In any case, we assume that the similarity score is normalized in the range [0, 1].

\textit{Problem statement.}
Given a graph $G$ and a query $q$, which may contain one or more similarity filters based on a given similarity function $sim$ and respective user-defined similarity thresholds $\theta_i$, our goal is to estimate the 
cardinality of results for~$q$.

\subsection{Graph queries}

We focus on the two most common types of queries: \emph{star} queries and \emph{chain} queries. Previous work~\cite{davitkova2021lmkg} has shown that
(a) building separate models for different types of queries increases performance, and (b) more complex queries (e.g., snowflake) can be decomposed into simpler ones. Star queries consist of triples centered around the same subject, whereas chain queries consist of a set of triples in which the object of the preceding triple is the subject of the next one. 

\begin{figure}[t]
\centering
\begin{minipage}[c]{0.4\linewidth}
    \centering
    \includegraphics[width=\linewidth]{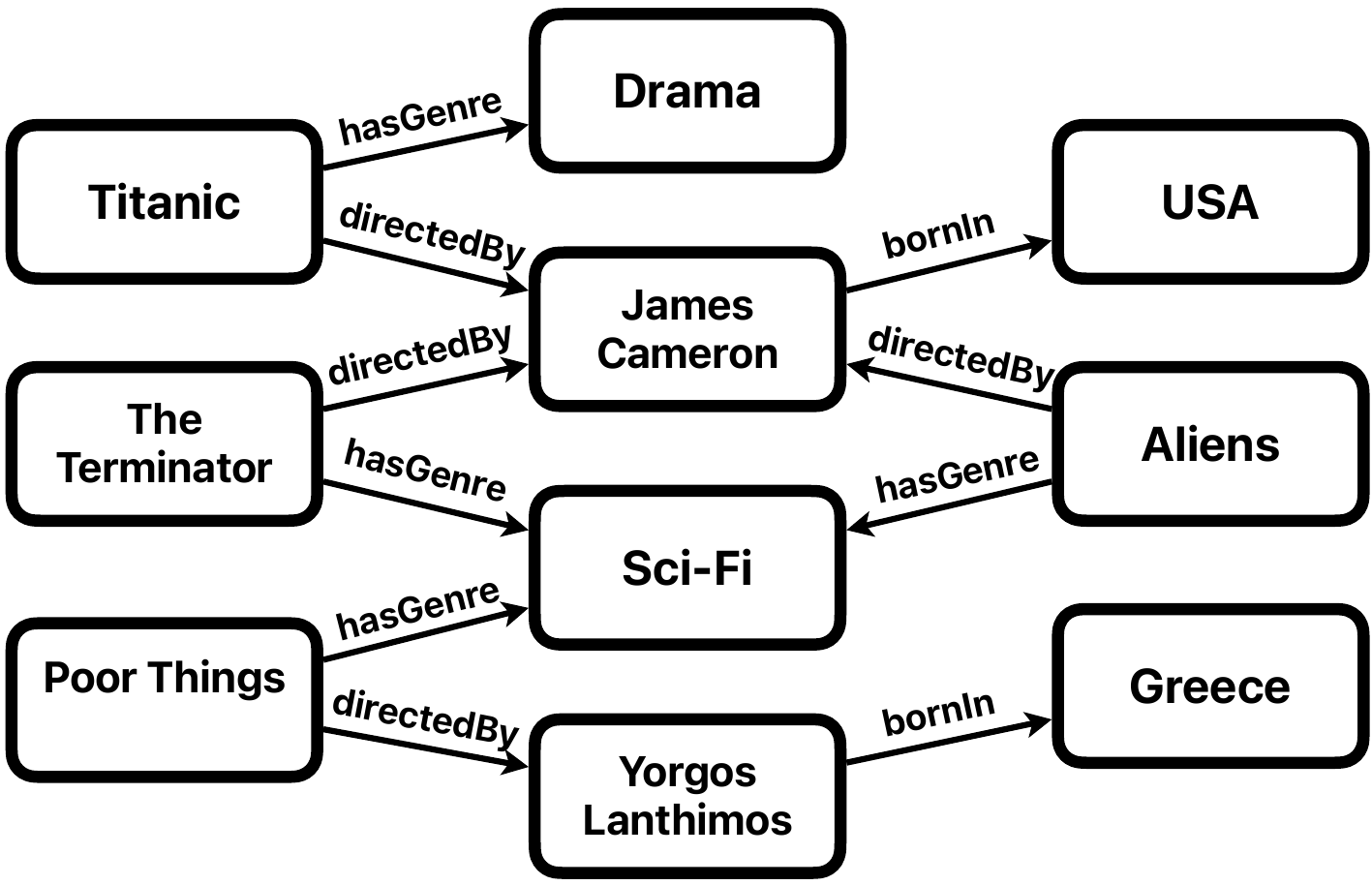}
\end{minipage}\hfill%
\begin{minipage}[c]{0.56\linewidth}
\footnotesize
\begin{verbatim}
SELECT ?Movie, ?Genre, ?Director WHERE {
    ?Movie :hasGenre ?Genre AND
    sim(:Sci-Fi, ?Genre) > 0.6;
    ?Movie :directedBy ?Director AND
    sim(:James Cameron, ?Director) > 0.8.
}
\end{verbatim}
\begin{verbatim}
SELECT ?Movie, ?Director, ?Country WHERE {
    ?Movie :directedBy ?Director AND
    sim(:Yorgos Lanthimos, ?Director) > 0.6;
    ?Director :bornIn ?Country AND
    sim(:Greece, ?Country) > 0.8.
}
\end{verbatim}
\end{minipage}
\caption{A knowledge graph (left) and example $2$-star (top right) and $2$-chain (bottom right) similarity queries over it}
\label{fig:graph}
\end{figure}

We refer to queries containing $n$ triples as $n$-star and $n$-chain queries.
\textit{Example.}
Consider the KG depicted in Figure~\ref{fig:graph} (left), together with an example \textit{$2$-star similarity query} (top right) and an example \textit{$2$-chain similarity query} (bottom right) over it.

\noindent The $2$-star query seeks all movies whose genre has similarity to \textit{Sci-Fi} at least 0.6 and that are directed by a director with similarity to \textit{James Cameron} at least 0.8. The $2$-chain query seeks all movies directed by \textit{Yorgos Lanthimos} or a similar director (similarity at least 0.6), where that director was born in \textit{Greece} or a similar country (similarity at least 0.8).

\subsection{Multilayer perceptron (MLP) neural networks}

In this work, we follow a supervised learning approach to the problem at hand.
Before we present our solution, we discuss basic background concepts.
Supervised learning methods for graph data vary.
However, lightweight MLP models have the advantages of faster speed and lower memory usage while being quite effective for estimating the number of rows in a relational database query~\cite{TzDJ11}. Such an MLP model takes as input workloads of star and chain queries.
As an example, for star queries containing two joins, the MLP input would be:

\begin{equation}
    p1-s1,p2-s2:c
\end{equation}

\noindent
where $p1$ is the first predicate of the star query that connects the query node to node $s1$, $p2$ and $s2$ are the second predicate and node respectively, and $c$ is the resulting cardinality given these predicates and nodes.

Each such training element is encoded and fed into the MLP for training using \textit{q-loss} as a loss function. The q-loss function is derived from the \textit{q-error}, which is often used in cardinality estimation, and it can be formulated as follows:

\begin{equation}
    \mathcal{L}(c, \hat{c}) = \max{\left(\frac{\hat{c}}{c}, \frac{c}{\hat{c}}\right)}
\end{equation}

\noindent
where $c$ is the actual cardinality and $\hat{c}$ is the estimated one. The intermediate fully connected layers employ the rectifier (ReLU) activation function, defined as:

\begin{equation}
    f(x) =\begin{cases}
      x & \text{if $x > 0$}\\
      0 & \text{otherwise}\\
    \end{cases}
\end{equation}

\noindent
where $x$ is the input to a neuron. The cardinalities are log-scaled to alleviate the effect of skewed data and min-max-scaled to allow a sigmoid function at the output. Thus, since cardinality estimation is a regression task, a sigmoid activation function is employed for the output layer that outputs a number in the $[0, 1]$ interval. The sigmoid activation function is defined as:

\begin{equation}
    f(x) = \frac{1}{1 + e^{-x}}
\end{equation}

\noindent
where $x$ is the input to a neuron.

%% file: Methods.tex
\section{Similarity-Aware Cardinality Estimation}
\label{sec:similarity}

\begin{figure*}[!t]
    \raggedright
    \includegraphics[width=\linewidth]{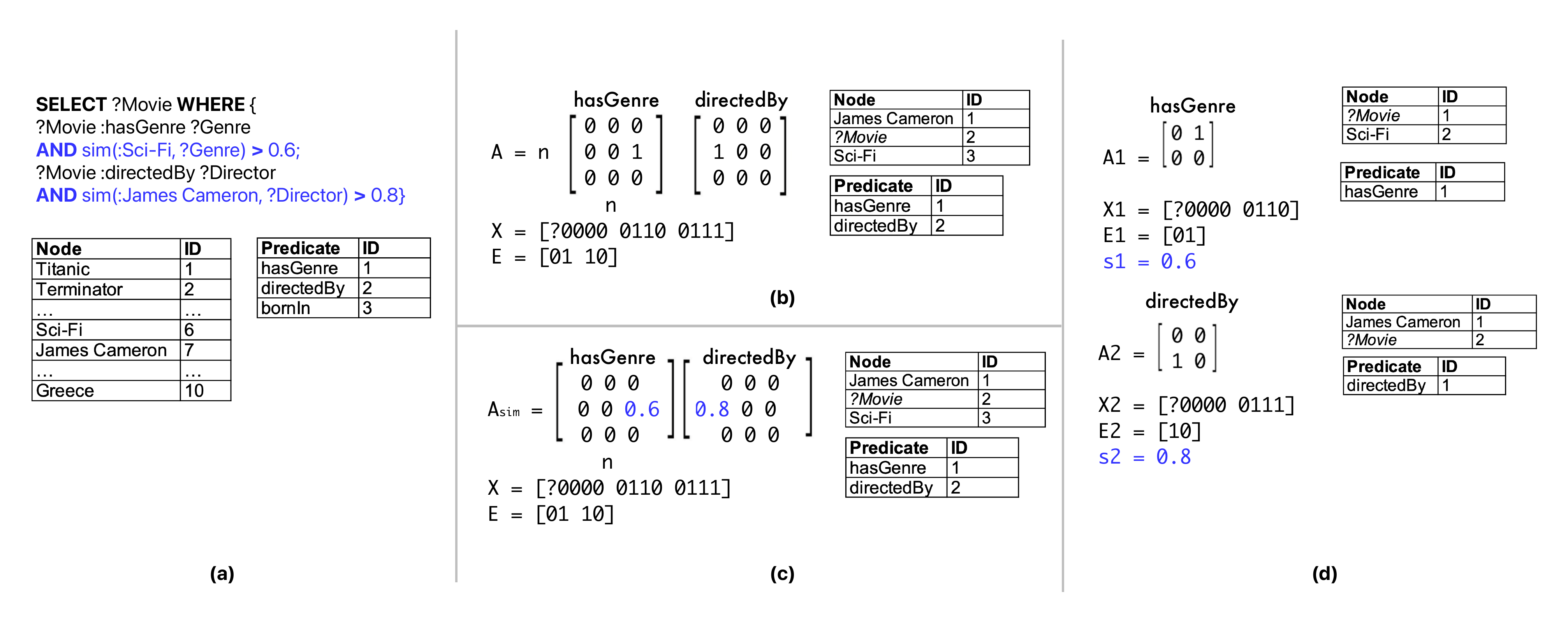}
    \caption{(a) Query/term mapping, (b) \baseline encoding, (c) \SGSIM encoding, (d) \SGSPLIT encoding}
    \label{fig:encodings}
\end{figure*}

\mnote{As discussed, \baseline does not consider similarity queries. Hence, a straightforward attempt to solve the problem would be to extend the SG encoding (see also Section~\ref{sec:related}) to capture such queries. 
To this end, we can incorporate the similarity corresponding to each object of a query's triplets, in order to enable the training of an MLP model on such data.
Conceptually, in star and chain similarity queries, similarities relate to the subject, predicate, and object of the triplet.}

\textit{Example}. 
Consider the query of Figure~\ref{fig:encodings}(a) defined over the KG of Figure~\ref{fig:graph}. This query seeks all Sci-Fi movies directed by James Cameron. It also contains similarity predicates. The similarity $0.6$ of the first triplet relates directly to the object, i.e., $?Genre$, since we seek genres similar to Sci-Fi and indirectly to the subject of the query, i.e., $?Movie$, since we seek movies that have a genre similar to the given one. 
 
We proceed as follows. Assume that an ID has been assigned to each node and predicate of the knowledge graph. For example, these could be the integer IDs assigned while building the knowledge graph. Then, all the nodes and predicates within a query are sorted based on their initial IDs and assigned new IDs based on their new order. 

In the SG encoding --see also Figure~\ref{fig:encodings}(b)-- a tensor $A$ is derived by assigning `$1$' for each predicate to the corresponding position of the nodes it connects. Vectors $X$ and $E$ contain the binary encoding of the nodes and predicates according to their initial IDs, respectively (i.e., the unbound variables are represented with zeroes).
Still, the SG encoding does not support similarities. 

Apparently, however, tensor $A$ is the only part of the SG encoding that combines information about subjects, predicates, and objects of query triplets. 
\mnote{Thus, to enable a model to be trained on data containing similarities, we differentiate the $A$ tensor by swapping the `$1$'s --which previously denoted the presence of a relationship-- with the corresponding similarity.}
The new tensor is termed $A_{sim}$. We retain the format of vectors $X$ and $E$. \mnote{Figure~\ref{fig:encodings}(c) illustrates this baseline approach}. Notice the swapped similarities in tensor $A_{sim}$, annotated with the blue color.

\mnote{To apply cardinality estimation on star and chain queries, using this extended SG encoding, any MLP that is able to receive $X$, $A_{sim}$ and $E$ as input can be used.}

\noindent Figure~\ref{fig:supervised} depicts a generic architecture of such a model, defined as follows:

\begin{equation}
    \hat{y} = F_2(F_{1,1}(X), F_{1,2}(A_{sim}), F_{1,3}(E))
\end{equation}

\noindent where $F_{1,1}, F_{1,2}, F_{1,3}$ are fully connected layers and $\hat{y}$ is the estimated cardinality.
$X$, $A_{sim}$ and $E$ are initially passed from three fully connected layers and then are concatenated before passing from an output fully connected layer.

The case of similarity-aware cardinality estimation is a superset of the cardinality estimation for queries without similarity predicates (i.e., the similarities would always be equal to $1$).

Thus, this generic model using the extended SG encoding can answer both.

\eat{
\begin{figure}[!ht]
    \centering
    \includegraphics[width=1\linewidth]{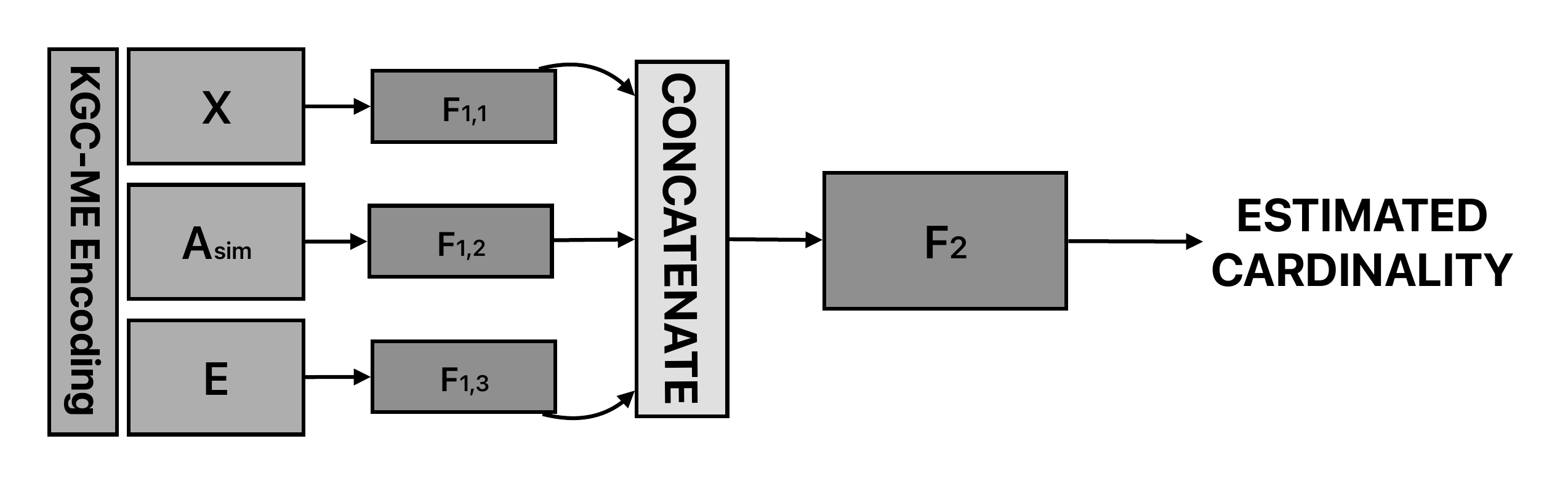}
    \caption{Generic MLP architecture for similarity-aware queries}
    \label{fig:supervised}
\end{figure}
}

{\color{red}
\begin{figure}[!ht]
    \centering
    \includegraphics[width=0.6\linewidth]{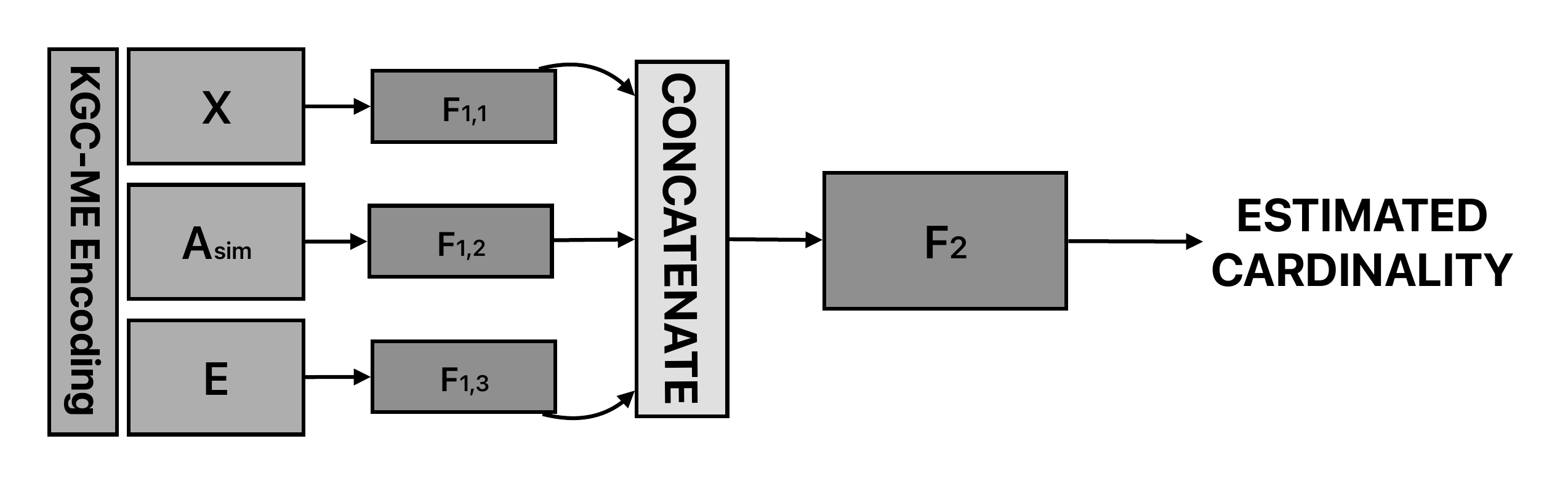}
    \caption{Generic MLP architecture for similarity-aware queries}
    \label{fig:supervised}
\end{figure}
}

\subsection{The \MODELb Model}
\label{subsec:model2}

\mnote{By considering only the nodes of the query, the extended SG encoding avoids large adjacency matrices and utilizes a size-optimized tensor $A_{sim}$. Still, $A_{sim}$ is a sparse matrix, which usually contains many zeros and only a few values signifying predicate similarity.}
For example, in Figure~\ref{fig:encodings}c, for each predicate, the corresponding similarity is a single value in an $n \times n$ matrix. 
\mnote{This is a significant drawback, as the impact of the similarity values on the computation of the cardinality estimation gets diminished with the significant number of linear algebra computations taking place during training and the inherent sparsity of the matrix.}
To mitigate this, we introduce \MODELb, a novel model for similarity-aware cardinality estimation. \MODELb employs the \SGSPLIT encoding, which splits the encoding process into multiple simple sub-encodings, enabling similarities to be handled outside the encoding and treated as first-class citizens in the estimation process.

\subsubsection{\SGSPLIT Encoding}
Figure~\ref{fig:encodings}d depicts an example of \SGSPLIT encoding computation for the query in Figure~\ref{fig:encodings}a, which seeks all Sci-Fi (or similar genre) movies directed by James Cameron (or similar director). Ignoring the similarities, this query contains the following two triples:

\footnotesize
\begin{verbatim}
    <?Movie :hasGenre ?Genre>
    <?Movie :directedBy ?Director>
\end{verbatim}
\normalsize

\SGSPLIT encoding considers separate $X_1, X_2$ $E_1, E_2$ vectors and $A_1, A_2$ tensors for each triple of the query. This further reduces the sparsity and dimensionality of $A_i$ tensors, i.e., $A_i \in \{0,1\}^{n*n}, 0 \leq i \leq |\textbf{t}|$, where $\textbf{t}$ is the set of triples (joins) in the query. The dimensionality of each tensor $A_i, 0 \leq i \leq |\textbf{t}|$ will be $2 \times 2$ for each triple and vectors $X_i$ and $E_i, 0 \leq i \leq |\textbf{t}|$ will always have a length of $2$ and $1$ respectively.

\subsubsection{Model Architecture}
\begin{figure}[t]
\centering
\begin{minipage}[t]{0.48\linewidth}
    \centering
    \includegraphics[width=\linewidth,height=4cm,keepaspectratio]{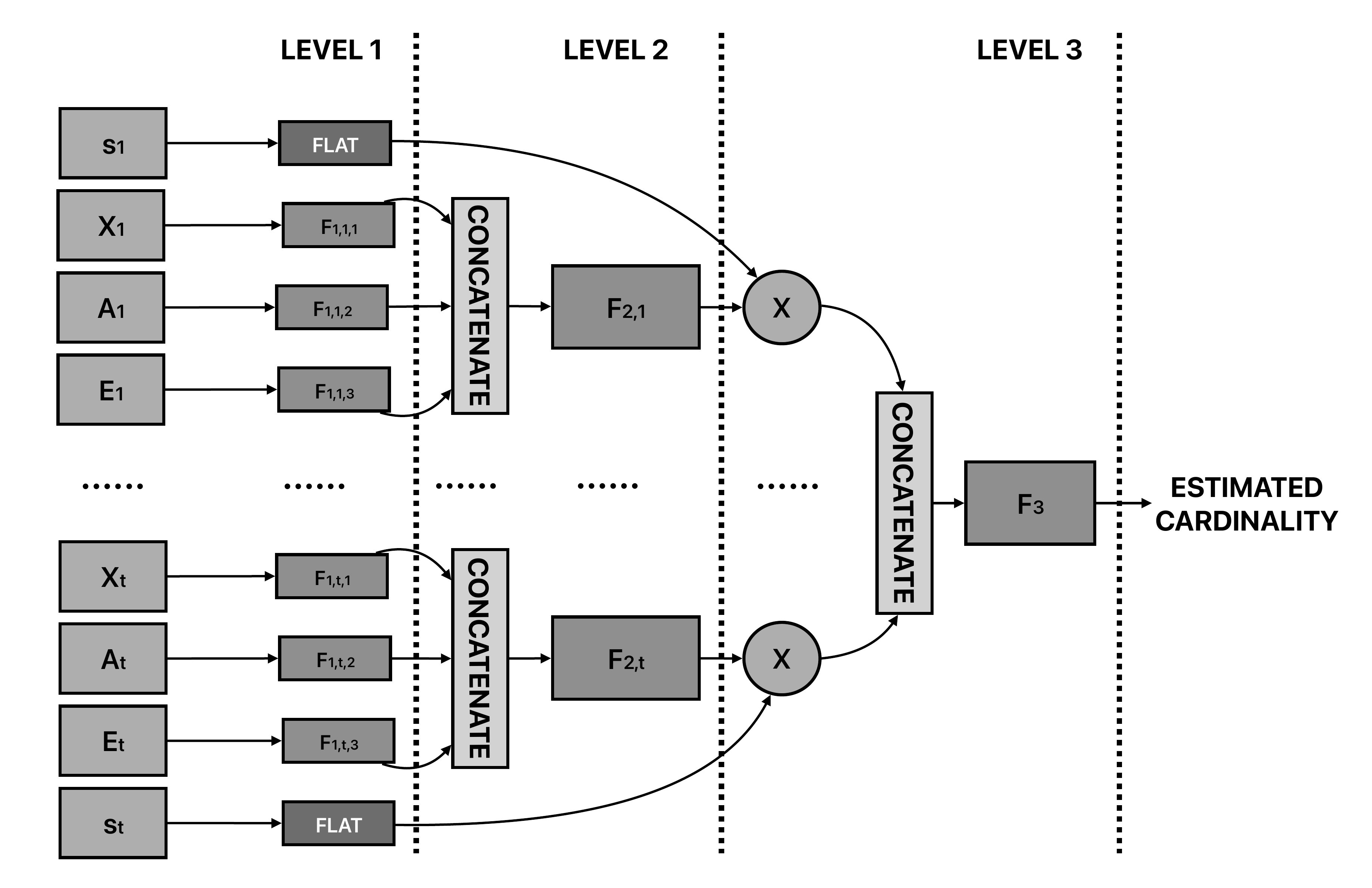}
    \caption{\MODELb architecture}
    \label{fig:simcard_kg}
\end{minipage}\hfill%
\begin{minipage}[t]{0.48\linewidth}
    \centering
    \includegraphics[width=\linewidth,height=4cm,keepaspectratio]{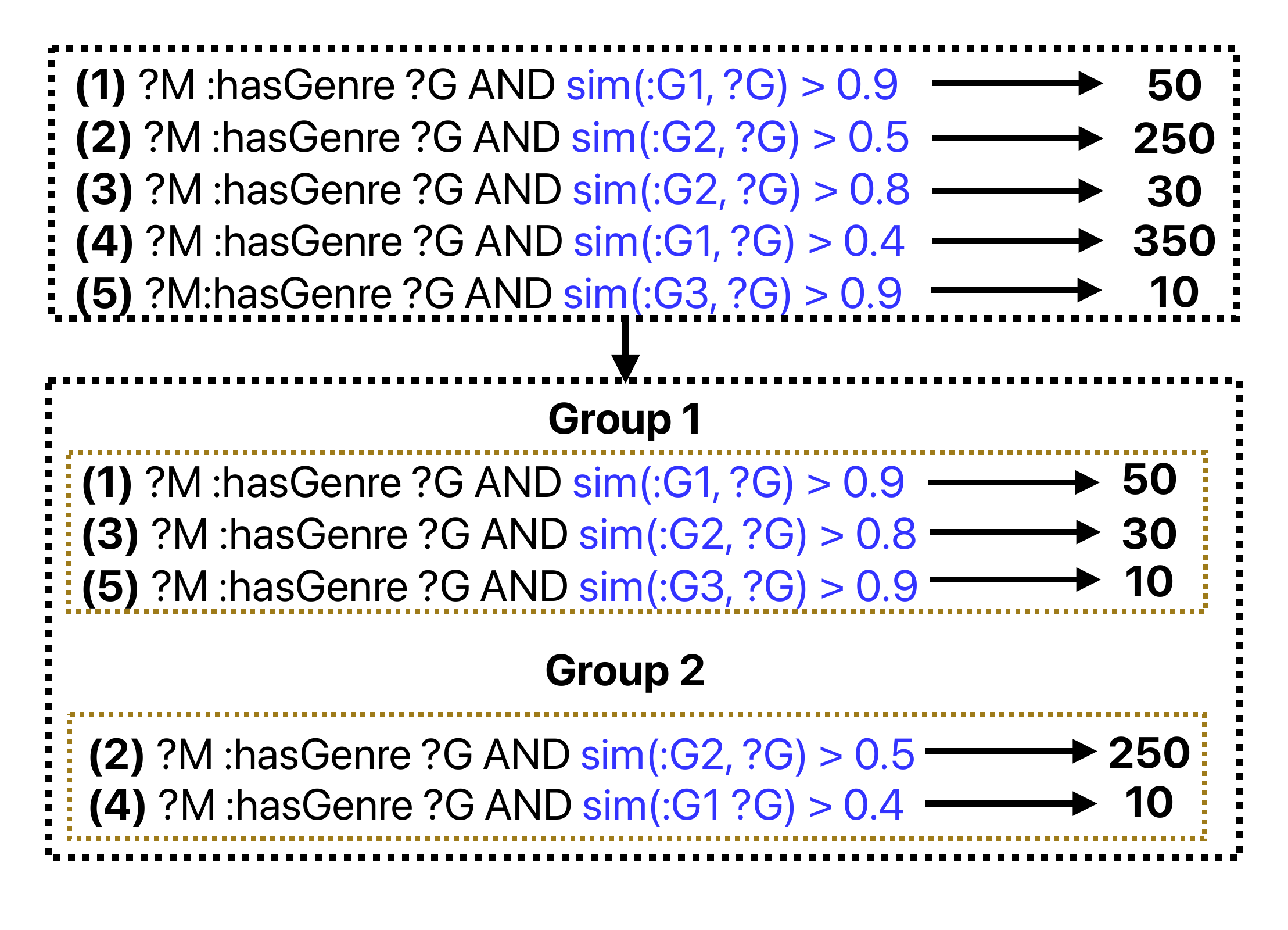}
    \caption{Clustering based on similarities}
    \label{fig:clustering}
\end{minipage}
\end{figure}

\MODELb is a 3-level multilayer perceptron (MLP) that performs regression tasks. Figure~\ref{fig:simcard_kg} depicts the architecture of \MODELb for the cardinality estimation of n-star queries. The idea behind it is that a $n$-star query can be broken into $n$ separate estimation sub-tasks \mnote{(LEVEL1 in Figure~\ref{fig:simcard_kg})} since each triple of the query attributes to the final cardinality \mnote{(LEVEL2 in Figure~\ref{fig:simcard_kg})}. \mnote{To incorporate the contribution of the query's similarities to each sub-estimation, the sub-estimations are multiplied by the corresponding similarity. Then, the latent vectors produced are concatenated and passed from a fully connected layer that determines the contribution of each triple to the cardinality and attempts to produce a final estimation (LEVEL3 in Figure~\ref{fig:simcard_kg})}. More formally, the model is defined as follows:

\footnotesize
\begin{equation}
\begin{aligned}
    \hat{y} = F_3(F_{2,1}(F_{1,1,1}(X_1), F_{1,1,2}(A_1)), F_{1,1,3}(E_1)))) \times s_1, ... \\
    ..., F_{2,t}(F_{1,t,1}(X_1), F_{1,t,2}(A_1)), F_{1,t,3}(E_1)) \times s_n)
\end{aligned}
\end{equation}
\normalsize

\noindent where $F_{i,x,x}, i\in[1,3]$ are the fully connected layers of levels $1$-$3$, $F_{x,i,x}, 0 \leq i \leq |\textbf{t}|$ are the fully connected layers that correspond to each triple of the query and $F_{x,x,i}, i \in [1,3]$ are the fully connected layers for $X_i, E_i, A_i, 0 \leq i \leq |\textbf{t}|$. First, the input is encoded using \SGSPLIT encoding. Then, each $X_i$, $E_i$ vectors, and $A_i$ tensor are passed through a fully connected layer and then concatenated per query triple before being passed again from a second group of fully connected layers. The output of each of these layers is then multiplied by the corresponding similarity (i.e., flattening is applied to the similarities to be compatible with the outputs of each fully connected layer). Finally, the outputs of the fully connected layers are concatenated again and passed by a third layer, which estimates the final cardinality.

The architecture of \MODELb allows us to treat similarity as a first-class citizen, by multiplying each similarity with the output of the corresponding triple estimation (i.e., the output of $F_{2,1}...F_{2,t}$, as shown in the figure). This way, since the outputs of these fully connected layers are not sparse, 
the impact of similarity in cardinality estimation is presumably higher.
\mnote{The advantage of \MODELb over the baseline approach becomes more apparent for queries 
that contain a higher amount of triples.}
A 5-star query would result in a $6\times6$-dimensional $A$ vector in \SGSIM, which is significantly more sparse than the one for 2-star queries. 
Our experiments (discussed shortly) also corroborate this observation. 

\subsection{Improving Accuracy with Clustering}
\label{subsec:clustering}

When performing similarity selection, the actual cardinality is increasing monotonically as the threshold $\theta$ is relaxed (see Section~\ref{sec:problem}). Overall, lower similarity thresholds tend to produce more results. Motivated by this,
the authors in~\cite{Yao+21}
consider a maximum possible threshold and map the thresholds to an integer space, thus generating a finite number of thresholds. Then, for each threshold group, they train a separate model. In our case, this is not possible, since we are dealing with more than one similarity thresholds per training sample; except for the case of queries that contain one triple. To tackle this issue, we apply multidimensional clustering of the training data based on the given similarities on each literal. This way, we end up with a set of training sets. For each set, we train a separate model. Upon estimation, we classify each testing element into one of the available clusters and run the corresponding model.

To provide an intuition of our clustering approach, consider
the example of a training set consisting of five 1-star queries illustrated in Figure~\ref{fig:clustering}. Notice that, there are various similarities for each triple, ranging from $0.4$ to $0.9$.
As already discussed and also shown in our experiments, as we keep relaxing the threshold the cardinality also increases.
Thus, a model that will be trained on these data will have to learn
that queries involving genre $G1$ have cardinalities $50$ (query (1)) and $350$ (query (4)), and those with genre $G2$ have cardinalities $30$ (query (3)) and $250$ (query (2)).
This imposes a challenge for an MLP, which is an inherent difficulty of the problem that we tackle in this paper:
correlating these genres with both low and high cardinalities would require more training data.
After clustering the training elements into two groups, notice that the grouped cardinalities are now similar, which is expected to facilitate the training process.

\subsection{Answering Complex Queries}

As discussed in Section~\ref{sec:problem}, without loss of generality, we presented our approach considering two query types: star and chain queries. However, \MODELb can answer more complex similarity-aware queries on knowledge graphs, such as tree, cycle, clique, petal, and flower queries. This is straightforwardly achieved as the queries can be broken down into triples and fed into the models via the \SGSPLIT encoding.
For space considerations, we omit here a detailed description.

\subsection{Training Data Creation}
\label{subsec:training_creation}

\cite{leskovec2006sampling} argue that generating ``realistic'' samples of graphs is essential. In our problem, the training set must cover as many graph patterns as possible (the more so given the added complexity of similarities) to let models learn the relationships between node/predicate IDs and cardinalities.
As suggested in~\cite{leskovec2006sampling} and also implemented in~\cite{davitkova2021lmkg}, an optimal way to sample the graph is via random walks. 
In our case, to simulate this we pick random sub-graphs from within a graph, insert unbound terms where necessary 
(i.e., to support queries containing one or more unbound variables), 
and issue the result as a query to the graph. 
For example, for a 2-star similarity query the 
obtained cardinalities along with the node/predicate IDs and corresponding similarities can then be formatted as follows:

\begin{equation}
p1-s1-sim1,p2-s2-sim2:c
\end{equation}

\noindent where $sim1$ and $sim2$ are the corresponding similarities.

\subsubsection{Using an LLM to generate training data}
\label{subsec:llm_training}

The random walk approach above uniformly covers the graph, but does not necessarily reflect how users phrase queries in practice. We thus also explore generating training data with a Large Language Model (LLM), composing query patterns that resemble realistic user workloads. Given the predicate vocabulary and a pool of real object identifiers from the graph, we prompt the LLM to produce star query patterns directly in the canonical training format $p_i\textnormal{-}s_i\textnormal{-}sim_i$. The core part of the prompt is the following:

\scriptsize
\begin{verbatim}
A STAR-3 query has 3 outgoing edges from a common source: p1-s1-sim1, p2-s2-sim2,p3-s3-sim3. 
p_i is a predicate ID; s_i is a real object ID or "*"; sim_i in {0.2, 0.4, 0.6, 0.8, 1.0}. 
Choose sim values to reflect realistic user search: sim=1.0 = exact match; lower = more tolerance.
\end{verbatim}
\normalsize

The generated patterns are labelled with their actual cardinality on the graph, as the random walk samples are. Since the two reflect complementary views (realistic phrasing vs.\ uniform graph coverage) they can also be combined into a hybrid training set, as studied in Section~\ref{subsec:llm_eval}.

%% file: Evaluation.tex
\section{Experimental Evaluation}
\label{sec:exp}

In this section, we present the results of our experimental analysis. We start with a description of our setup and then proceed with the evaluation of the methods proposed. Our code is available online\footnote{\url{https://github.com/chgeorgakidis/kgc}}.

\subsection{Experimental Setup}
\label{subsec:exp_setup}

\paragraph{System}
The KGs were built using neo4j 5.13, with our methods implemented in Python and Tensorflow 2.7.1. \mnote{Experiments ran on a server with an AMD Ryzen Threadripper 3960X (24 cores), 256GB RAM, and an NVIDIA GeForce RTX 2080 Ti GPU.}

\begin{figure}[t]
\centering
\begin{minipage}[c]{0.40\linewidth}
    \centering
    \scriptsize
    \setlength{\tabcolsep}{3pt}
    \captionof{table}{Datasets}\label{tab:datasets}
    \begin{tabular}{l|cccc}
    {\em \textbf{Dataset}} & Nodes & Rels. & Triples & Size \\
    \hline
    SWDF     & 106K & 65   & 347K  & 24MB  \\
    Yelp     & 9.2M & 6    & 14.4M & 14GB  \\
    Wikidata & 739M & 8605 & 1.25B & 99GB  \\
    LUBM     & 545K & 13   & 1.6M  & 331MB \\
    \end{tabular}
\end{minipage}\hfill%
\begin{minipage}[c]{0.28\linewidth}
    \centering
    \scriptsize
    \setlength{\tabcolsep}{3pt}
    \captionof{table}{Parameters}\label{tab:parameters}
    \begin{tabular}{l|l}
    {\em \textbf{Parameter}} & Values \\
    \hline
    Query type & \textbf{star}, chain \\
    {Clusters} & {\textbf{1}, 2, 3} \\
    Joins & 2, 3, \textbf{5}, 7 \\
    Epochs & 20, \textbf{50}, 100, 200 \\
    {Binning} & {\textbf{equi-sized},} \\
             & {equi-width,} \\
             & {order-of-mag.} \\
    \end{tabular}
\end{minipage}\hfill%
\begin{minipage}[c]{0.28\linewidth}
    \centering
    \includegraphics[width=\linewidth]{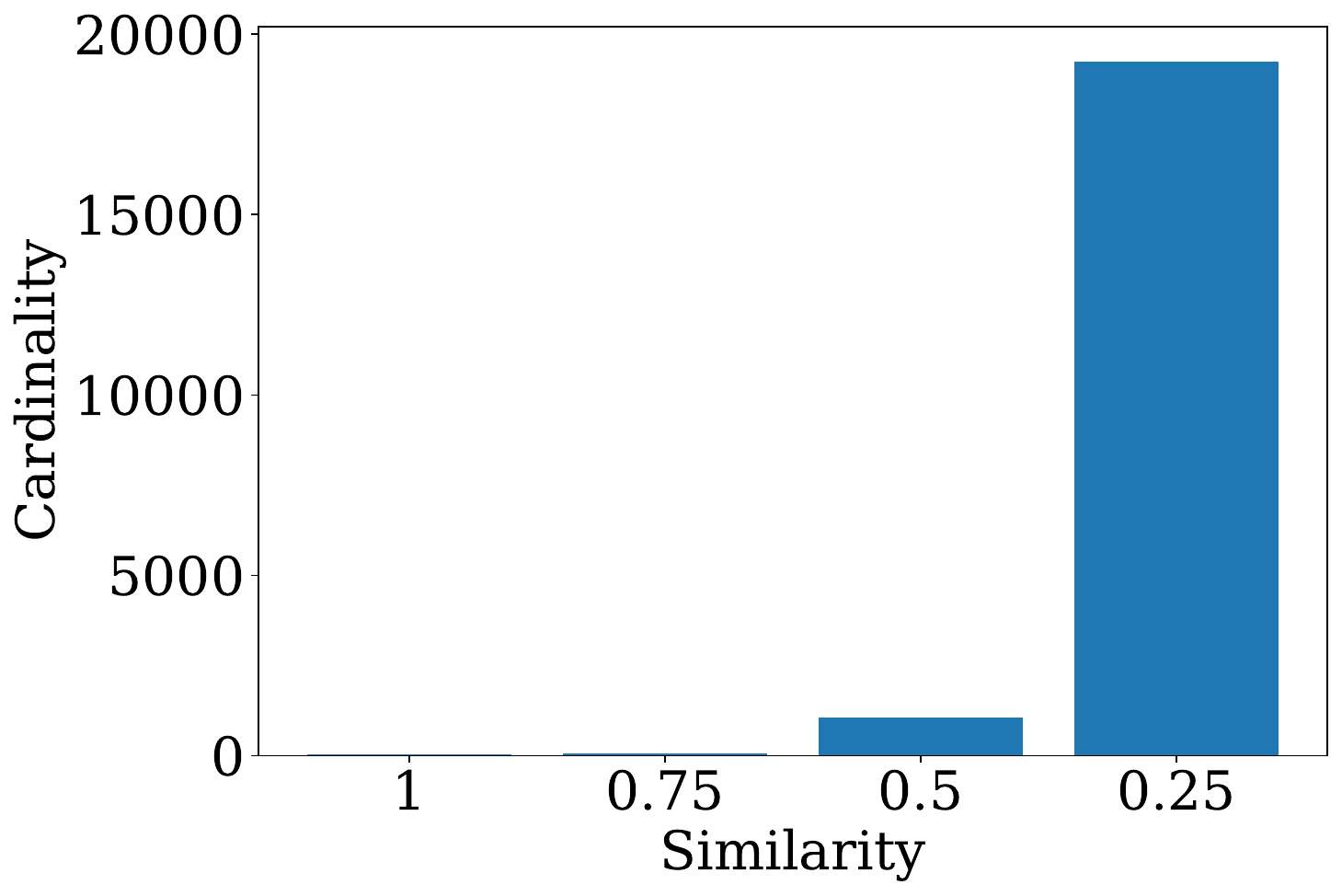}
    \caption{Cardinality for various thresholds}\label{fig:sim_effect}
\end{minipage}
\end{figure}

{\color{red}
\begin{figure*}[!ht]
\centering
\subfloat{\includegraphics[width=0.35\linewidth]{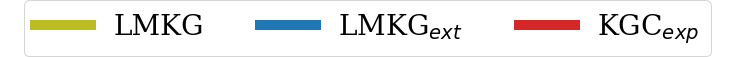}}
\\
\vspace{-12pt}
\setcounter{subfigure}{0}
\subfloat[Median q-err.]{\includegraphics[width=0.2\linewidth]{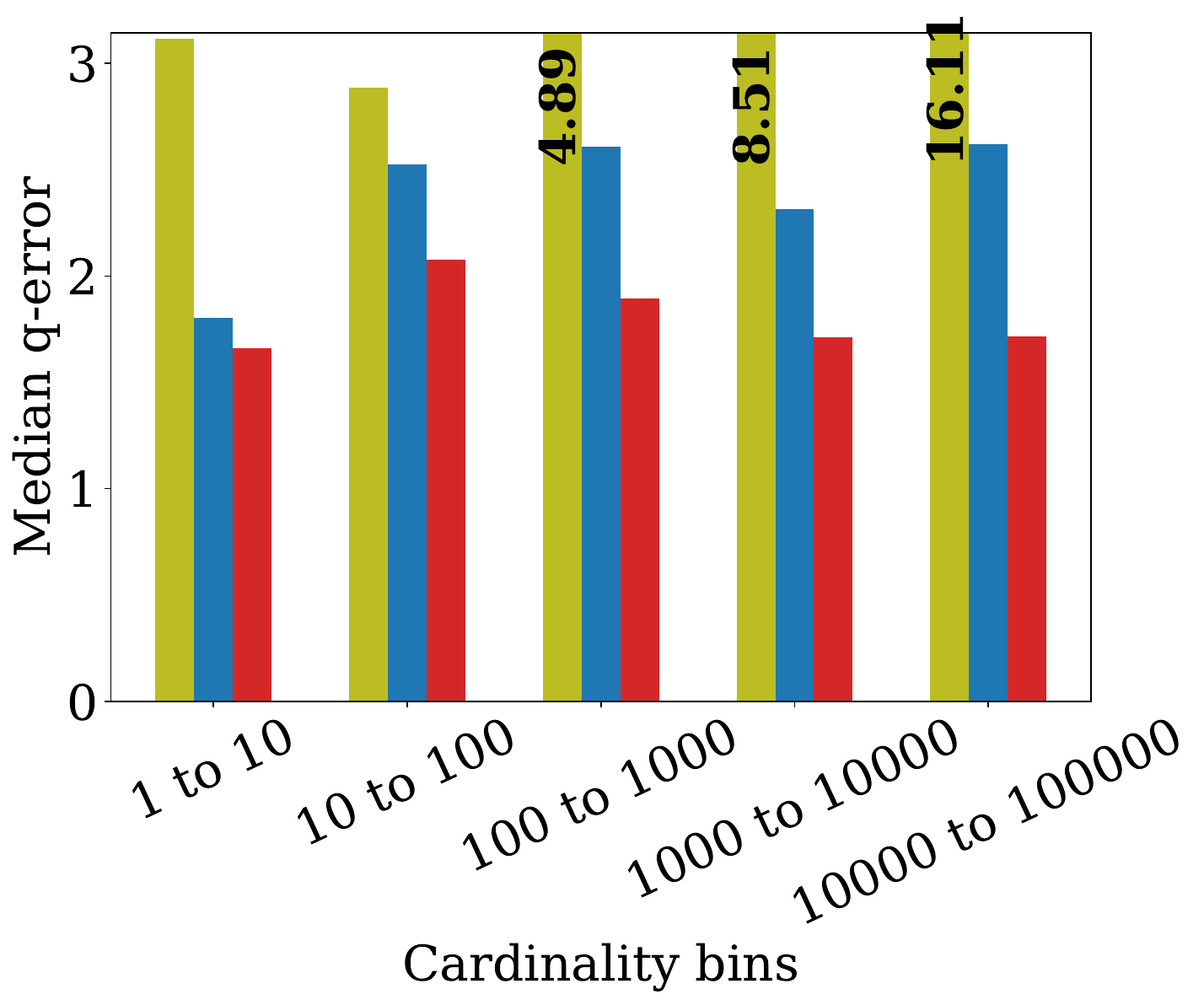}\label{subfig:order_error}}
\subfloat[Class. acc.]{\includegraphics[width=0.2\linewidth]{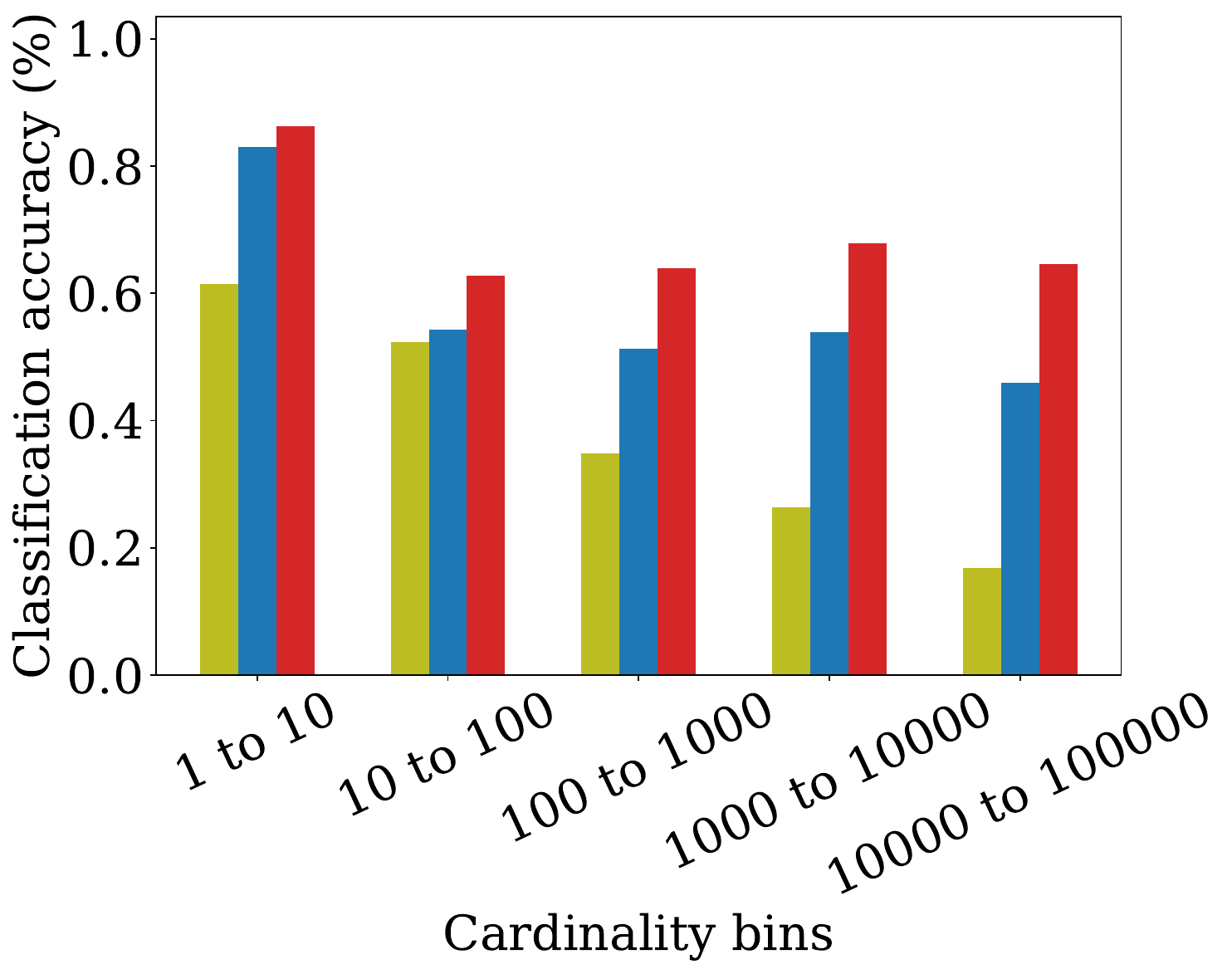}\label{subfig:order_acc}}
\subfloat[\MODELa]{\includegraphics[width=0.2\linewidth]{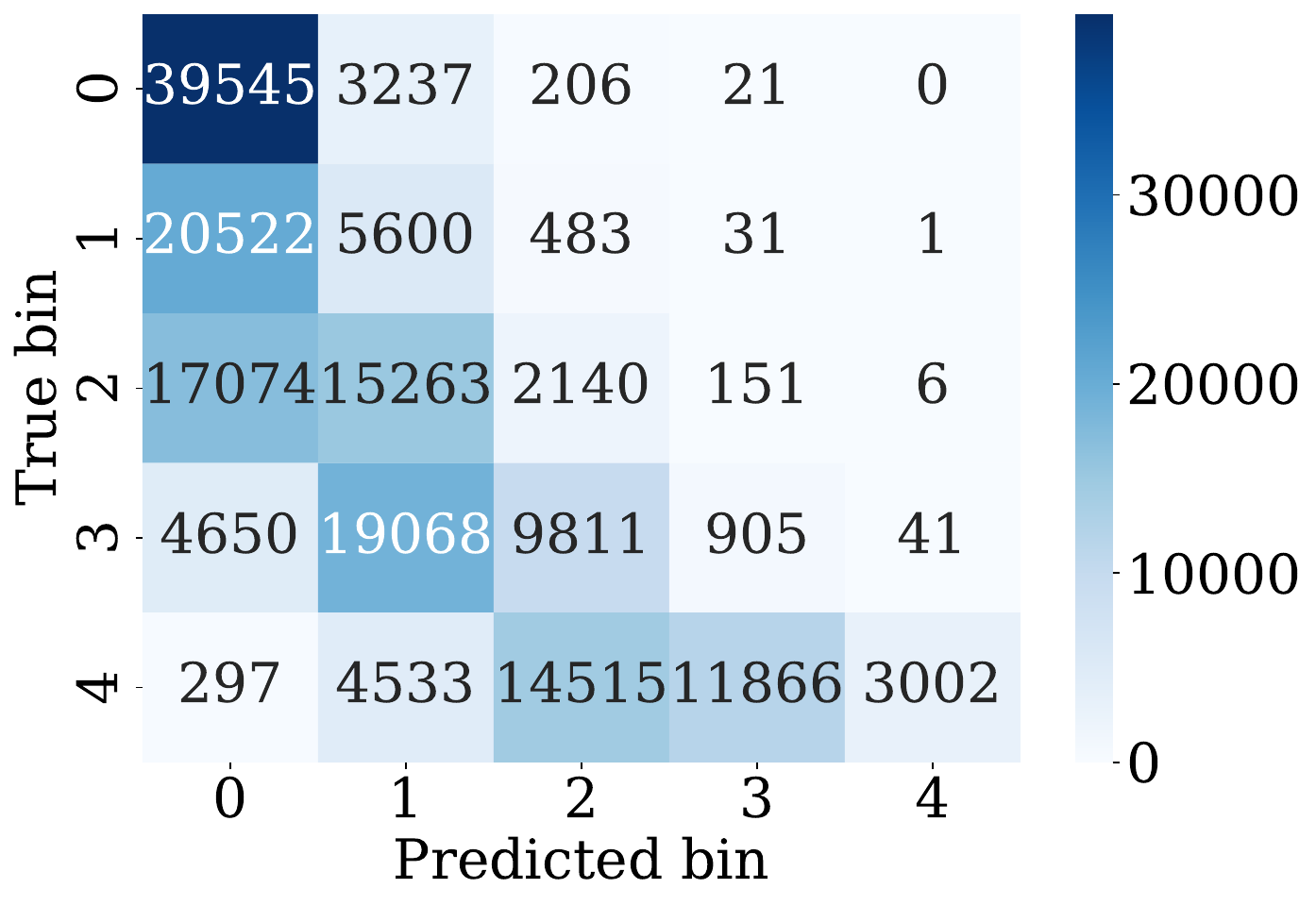}\label{subfig:order_conf_mlp}}
\subfloat[\baseline]{\includegraphics[width=0.2\linewidth]{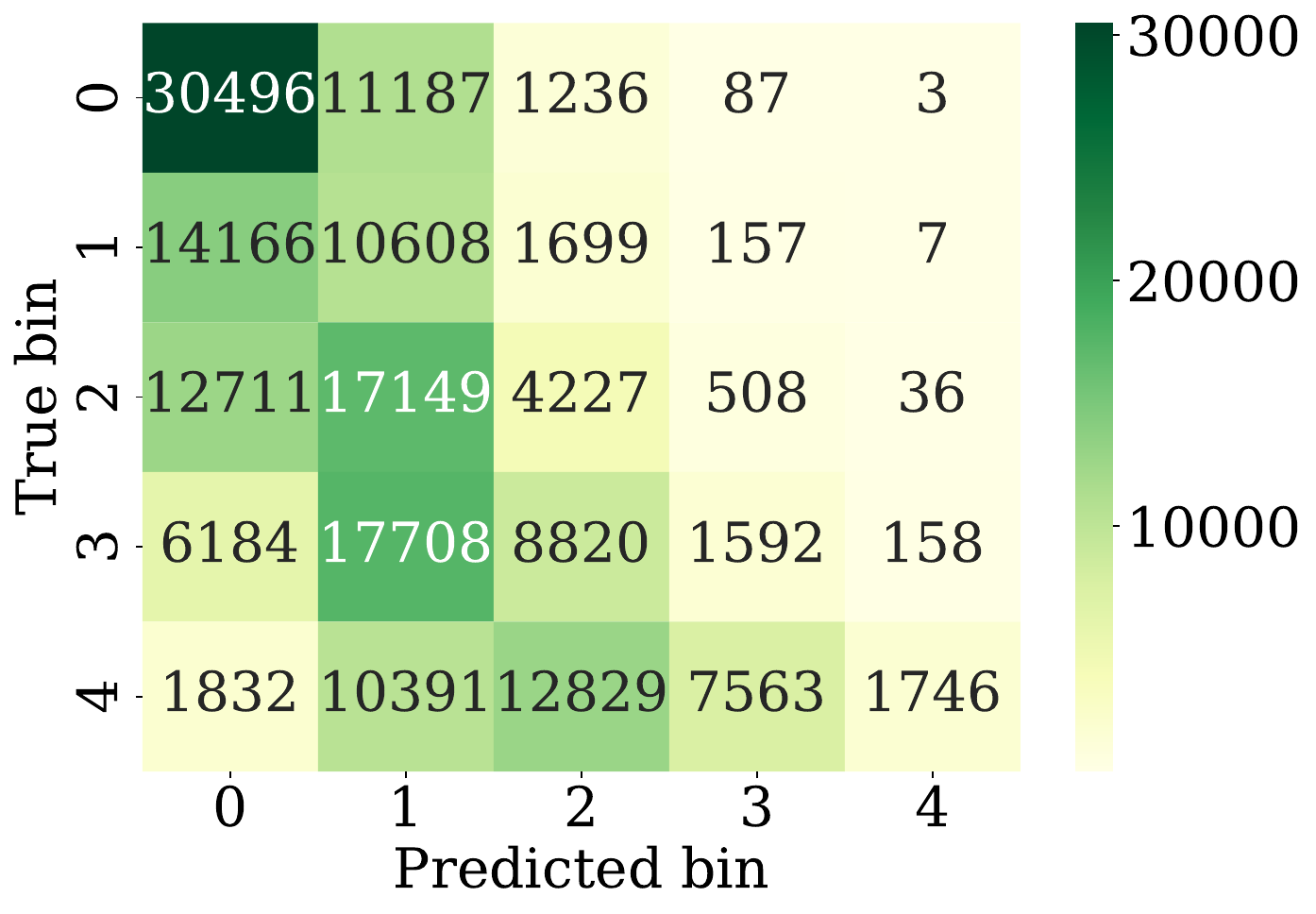}\label{subfig:order_conf_lmkg}}
\subfloat[\MODELb]{\includegraphics[width=0.2\linewidth]{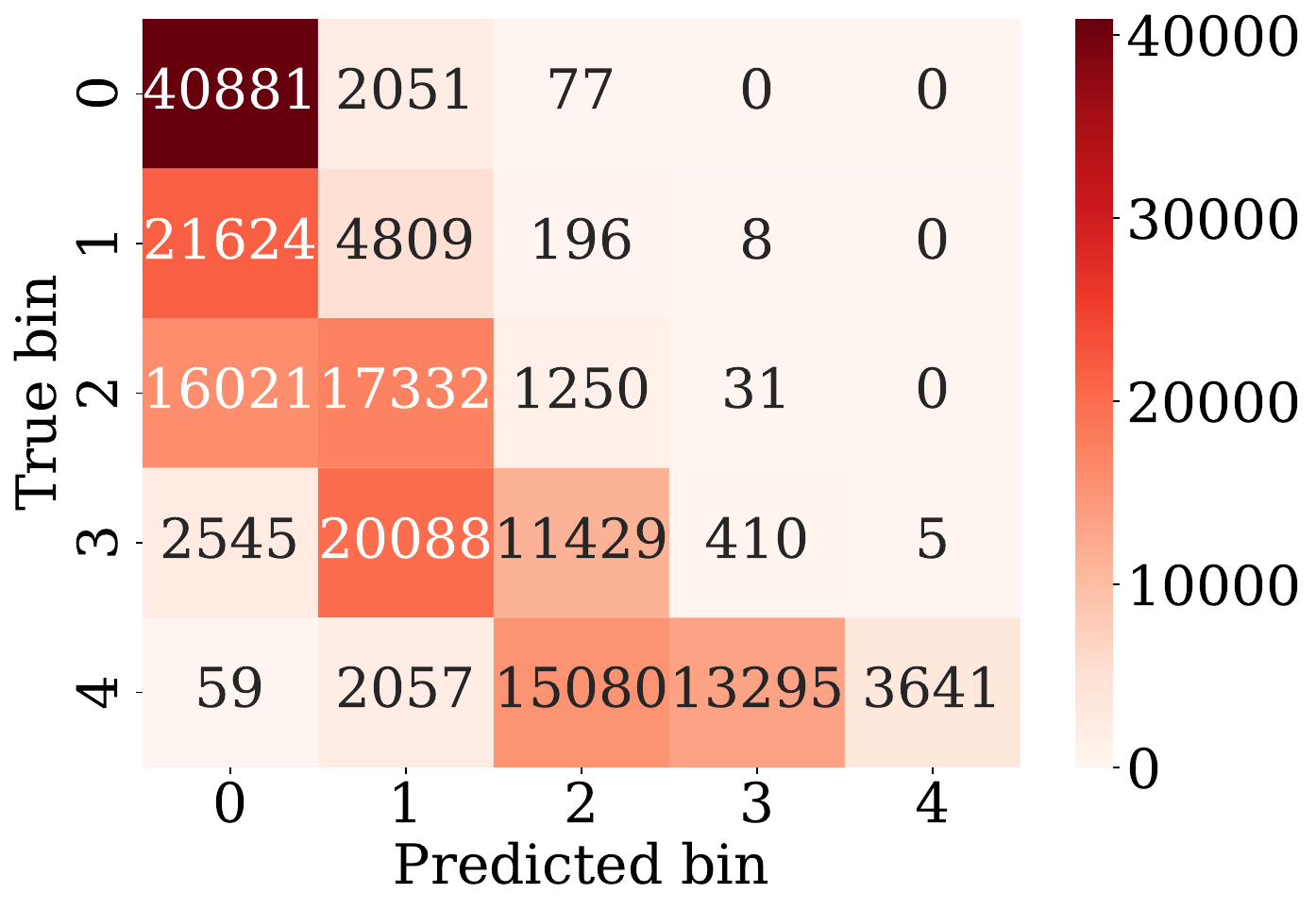}\label{subfig:order_conf_sce}} \\
\vspace{-4pt}
\subfloat[Median q-err.]{\includegraphics[width=0.2\linewidth]{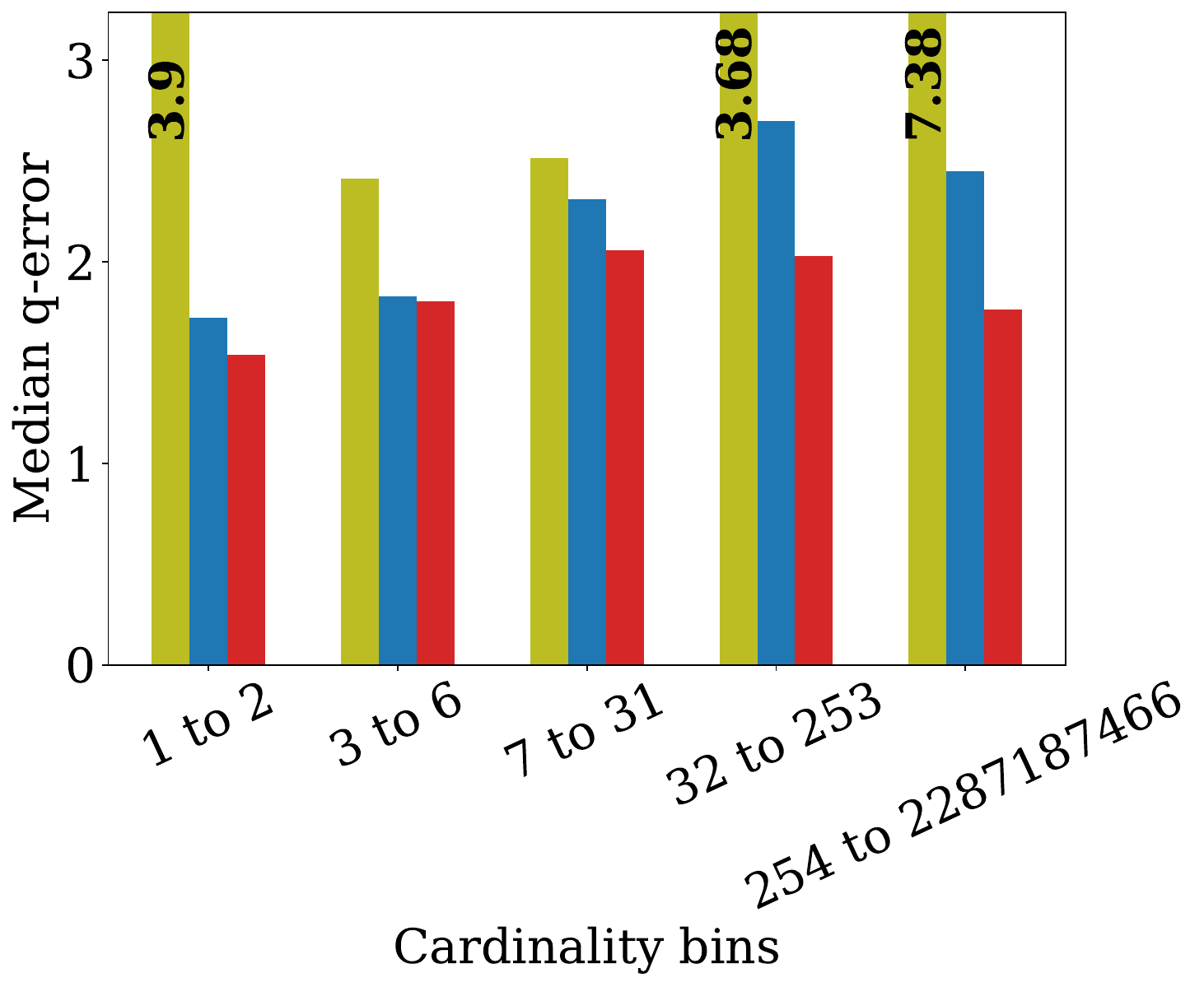}\label{subfig:equi_5_error}}
\subfloat[Class. acc.]{\includegraphics[width=0.2\linewidth]{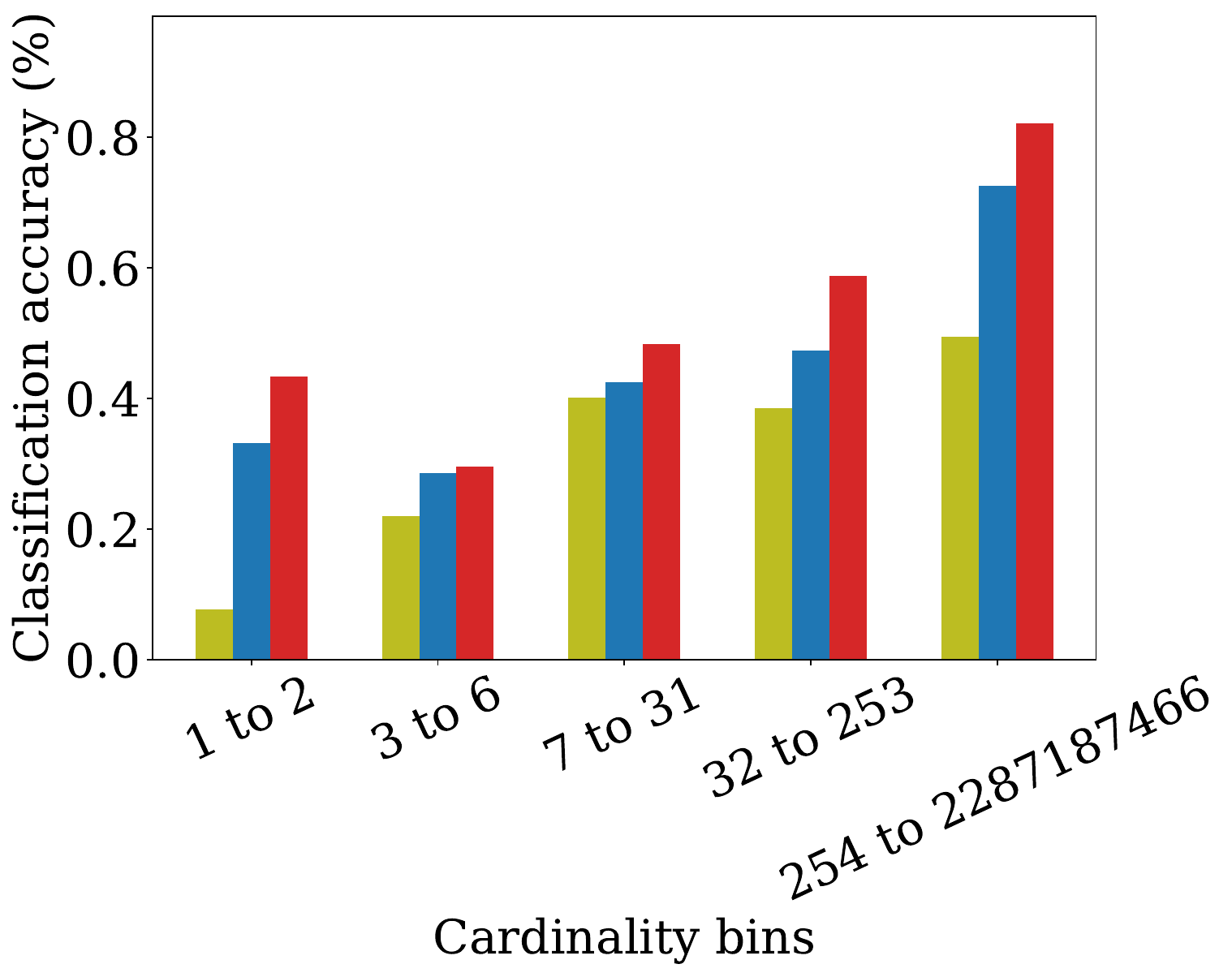}\label{subfig:equi_5_acc}}
\subfloat[\MODELa]{\includegraphics[width=0.2\linewidth]{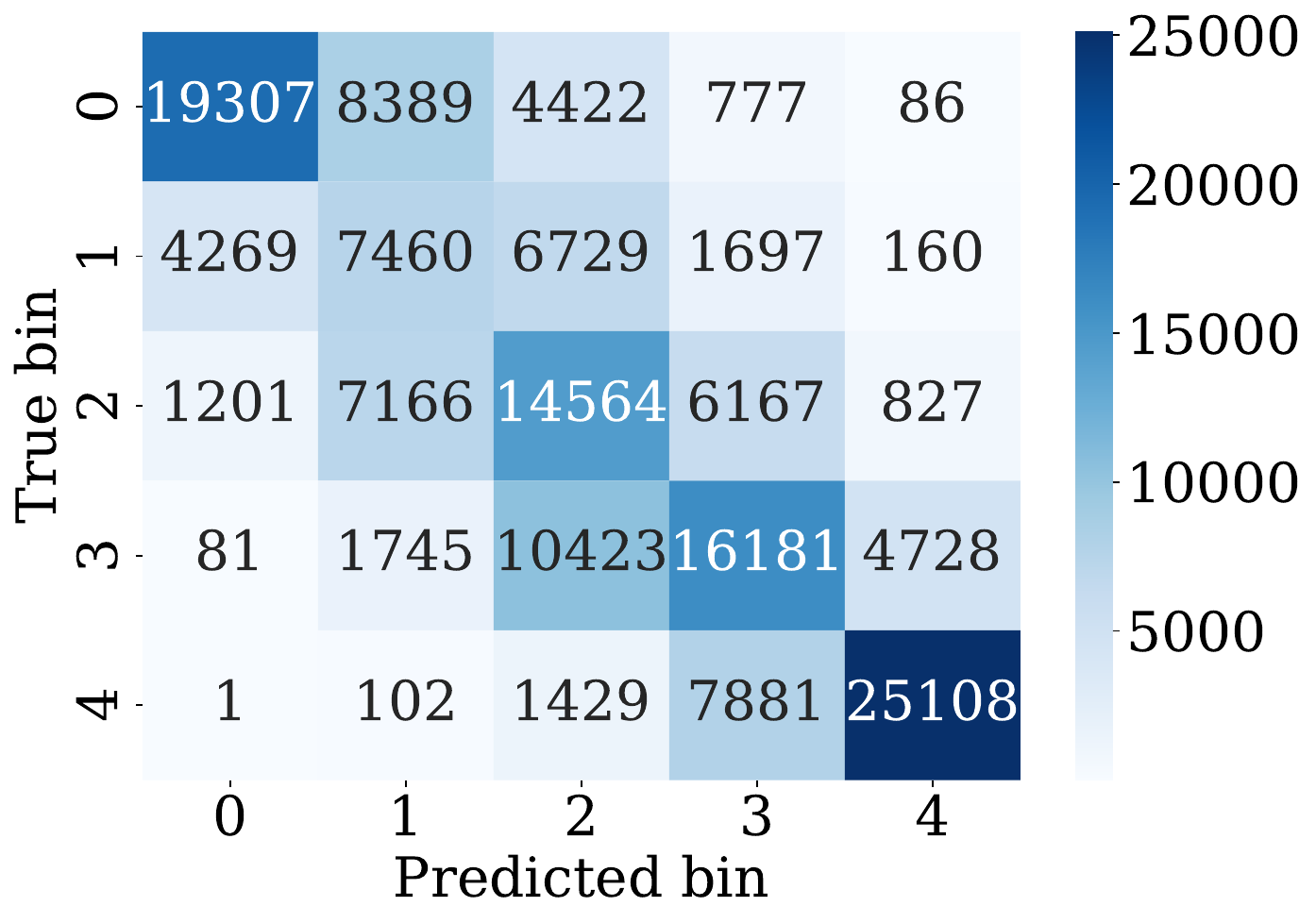}\label{subfig:equi_5_mlp}}
\subfloat[\baseline]{\includegraphics[width=0.2\linewidth]{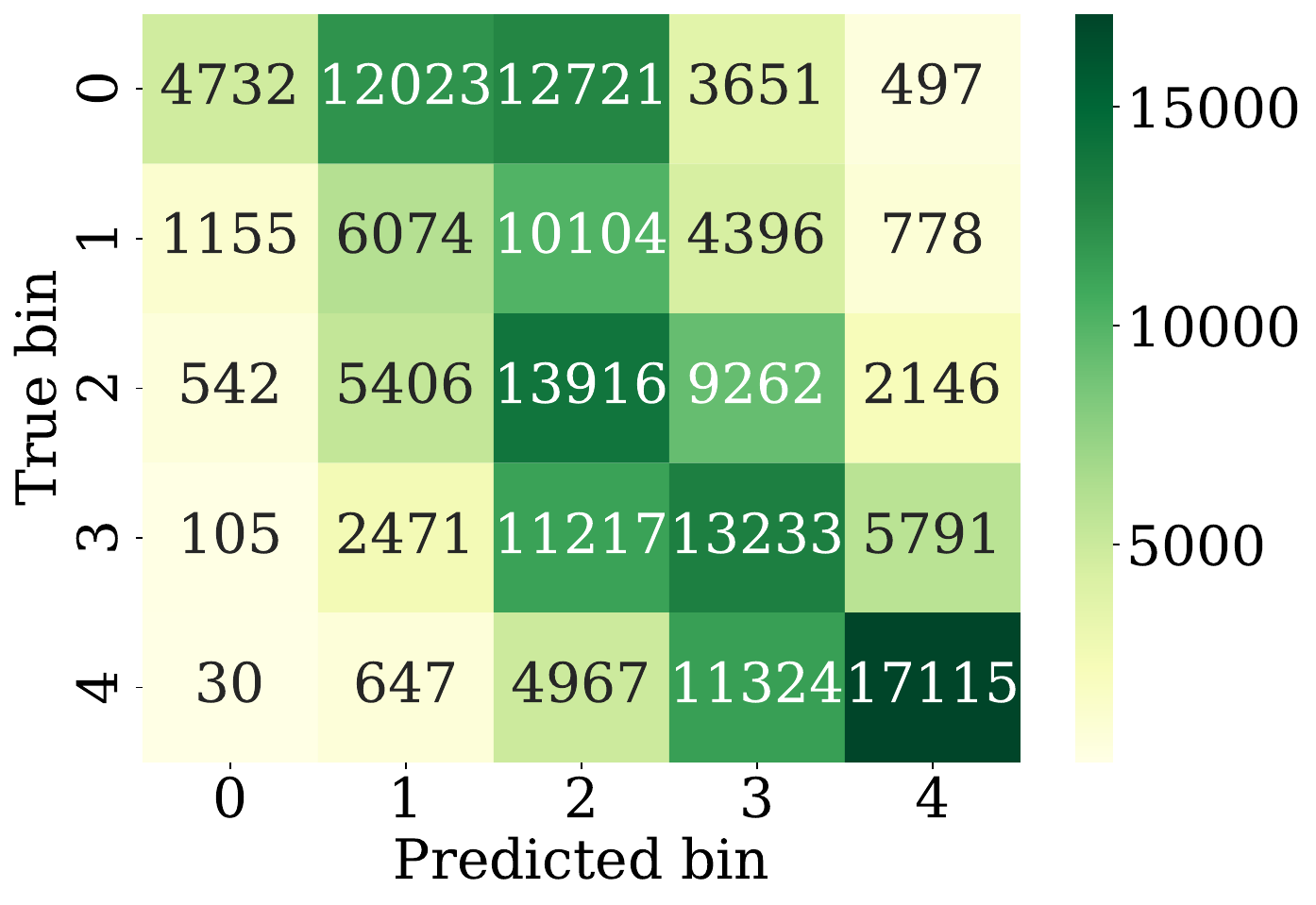}\label{subfig:equi_5_lmkg}}
\subfloat[\MODELb]{\includegraphics[width=0.2\linewidth]{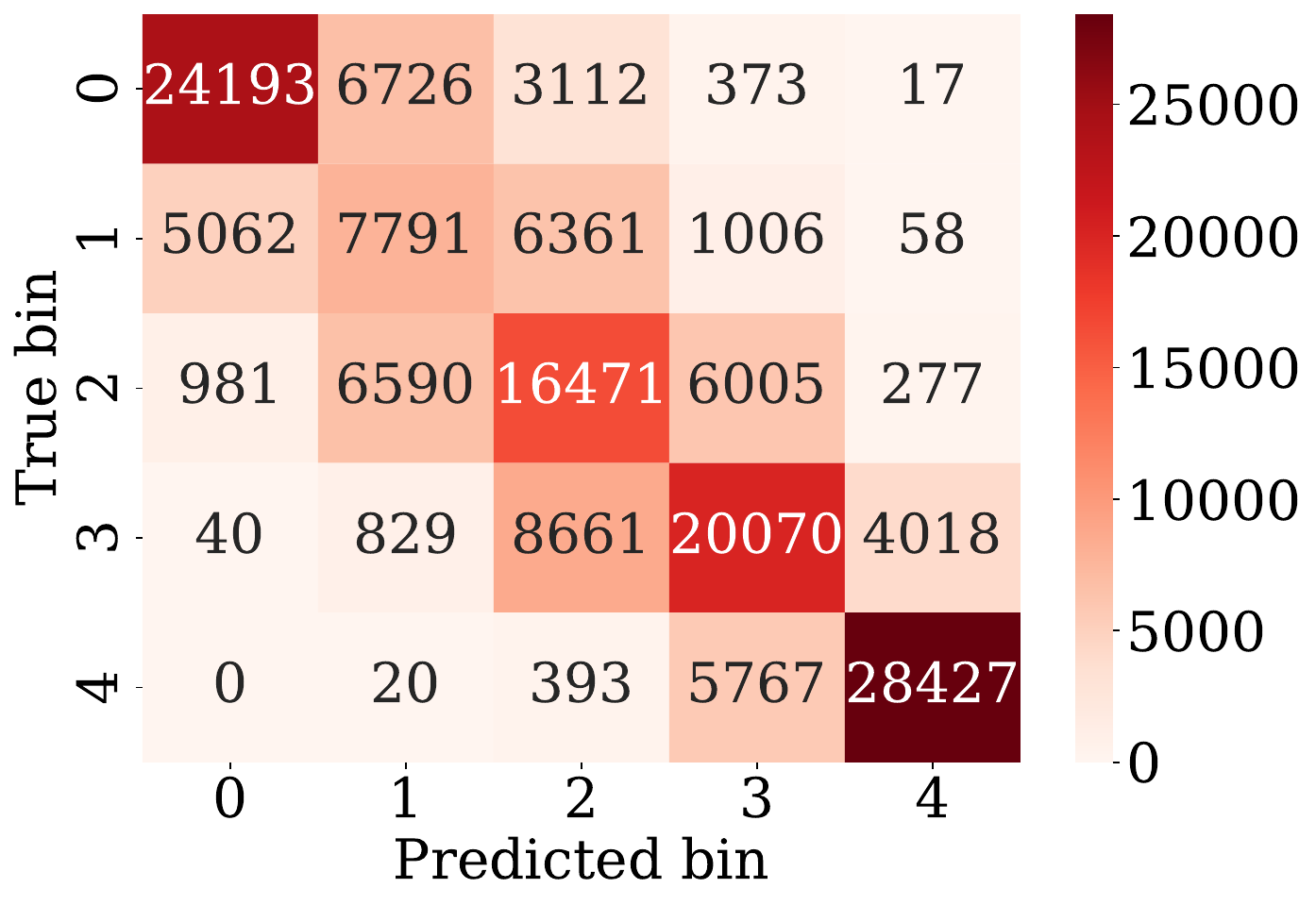}\label{subfig:equi_5_sce}} \\
\vspace{-4pt}
\subfloat[Median q-err.]{\includegraphics[width=0.2\linewidth]{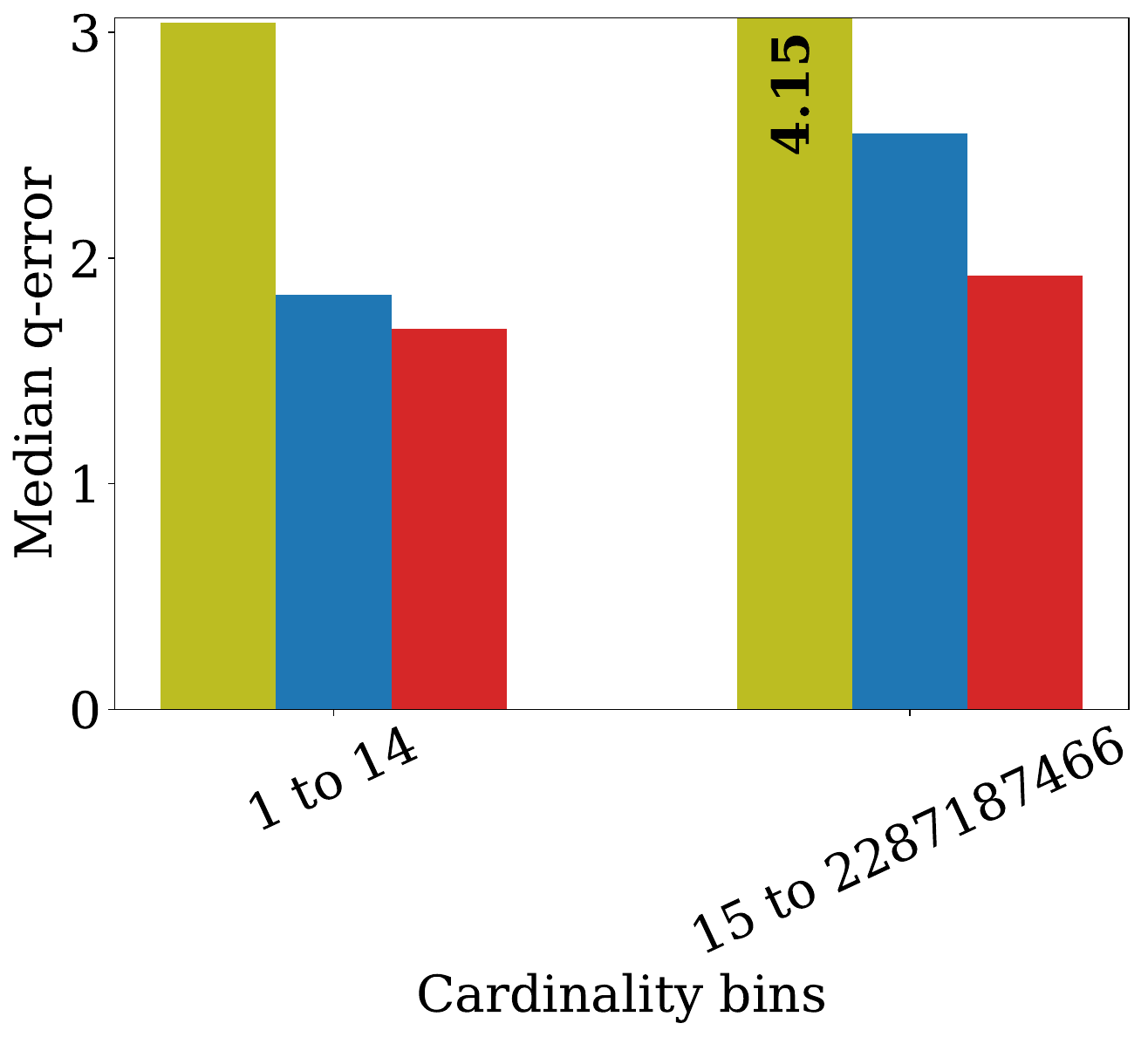}\label{subfig:equi_2_error}}
\subfloat[Class. acc.]{\includegraphics[width=0.2\linewidth]{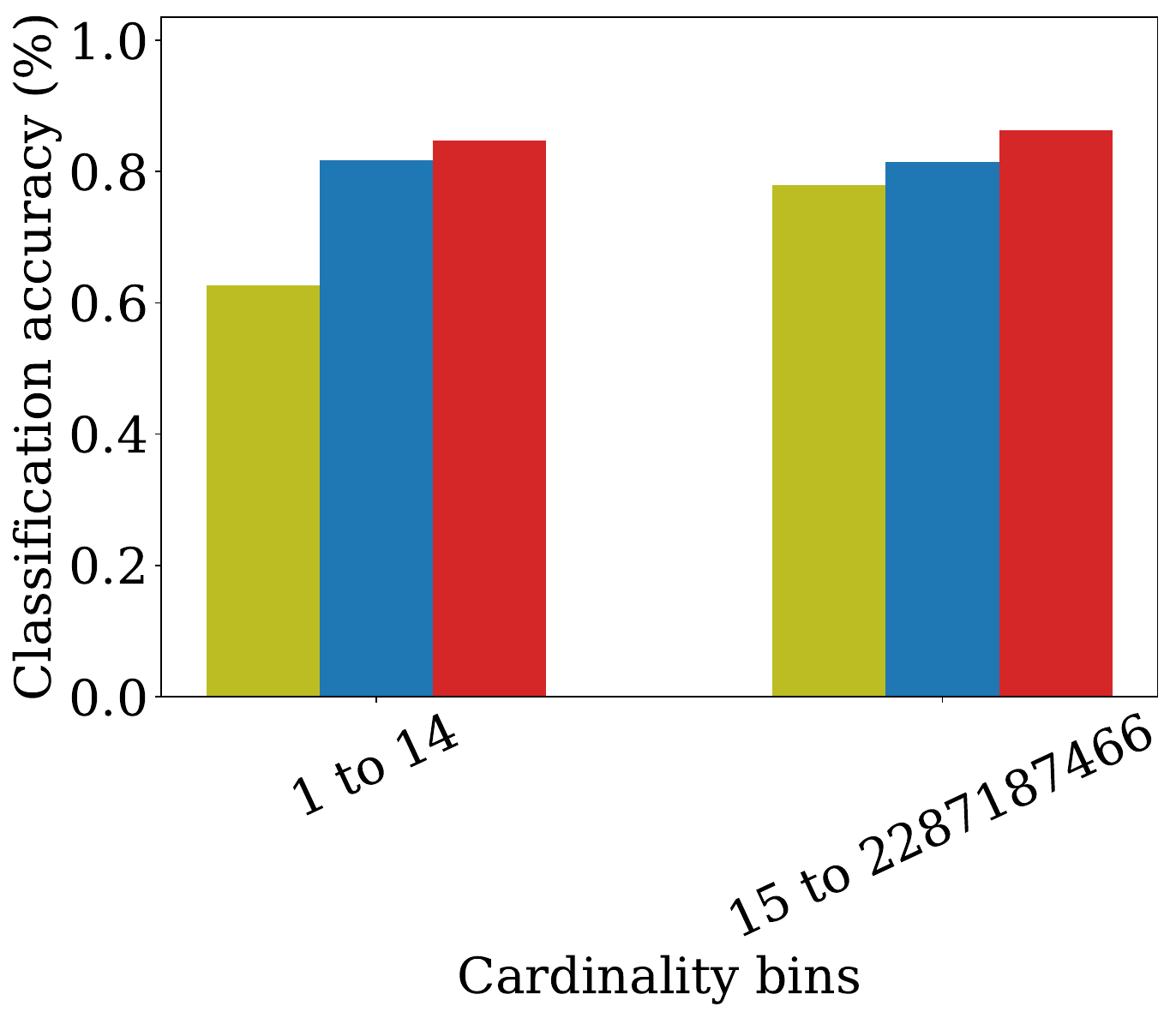}\label{subfig:equi_2_acc}}
\subfloat[\MODELa]{\includegraphics[width=0.2\linewidth]{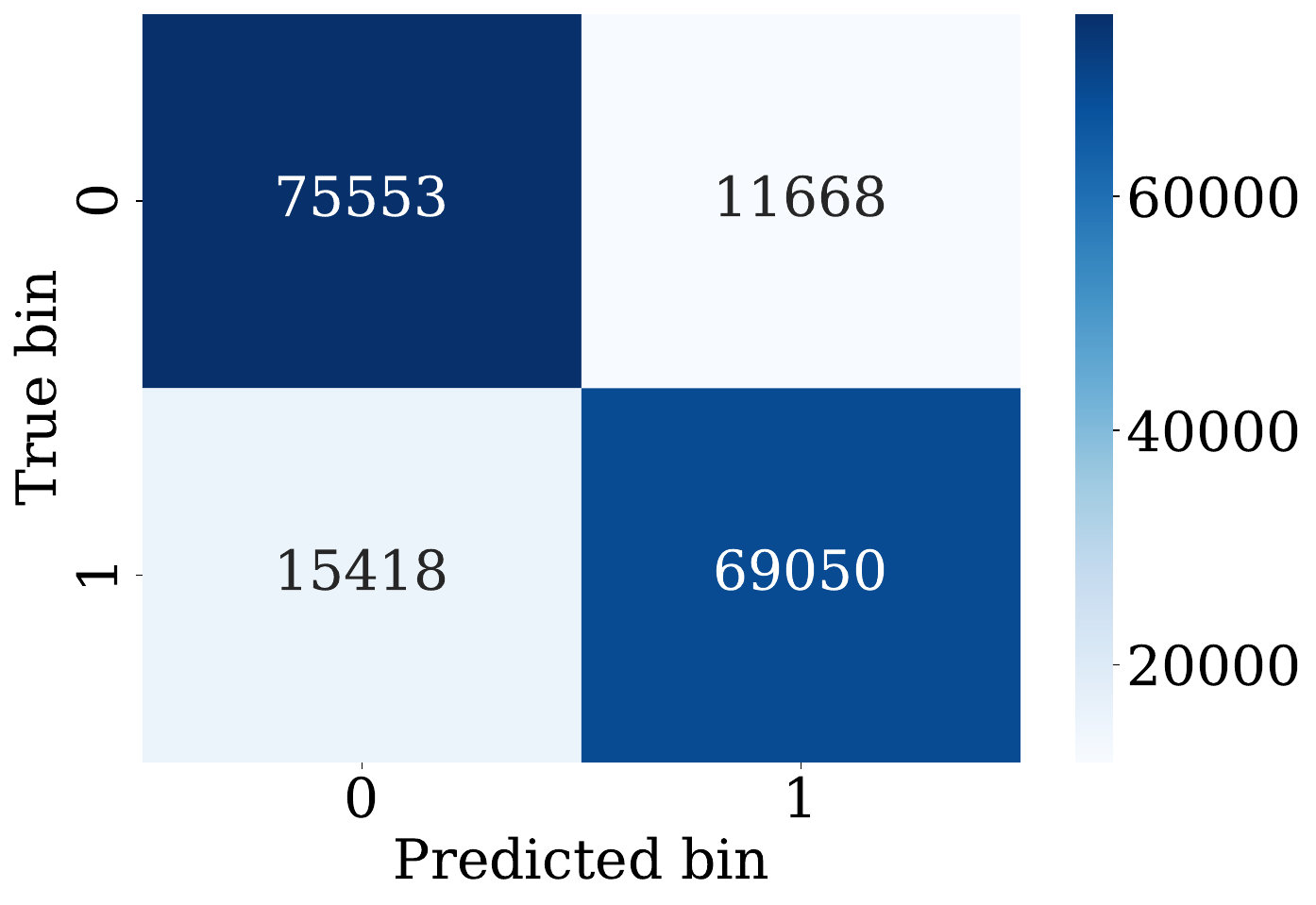}\label{subfig:equi_2_mlp}}
\subfloat[\baseline]{\includegraphics[width=0.2\linewidth]{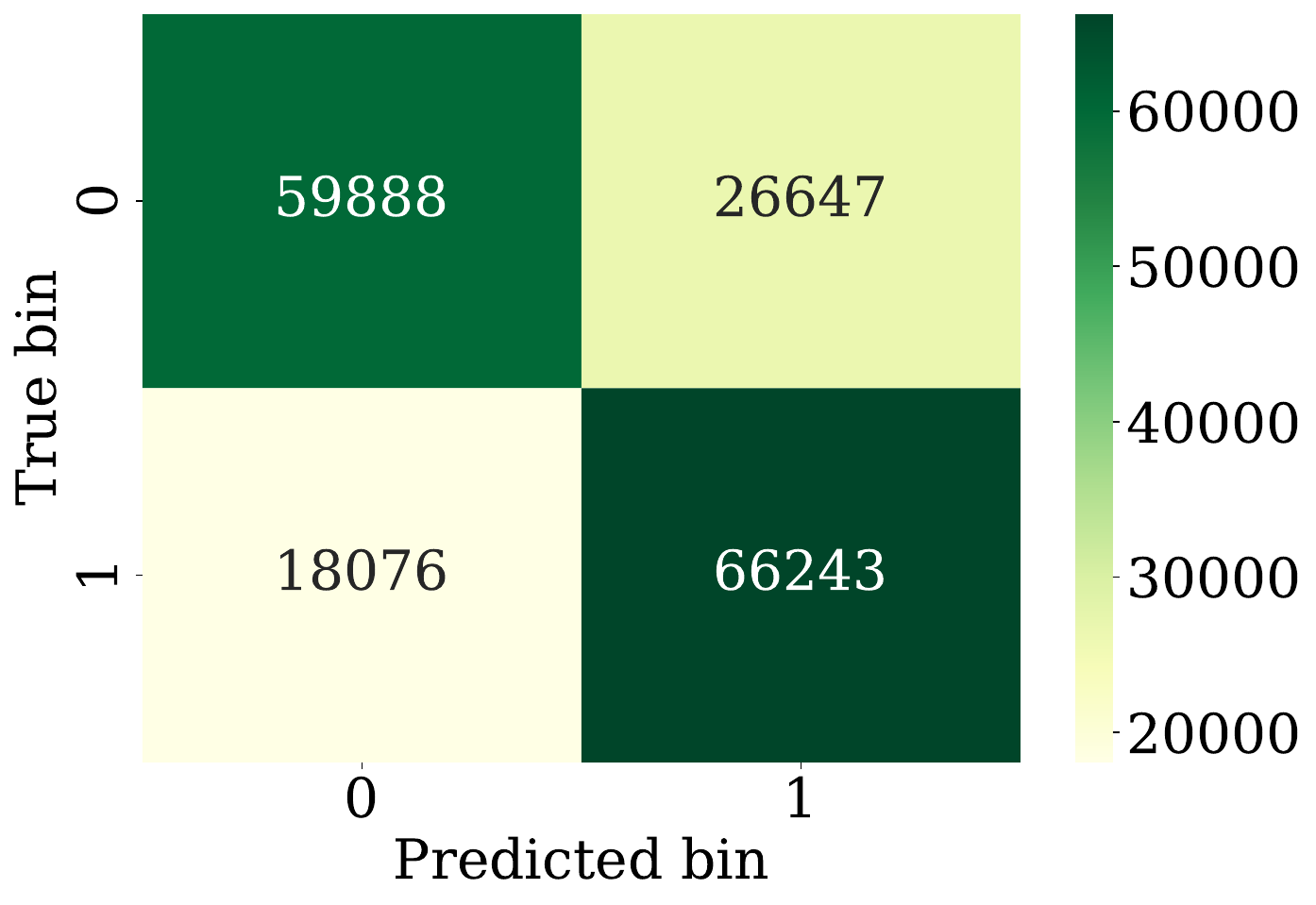}\label{subfig:equi_2_lmkg}}
\subfloat[\MODELb]{\includegraphics[width=0.2\linewidth]{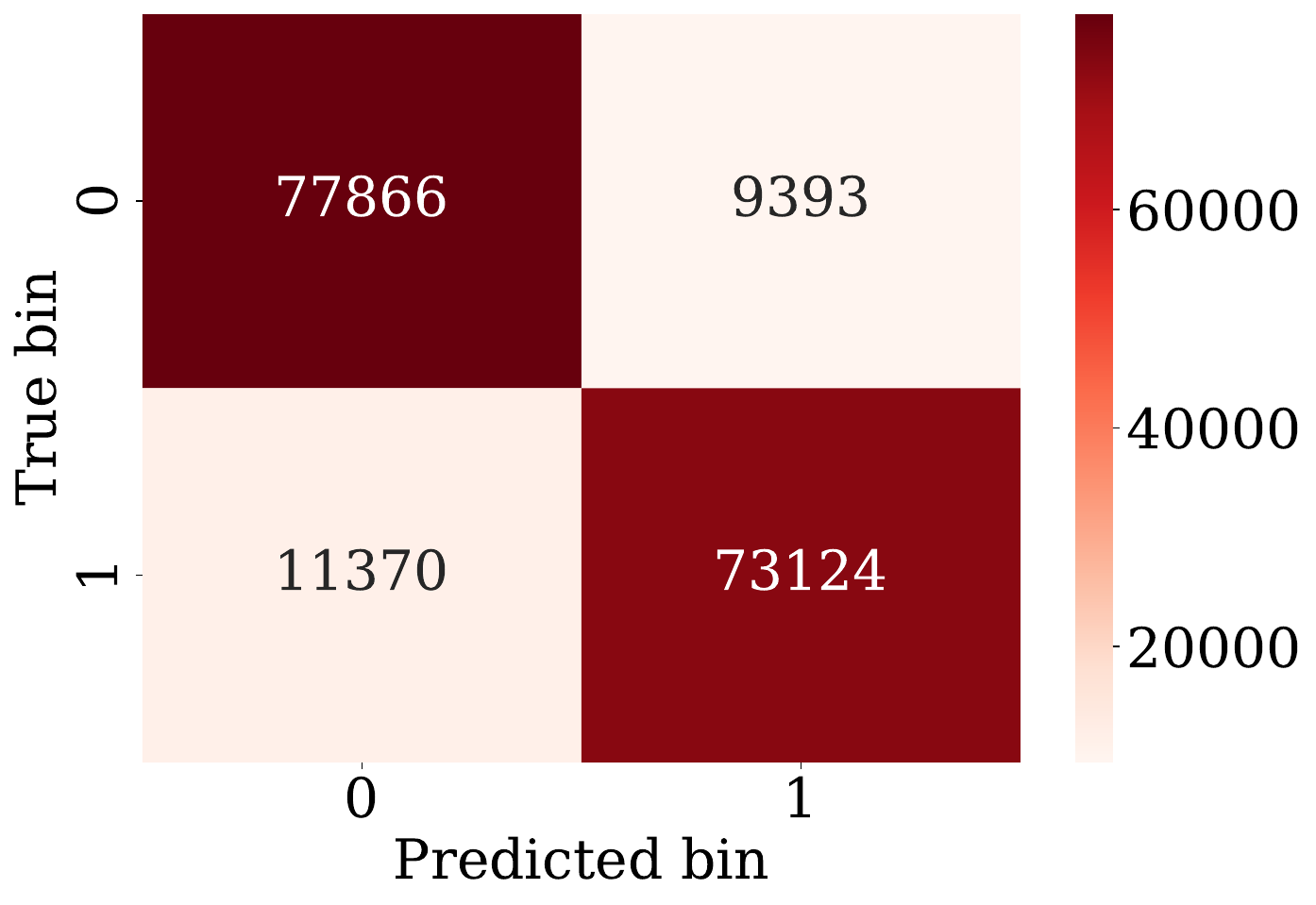}\label{subfig:equi_2_sce}}
\caption{(a-e) Order-of-magnitude binning approach; (f-j) equi-sized binning approach for 5 bins; (k-o) equi-sized binning approach for 2 bins}
\label{fig:equi_bins}
\vspace{-8pt}
\end{figure*}
}

\paragraph{Datasets}

We employ three real-world datasets (SWDF, Yelp and Wikidata) and a synthetic one (LUBM):
SWDF (Semantic Web Dog Food)~\cite{moller2007recipes}, a benchmark linked dataset of academic entities and relations (e.g., people, publications, organizations), built from Scholarlydata\footnote{\url{http://www.scholarlydata.org}} dump files;
Yelp\footnote{\url{https://www.yelp.com/dataset}}, an open U.S. business dataset including users, check-ins, reviews, and business metadata (e.g., state, city), where user reviews were omitted due to their length and the prohibitive query-generation cost driven mainly by Levenshtein distance calculations;
Wikidata\footnote{\url{https://www.wikidata.org/wiki/Wikidata:Main_Page}}, a large open knowledge base hosted by the Wikimedia Foundation, for which we used the WDBench~\cite{angles2022wdbench} benchmark and restricted training to a UK-people subgraph of approximately 160K nodes to control scale;
and LUBM (Lehigh University Benchmark)~\cite{guo2005lubm}, a synthetic dataset based on the Univ-Bench ontology, generated using their official tool\footnote{\url{http://swat.cse.lehigh.edu/projects/lubm/}} for 20 academic organizations.

For each dataset, we build a knowledge graph using the \textit{neo4j}\footnote{\url{https://neo4j.com}} graph database system.
The training samples are then generated as described in Section~\ref{subsec:training_creation}, using the Levenshtein distance as a similarity measure, which is supported by the neo4j \textit{APOC}\footnote{\url{https://neo4j.com/developer/neo4j-apoc/}} library.
Note that the choice of a similarity measure is orthogonal to how our methods operate; any similarity measure suitable for graphs would work. Each query element contains zero or more unbound terms (i.e., denoted by `$*$'), where their number and placement are picked randomly. Before training, each training sample is shuffled and  split into training and testing datasets. Table~\ref{tab:datasets} lists, for each dataset, the number of nodes, relationships, and triples in the initial graph, as well as the size of the corresponding training samples.

\paragraph{Methods}
We evaluate the following methods: 
(a)~\baseline: As it does not support similarity predicates, for each training element we remove the similarities and treat it as a graph query without similarity predicates.
\mnote{(b)~\MODELa: Our MLP based technique that employs the extended SG encoding.}
(c)~\MODELb: Our approach that treats similarities as first-class citizens, employing the \SGSPLIT encoding.

\paragraph{Parameters}
Table~\ref{tab:parameters} lists 
the parameters used in our experimental evaluation. We conducted 
tuning tests for each one to determine their ranges and default values, 
which are shown in bold in the table.
We selected star similarity queries as the default because the datasets lack long enough paths for chain queries.
To showcase the advantage of \MODELb in queries with a larger number of joins, we selected $5$ joins as default for star similarity queries and $2$ joins for chain similarity queries. 

As \MODELa does not cope well with clustering (discussed below), we report results with $1$ cluster as the default.

In several cases in query optimization, especially when the qualifying results are too many, the exact cardinality value is hard to be detected; it suffices, though, for the estimator to at least produce a value of the same order of magnitude as the actual result.
Hence, we consider two types of cardinality binning and treat the problem as a classification task. Specifically, the binning types are (i) \textit{equi-sized}, where each bin contains approximately the same number of cardinalities, and (ii) \textit{order-of-magnitude}, where the cardinalities differ by an order-of-magnitude.
Unless otherwise stated, the equi-sized binning method is used.

All the ML models used in this work are MLPs. 
For performing a fair comparison, we tried to configure them as even as possible.
Specifically, each MLP consists of four hidden layers, with $256$ neurons each (except the input and the output layer). We use the ReLU activation function for each layer except the output, where we use sigmoid. We use q-loss, which is based on \textit{q}-error, for a loss function. 
Finally, we used $50$ training epochs as default, a number that we found achieves a good estimation in our experiments.

\paragraph{LLM-generated test sets}
We explore two different LLM-generated test sets. Both are produced by prompting an LLM with a short natural-language description and asking it to phrase realistic multi-hop questions, each with its relations and a per-fact confidence level; the outputs are then mapped to the canonical $p_i\textnormal{-}s_i\textnormal{-}sim_i$ format and validated against the graph. The core part of the prompt is the following:

\scriptsize
\begin{verbatim}
You are a person searching a knowledge graph that catalogs a university system. Imagine you are a
real person -- a student looking up an advisor, a curious outsider browsing -- what multi-step 
questions would you ask? Produce one line per query: <NL query> | r1|r2|r3 | c1|c2|c3, where c_i
in {exact, near, loose, vague} = your confidence in the name spelling.
\end{verbatim}
\normalsize

The first test set is generated with the same LLM used for the training data (gemini-2.5-flash), but through the above natural-language prompt rather than the canonical-format one of Section~\ref{subsec:llm_training}; it thus probes a different query distribution while keeping the model fixed ($1{,}707$ queries). To rule out that any improvement is a product of training and testing with the same model, and to verify that the benefit generalises across LLM families, we employ a different LLM (gpt-4o) for a second test set ($1{,}404$ queries).

\paragraph{Similarity effect on cardinality}
\label{par:selectivity}

In similarity queries, the selectivity relates to the similarity provided for each triple. This motivates the need to employ a different approach than existing, non-similarity-aware methods such as \baseline. If the resulting cardinalities change by relaxing the similarity threshold, it becomes inherently harder to perform estimation solely based on node and predicate IDs. Indicatively, Figure~\ref{fig:sim_effect} shows the behavior of cardinality with varying similarity for a $1$-star query on the SWDF dataset. 
As expected, relaxing the similarity threshold causes the cardinality to increase exponentially, which justifies approaching similarity-aware cardinality estimation as a different problem.

\subsection{Similarity-Aware Cardinality Estimation}

\subsubsection{Cardinality estimation without similarity predicates}
\label{subsubsec:no_sims}

\begin{figure}[t]
    \centering
    \subfloat{\includegraphics[width=0.4\linewidth]{Figures/plots/legend.png}}
    \\
    \vspace{-12pt}
    \setcounter{subfigure}{0}
    \subfloat[Star (\textit{q}-error)]{\includegraphics[width=0.25\linewidth]{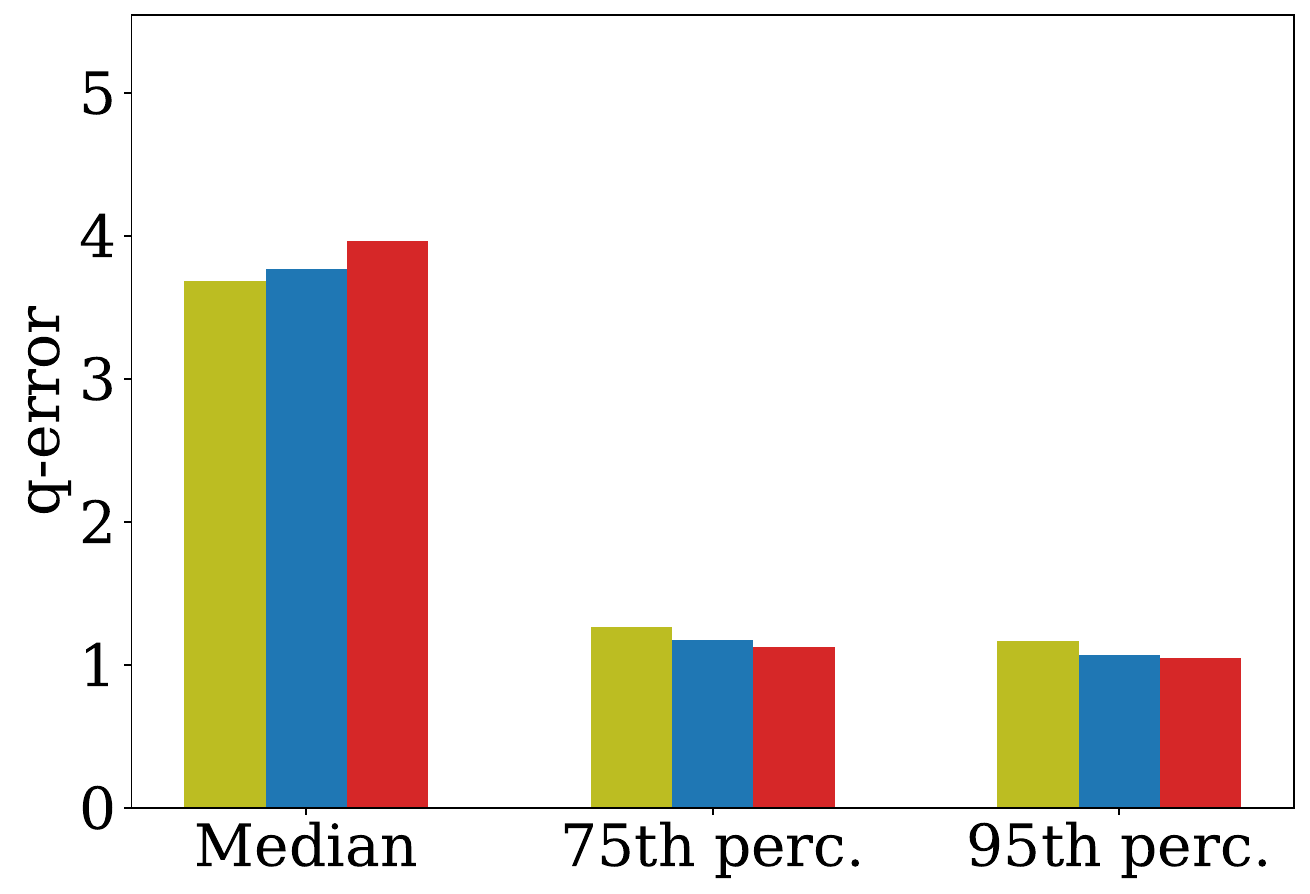}\label{subfig:star_lmkg}}
    \subfloat[Chain (\textit{q}-error)]{\includegraphics[width=0.25\linewidth]{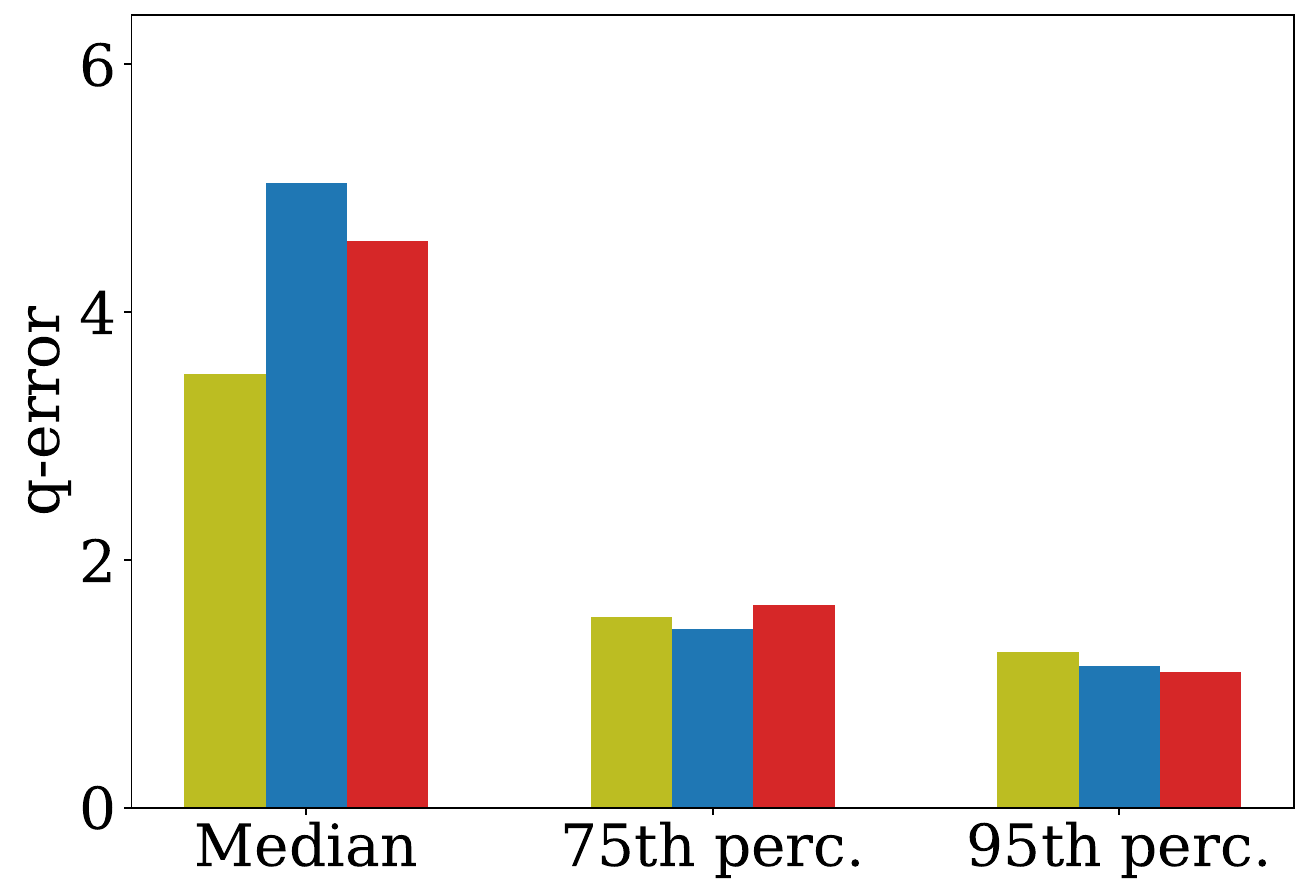}\label{subfig:chain_lmkg}}
    \subfloat[Star (accuracy)]{\includegraphics[width=0.25\linewidth]{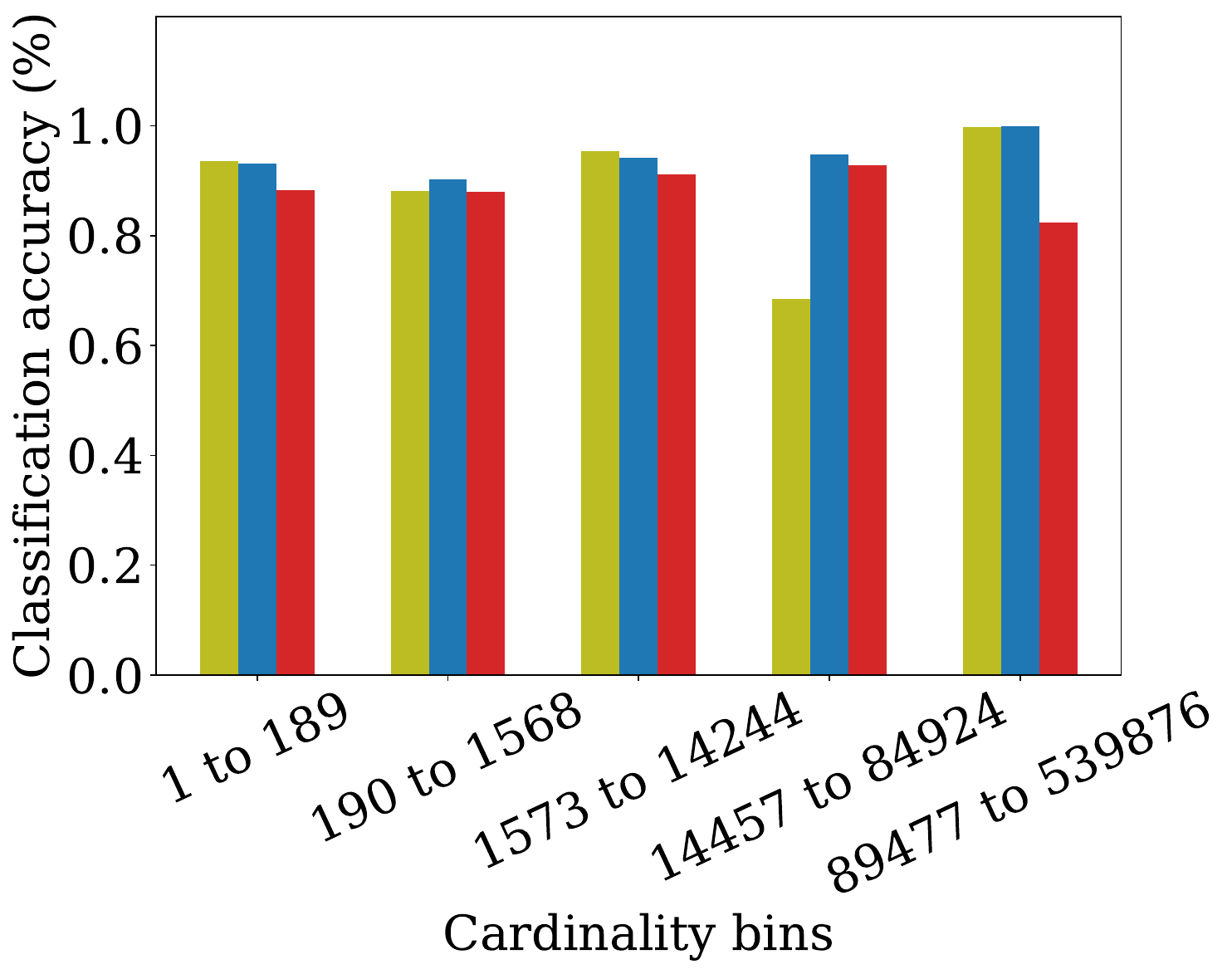}\label{subfig:star_lmkg_class}}
    \subfloat[Chain (accuracy)]{\includegraphics[width=0.25\linewidth]{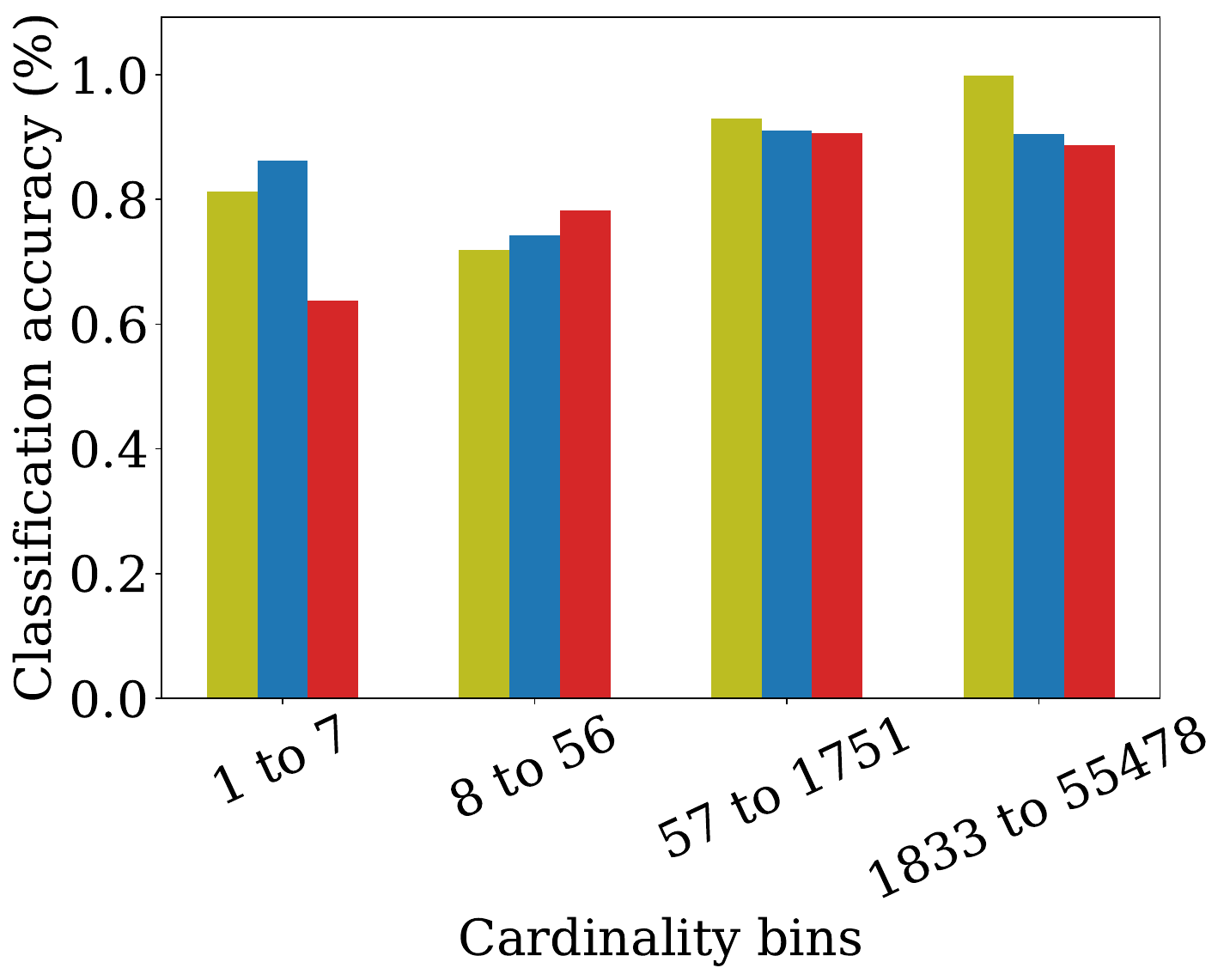}\label{subfig:chain_lmkg_class}}
    \vspace{-6pt}

    \caption{Cardinality estimation without similarity predicates: median, 75th and 95th percentile \textit{q}-errors}
    \label{fig:median_exact}
    \vspace{-6pt}
\end{figure}

A natural question that arises regarding the applicability of our approach 
is whether it can be applied for cardinality estimation of queries without similarity predicates. To test this, we generate training sets by only issuing queries where similarity equals $1$. Figure~\ref{fig:median_exact} shows the results.

The performance of \MODELb and \MODELa is similar to the one of \baseline for both star and chain similarity queries with minimal differences; see Figures~\ref{subfig:star_lmkg}, \ref{subfig:star_lmkg_class} and \ref{subfig:chain_lmkg}, \ref{subfig:chain_lmkg_class} respectively.

In other words, in the scenario that the related work excels (similarity=$1$), our methods are at least as good as the state of the art. In what follows, we show that our methods provide excellent results for a scenario (similarity$<=$1) that the related work has not dealt with so far.

\subsubsection{Effect of cardinality binning}
Depending on the application domain, the provided cardinality estimation can be evaluated separately for a cardinality interval of interest.

In a case where the cardinality of the answered queries is skewed to the left (i.e., towards $1$), it is preferable to assess the error and accuracy of the estimation by considering cardinality bins whose range increases in an order-of-magnitude manner. Figure~\ref{subfig:order_error} depicts the median \textit{q}-error for this case using the Wikidata dataset, which is rather heavily skewed towards cardinality $1$. Notice that, in all cases, \MODELb achieves the lowest median \textit{q}-error
(up to $40\%$ lower than \MODELa),
while \baseline cannot cope well due to its inability to consider the provided similarities (i.e., $8$x larger \textit{q}-error in the worst case). Another way to evaluate the quality of the estimation is by treating the problem as a classification task, where the classes are a set of selected bins on the cardinality axis. Figure~\ref{subfig:order_acc} depicts this accuracy for bins of order-of-magnitude-increasing size. As expected, the results are similar here, with \baseline yielding the worst accuracy. \MODELb is more accurate than \MODELa in all cases, reaching more than $80\%$ accuracy in some cases. Notice that, due to the skewness of the Wikidata dataset, the accuracy tends to be higher for cardinalities closer to $1$, since there are more training data for that specific range of cardinalities. Figures~\ref{subfig:order_conf_sce} \ref{subfig:order_conf_mlp} and \ref{subfig:order_conf_lmkg} illustrate the corresponding confusion matrices for the classification task.

Another way to generate the cardinality bins is by considering intervals containing approximately equal numbers of elements. This way of binning would be preferable for less skewed datasets. Figures~\ref{subfig:equi_5_error} and \ref{subfig:equi_5_acc} depict the \textit{q}-error and corresponding accuracies for a set of $5$ equi-sized bins. Notice that, in this case, even though the \textit{q}-error is again rather low in all cases (except \baseline), the classification accuracy is significantly worse due to the left-wise skewness of the dataset that causes the first three bins' intervals to be rather narrow. Consequently, a miss that may yield low \textit{q}-error, for example, estimating cardinality 3 instead of 2, might lead to falsely classifying the element and thus to lower accuracy. This can be verified by the corresponding confusion matrices in all cases (Figures~\ref{subfig:equi_5_sce}, \ref{subfig:equi_5_mlp} and \ref{subfig:equi_5_lmkg}) where the falsely predicted cardinalities tend to be for a neighboring bin.

Figures~\ref{subfig:equi_2_error} and \ref{subfig:equi_2_acc} show the results for a
commonly considered case
involving 2 equi-sized bins (i.e., binary classification). The classification accuracy is much better here for both classes, with \MODELb demonstrating the higher numbers. This can also be verified by the corresponding confusion matrices, depicted in Figures~\ref{subfig:equi_2_sce}, \ref{subfig:equi_2_mlp} and \ref{subfig:equi_2_lmkg}.

\begin{figure*}[!ht]
\centering
\subfloat{\includegraphics[width=0.35\linewidth]{Figures/plots/legend.png}}
\\
\vspace{-12pt}
\setcounter{subfigure}{0}
\subfloat[Wikidata]{\includegraphics[width=0.25\linewidth]{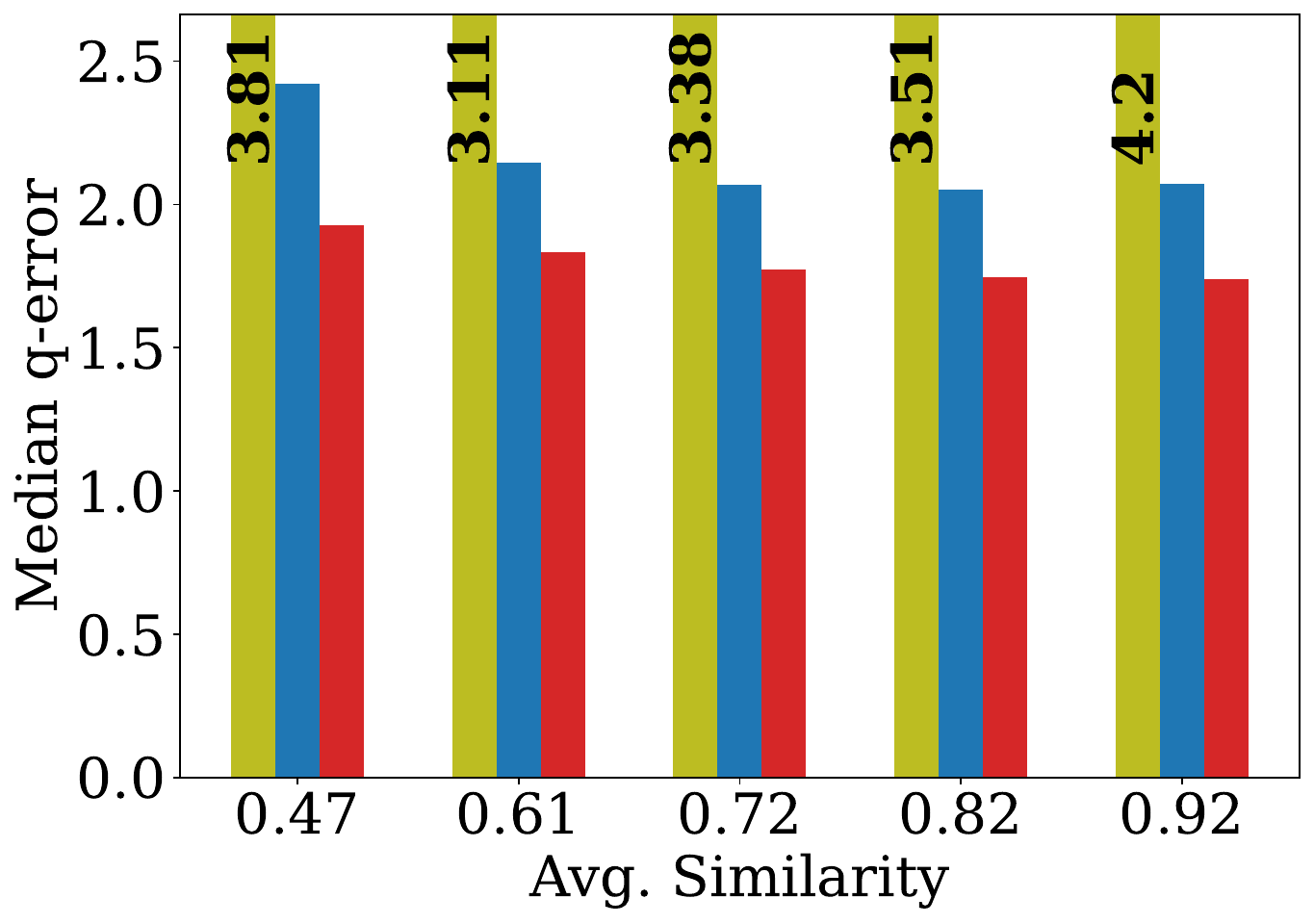}\label{subfig:wiki_selectivity}}
\subfloat[Yelp]{\includegraphics[width=0.25\linewidth]{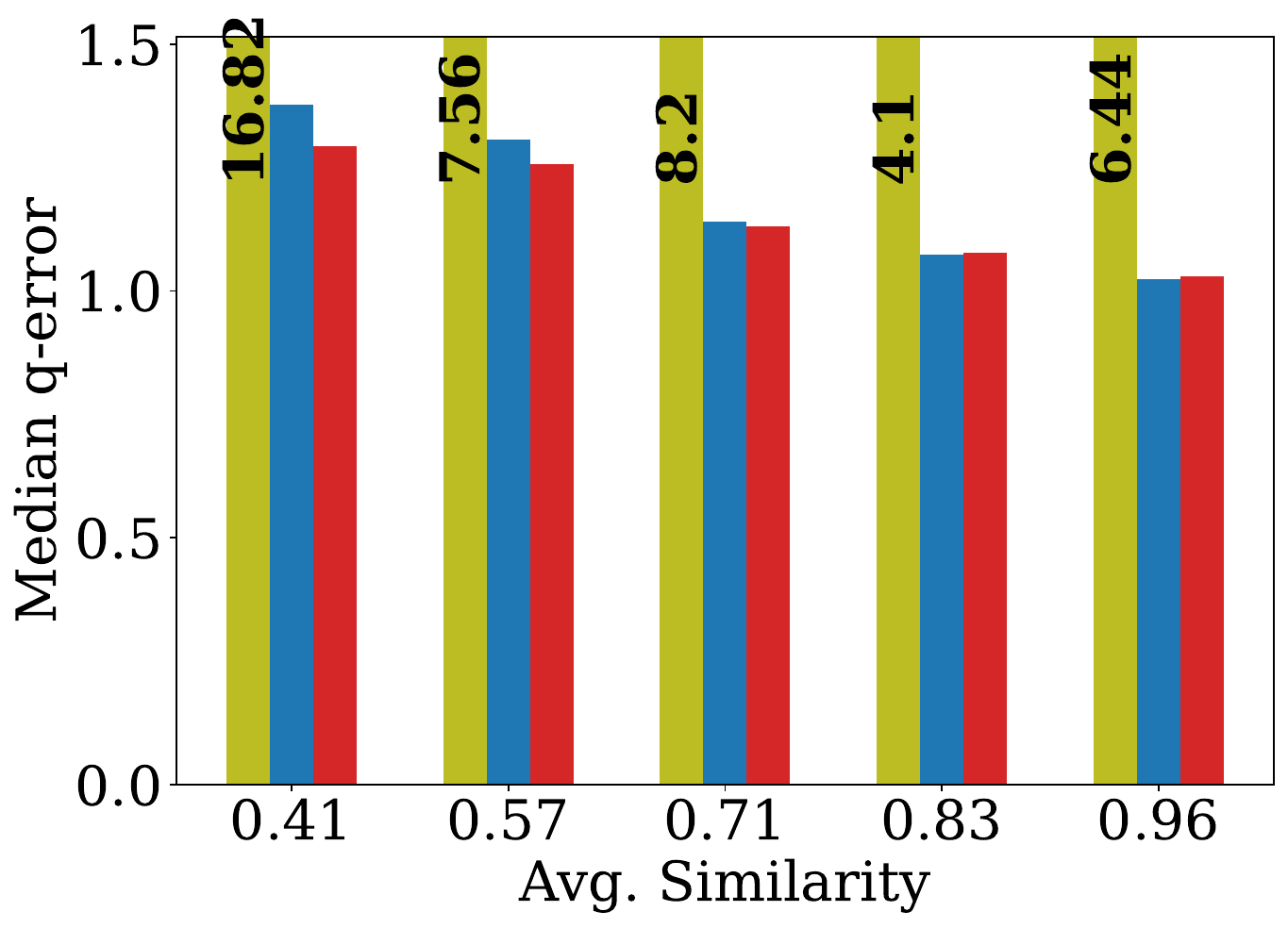}\label{subfig:yelp_selectivity}}
\subfloat[SWDF]{\includegraphics[width=0.25\linewidth]{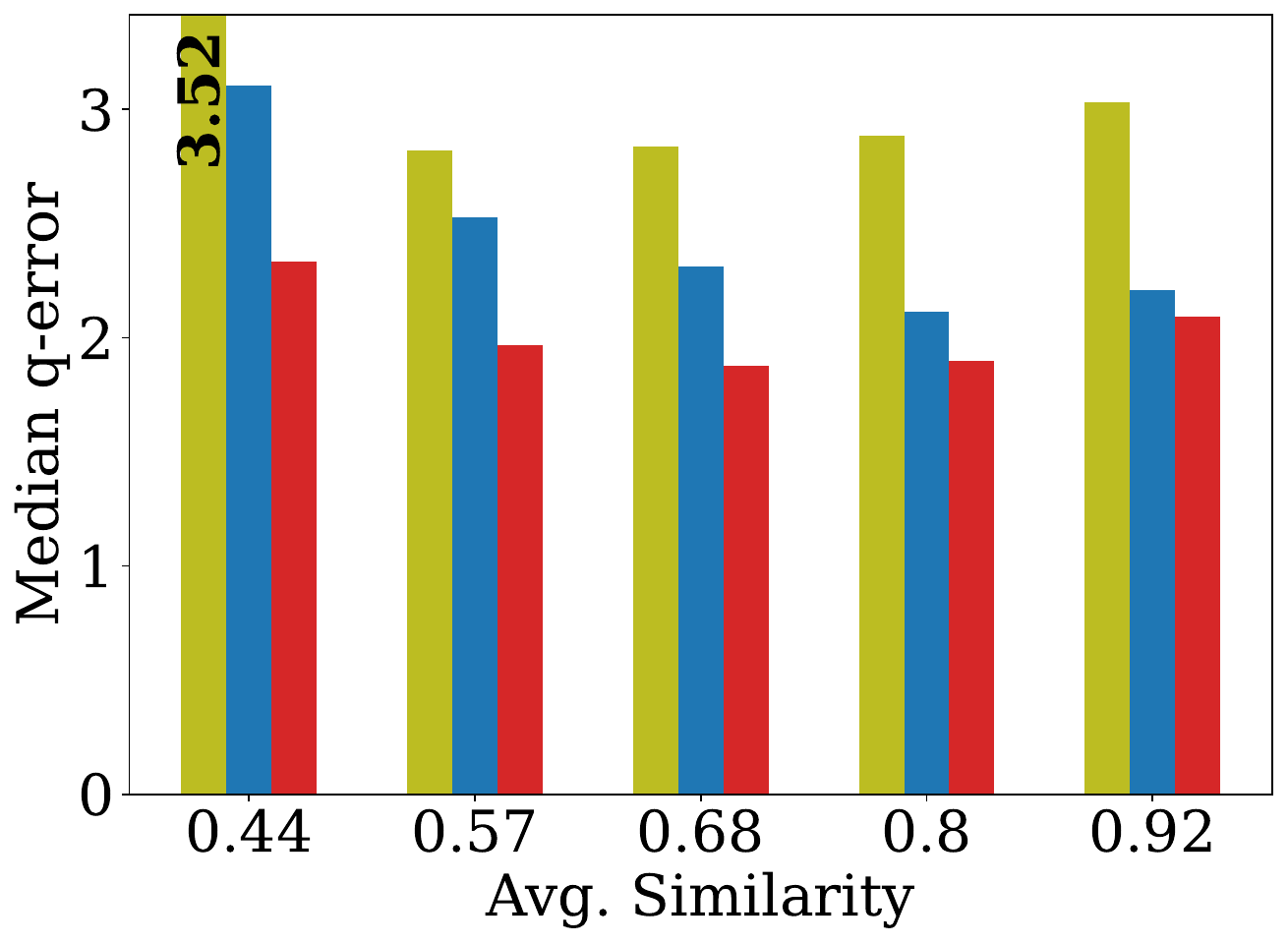}\label{subfig:swdf_selectivity}}
\subfloat[LUBM]{\includegraphics[width=0.25\linewidth]{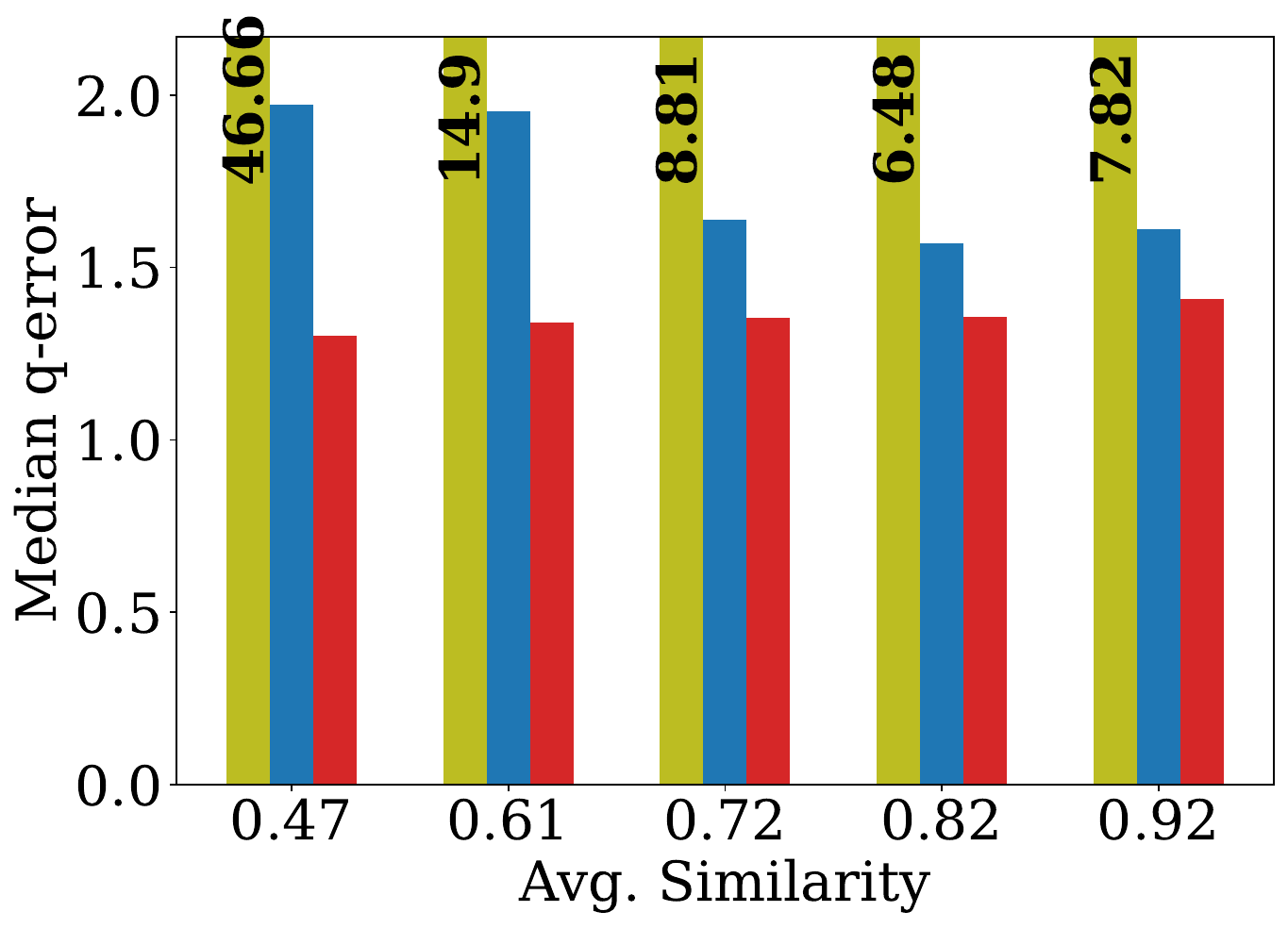}\label{subfig:lubm_selectivity}}
\vspace{-4pt}
\caption{Median \textit{q}-error with varying similarity}
\label{fig:selectivity}
\vspace{-4pt}
\end{figure*}

{\color{red}
\begin{figure*}[!ht]
\centering
\subfloat{\includegraphics[width=0.25\linewidth]{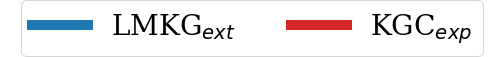}}
\\
\vspace{-12pt}
\setcounter{subfigure}{0}
\subfloat[Wikidata]{\includegraphics[width=0.25\linewidth]{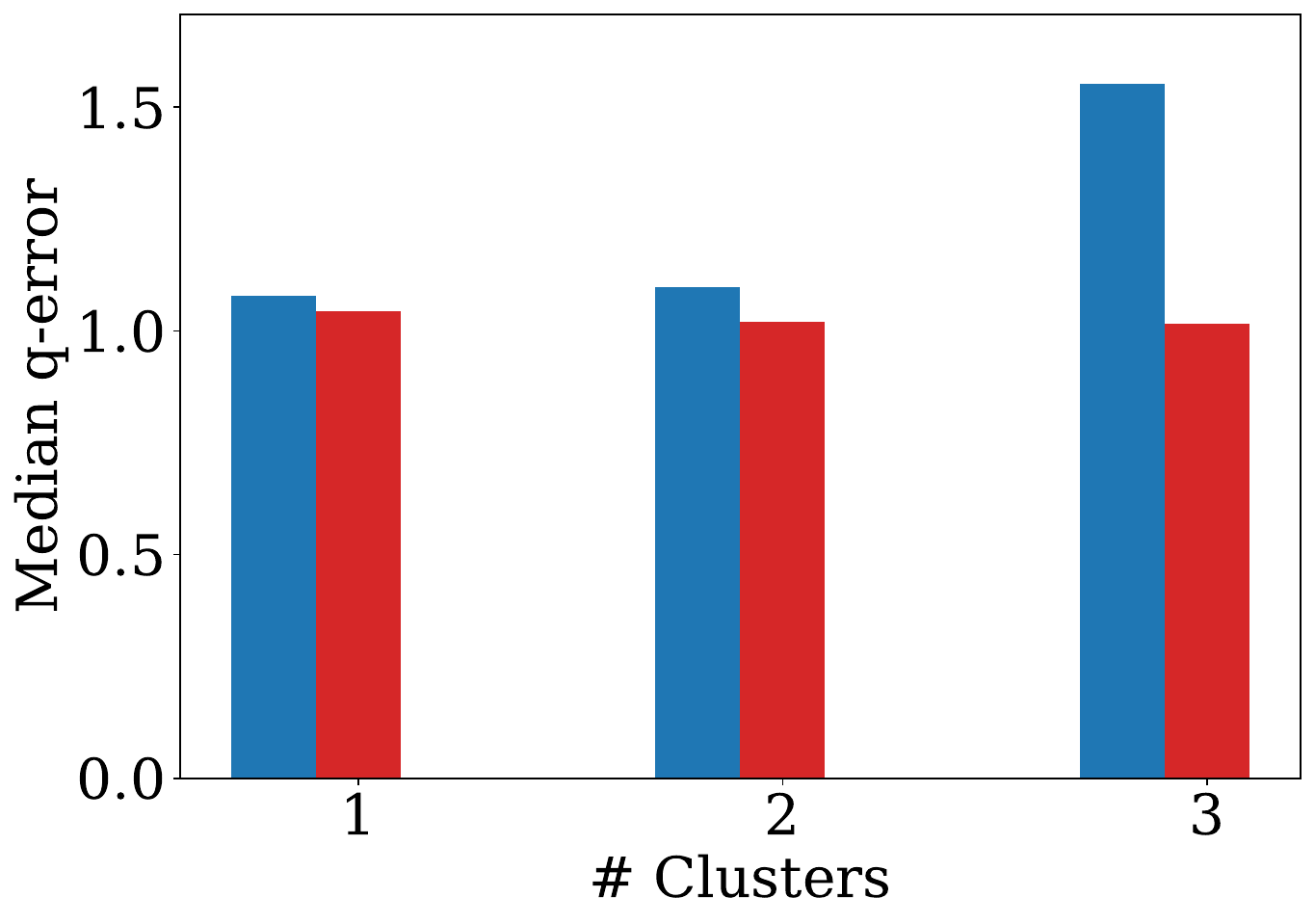}\label{subfig:wiki_clusters}}
\subfloat[Yelp]{\includegraphics[width=0.25\linewidth]{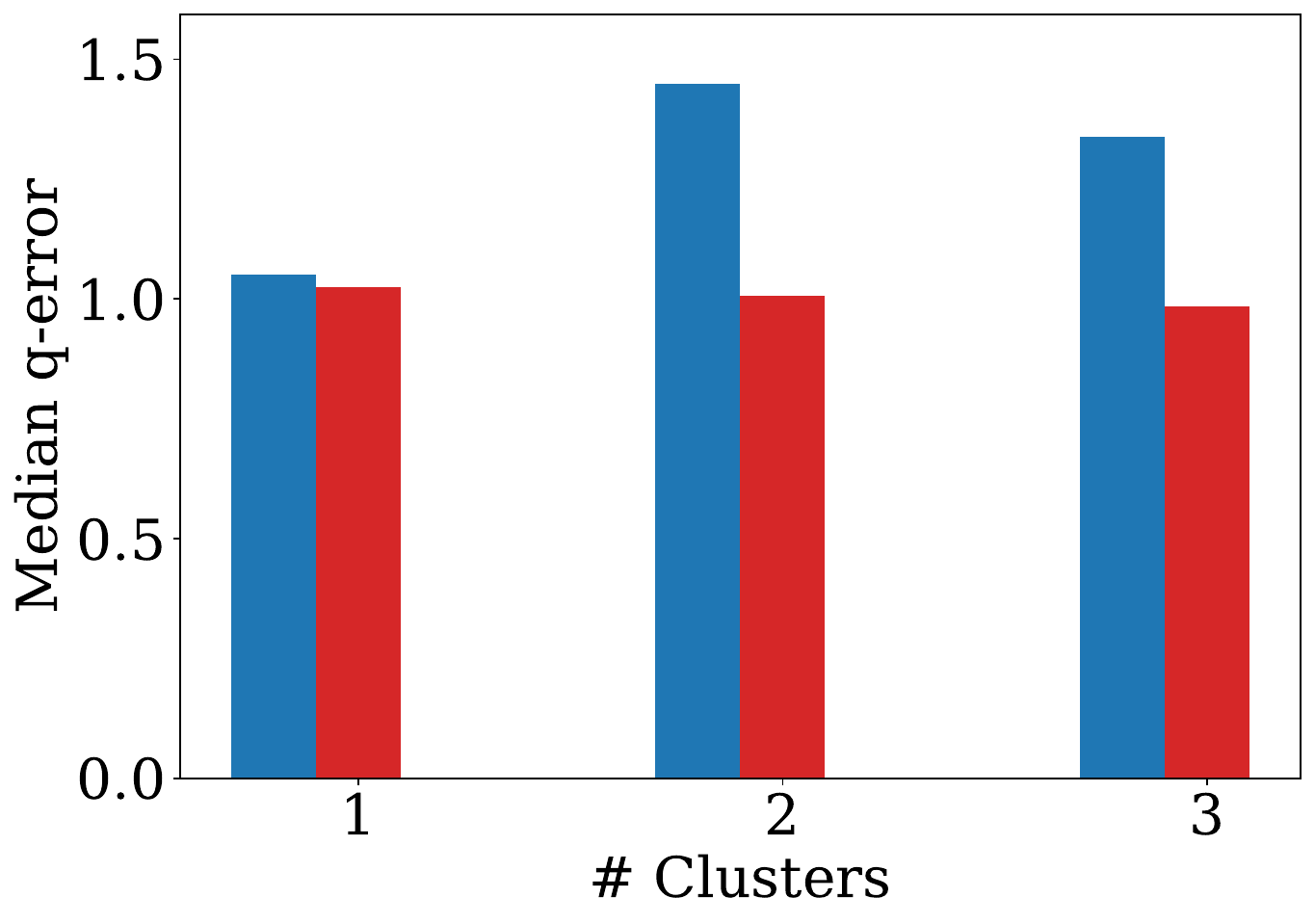}\label{subfig:yelp_clusters}}
\subfloat[SWDF]{\includegraphics[width=0.25\linewidth]{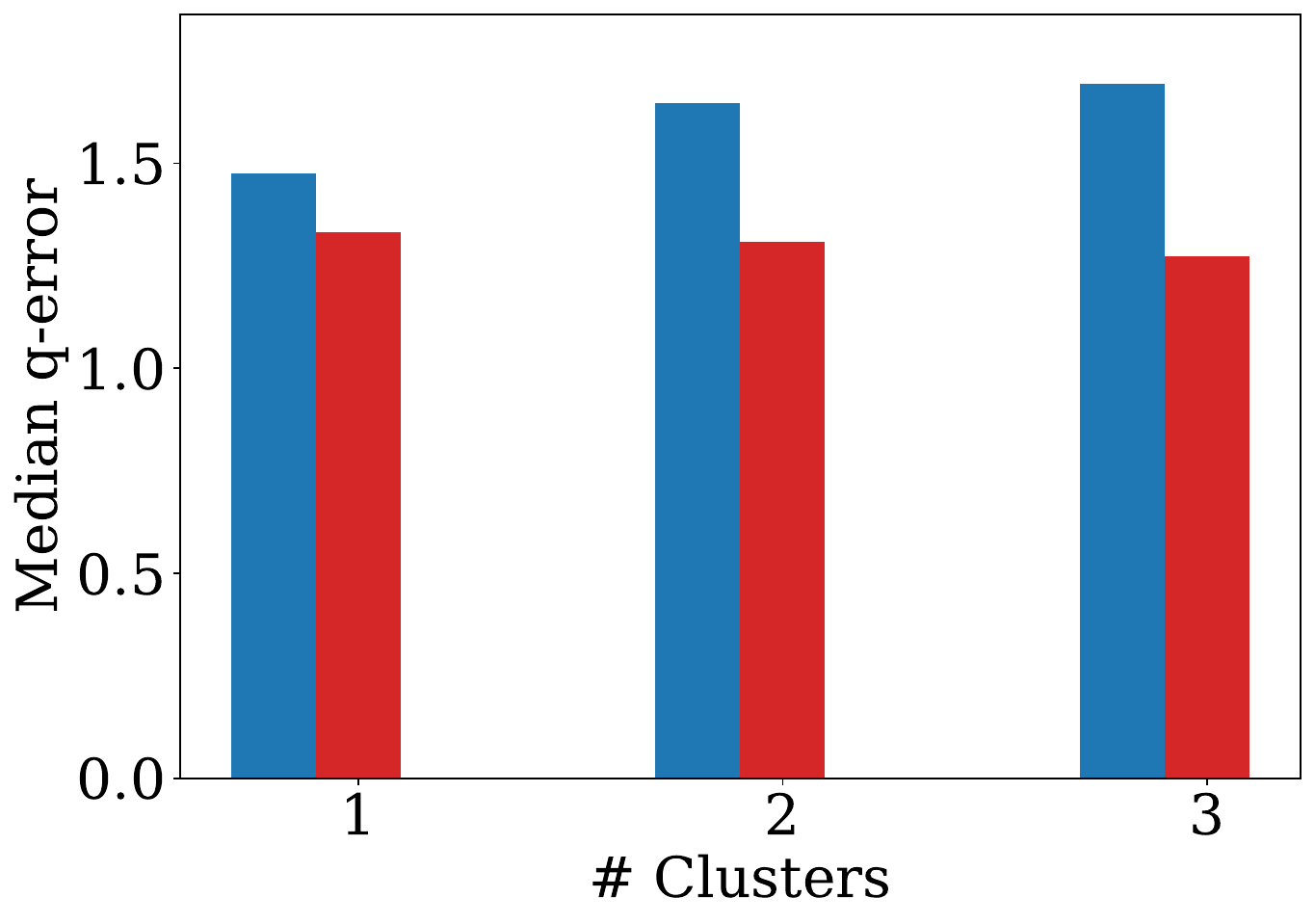}\label{subfig:swdf_clusters}}
\subfloat[LUBM]{\includegraphics[width=0.25\linewidth]{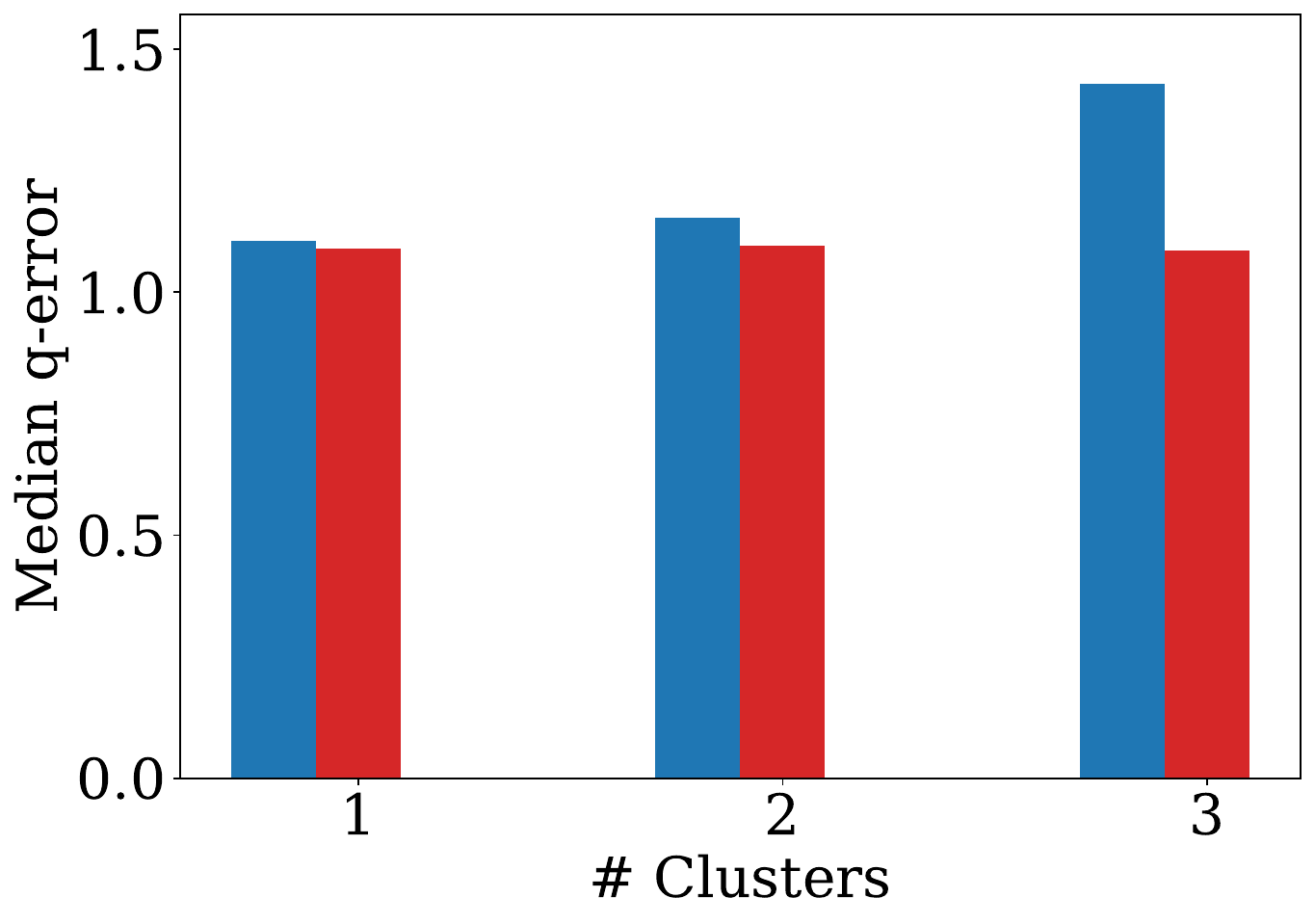}\label{subfig:lubm_clusters}}
\vspace{-4pt}
\caption{Median, 75th and 95th percentile \textit{q}-errors with varying number of clusters}
\vspace{-4pt}
\label{fig:no_clusters}
\end{figure*}
}

\begin{figure*}[!ht]
\centering
\subfloat{\includegraphics[width=0.35\linewidth]{Figures/plots/legend.png}}
\\
\vspace{-12pt}
\setcounter{subfigure}{0}
\subfloat[Wikidata (star)]{\includegraphics[width=0.25\linewidth]{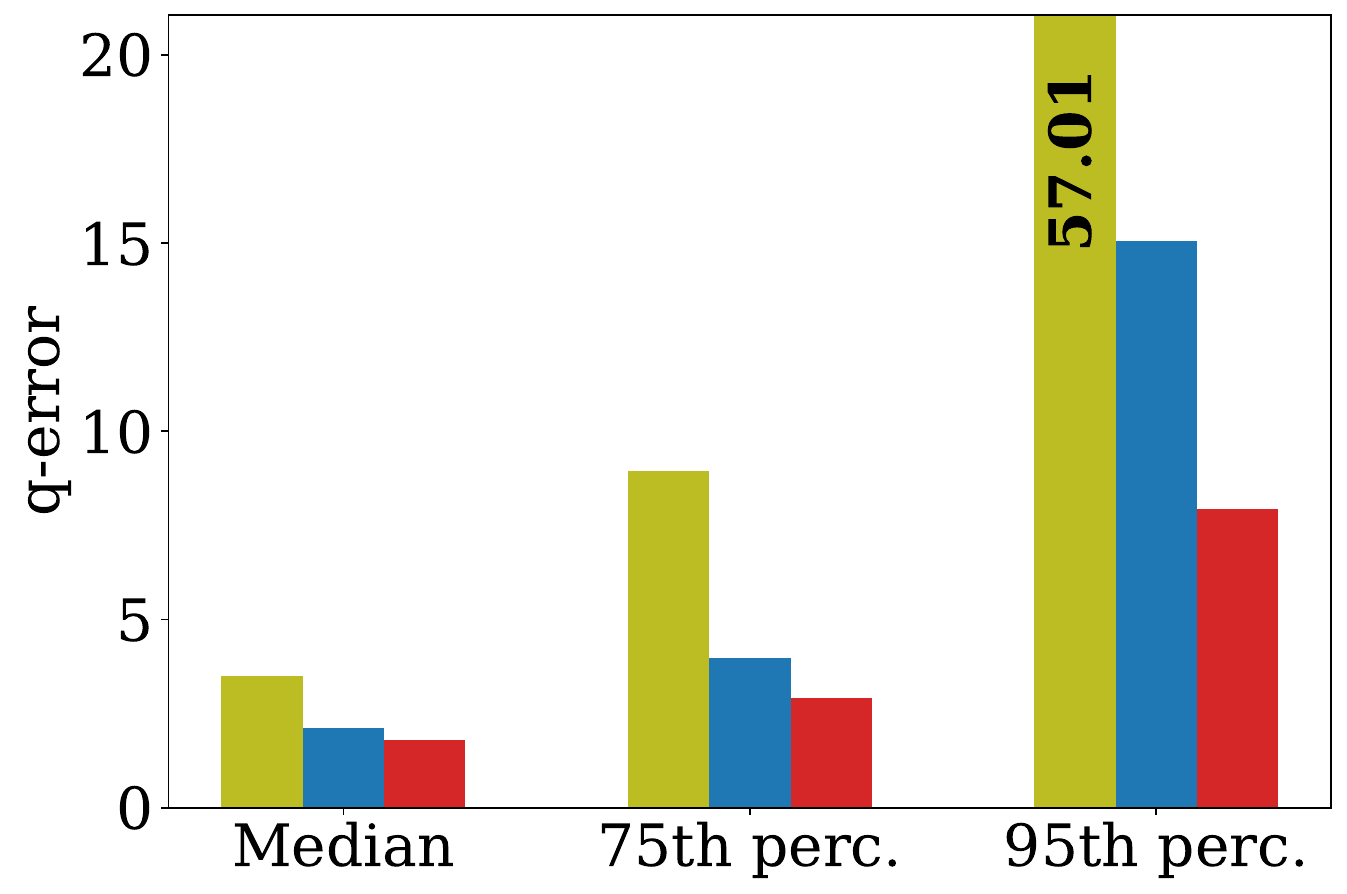}\label{subfig:wiki_star}}
\subfloat[Yelp (star)]{\includegraphics[width=0.25\linewidth]{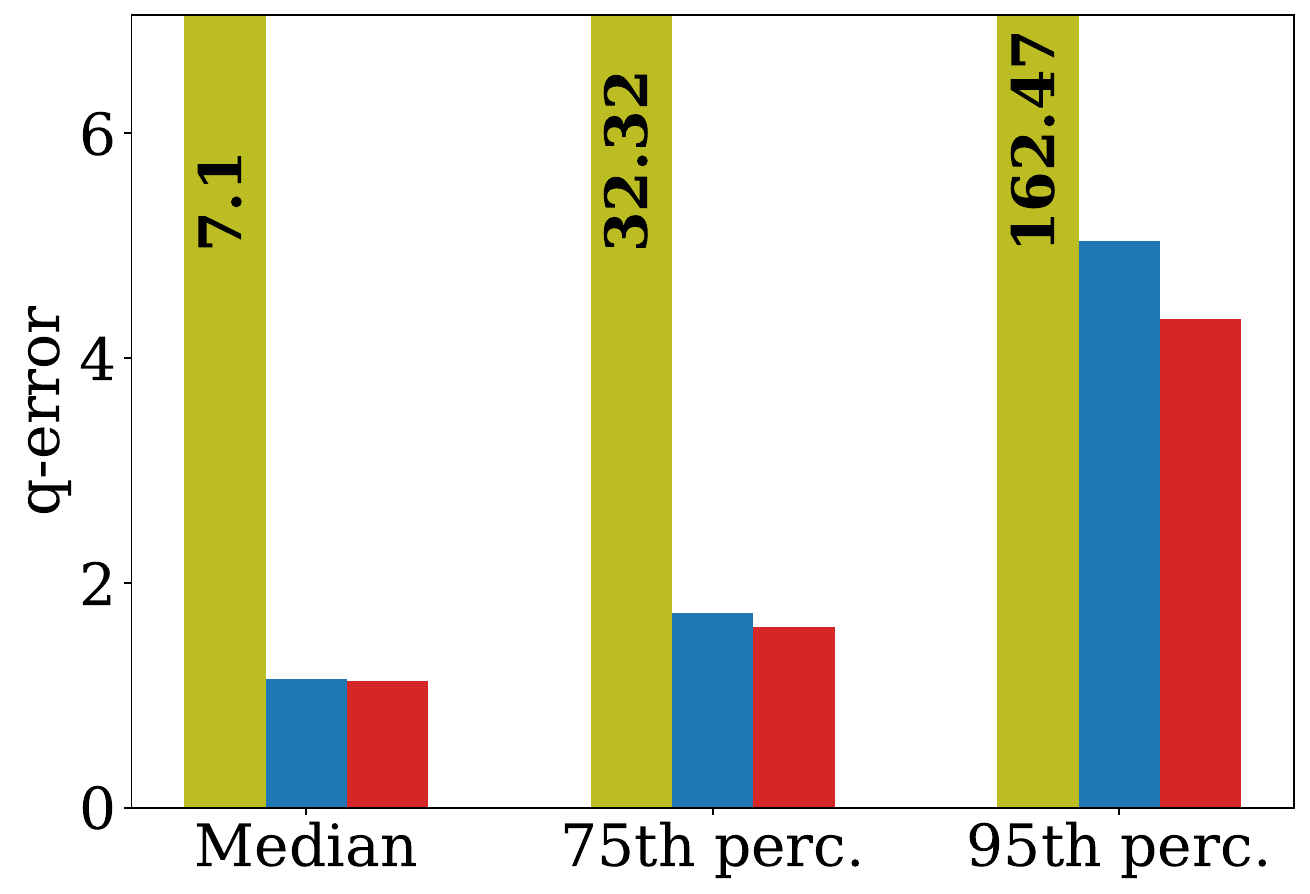}\label{subfig:yelp_star}}
\subfloat[SWDF (star)]{\includegraphics[width=0.25\linewidth]{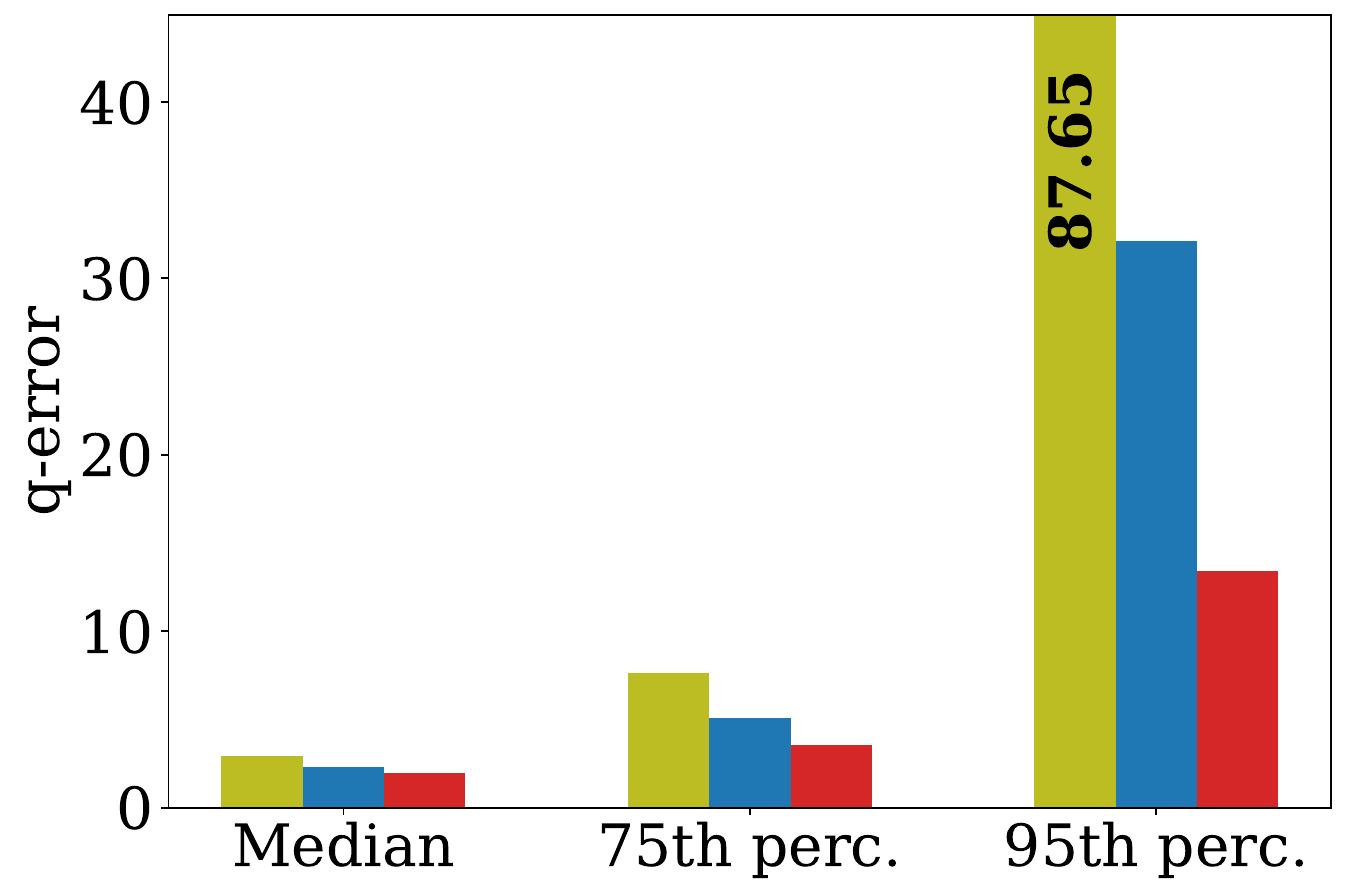}\label{subfig:swdf_star}}
\subfloat[LUBM (star)]{\includegraphics[width=0.25\linewidth]{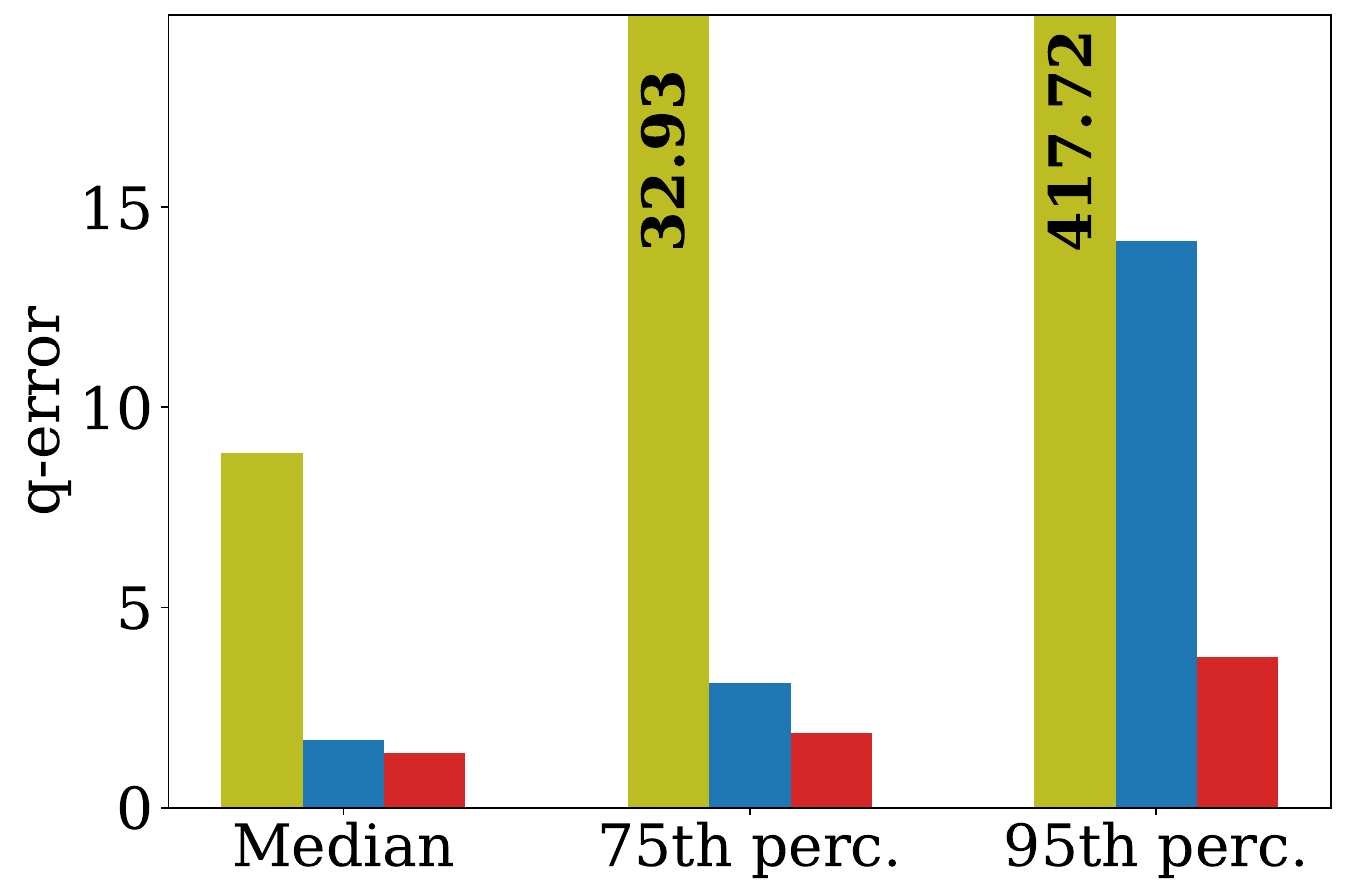}\label{subfig:lubm_star}} \\
\vspace{-8pt}
\subfloat[Wikidata (chain)]{\includegraphics[width=0.25\linewidth]{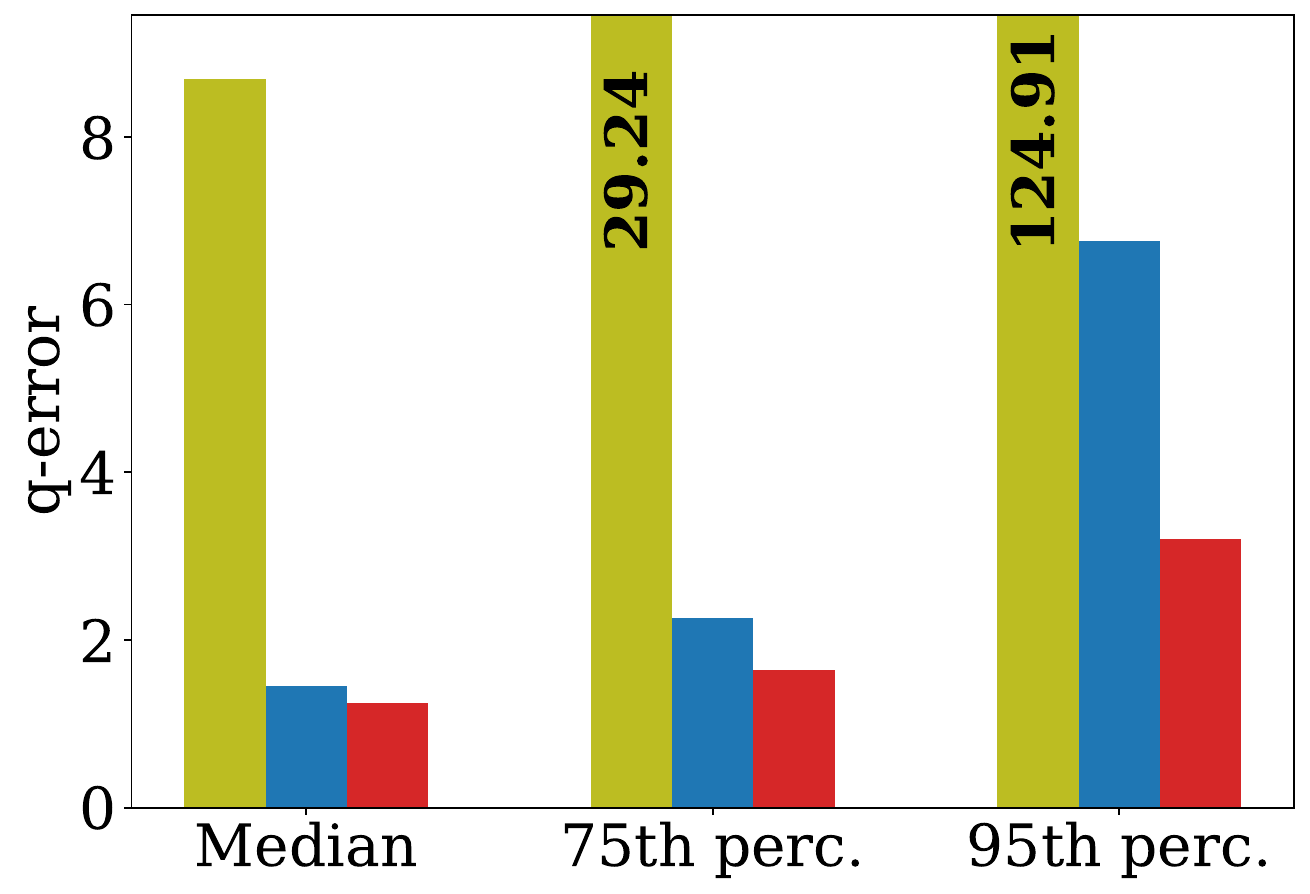}\label{subfig:wiki_chain}}
\subfloat[Yelp (chain)]{\includegraphics[width=0.25\linewidth]{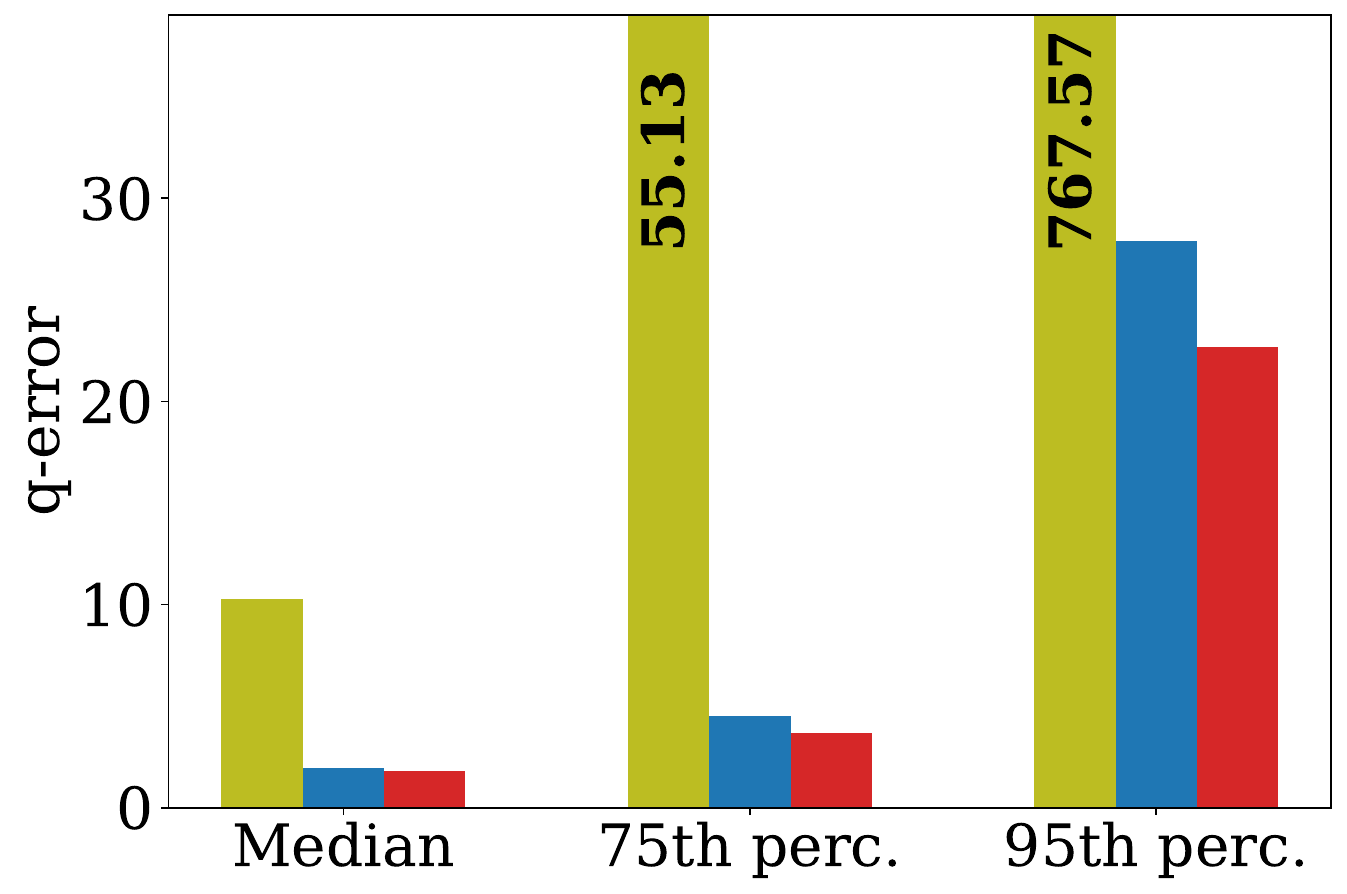}\label{subfig:yelp_chain}}
\subfloat[SWDF (chain)]{\includegraphics[width=0.25\linewidth]{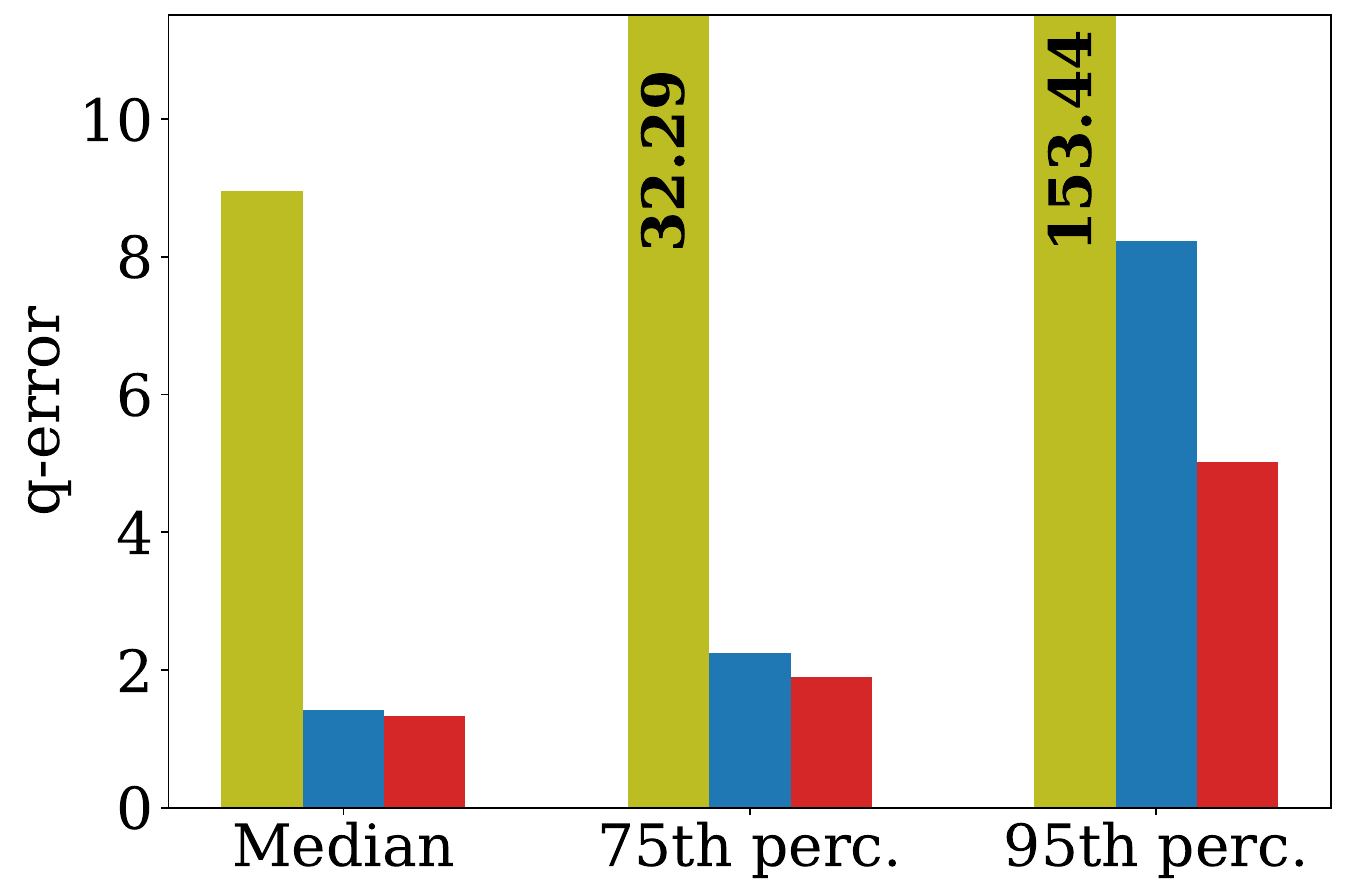}\label{subfig:swdf_chain}}
\subfloat[LUBM (chain)]{\includegraphics[width=0.25\linewidth]{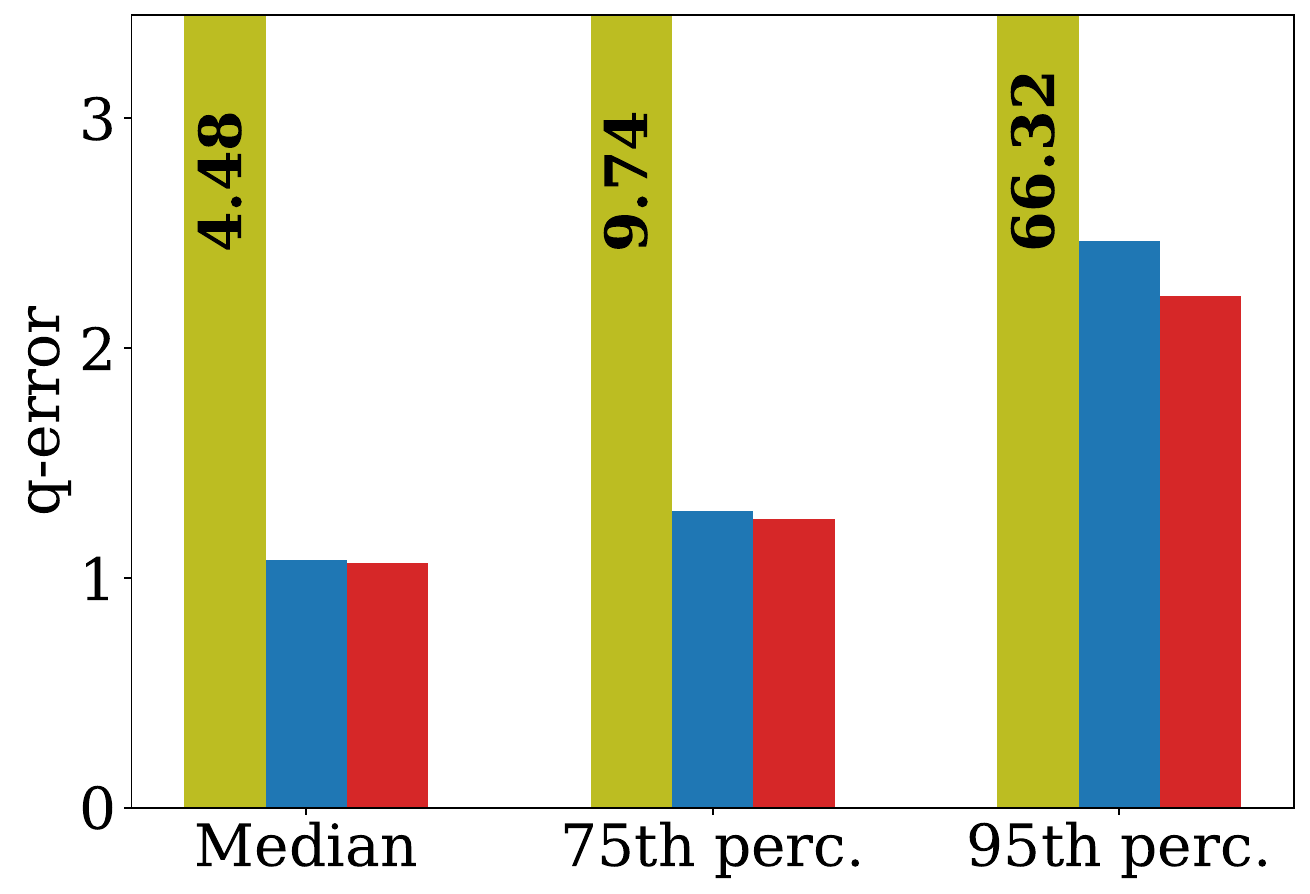}\label{subfig:lubm_chain}}
\vspace{-4pt}
\caption{Median, 75th and 95th percentile \textit{q}-errors for star (a-d) and chain similarity queries (e-h)}
\vspace{-5pt}
\label{fig:query_type}
\end{figure*}

{\color{red}
\begin{figure*}[!ht]
\centering
\subfloat{\includegraphics[width=0.35\linewidth]{Figures/plots/legend.png}}
\\
\vspace{-12pt}
\setcounter{subfigure}{0}
\subfloat[Wikidata]{\includegraphics[width=0.25\linewidth]{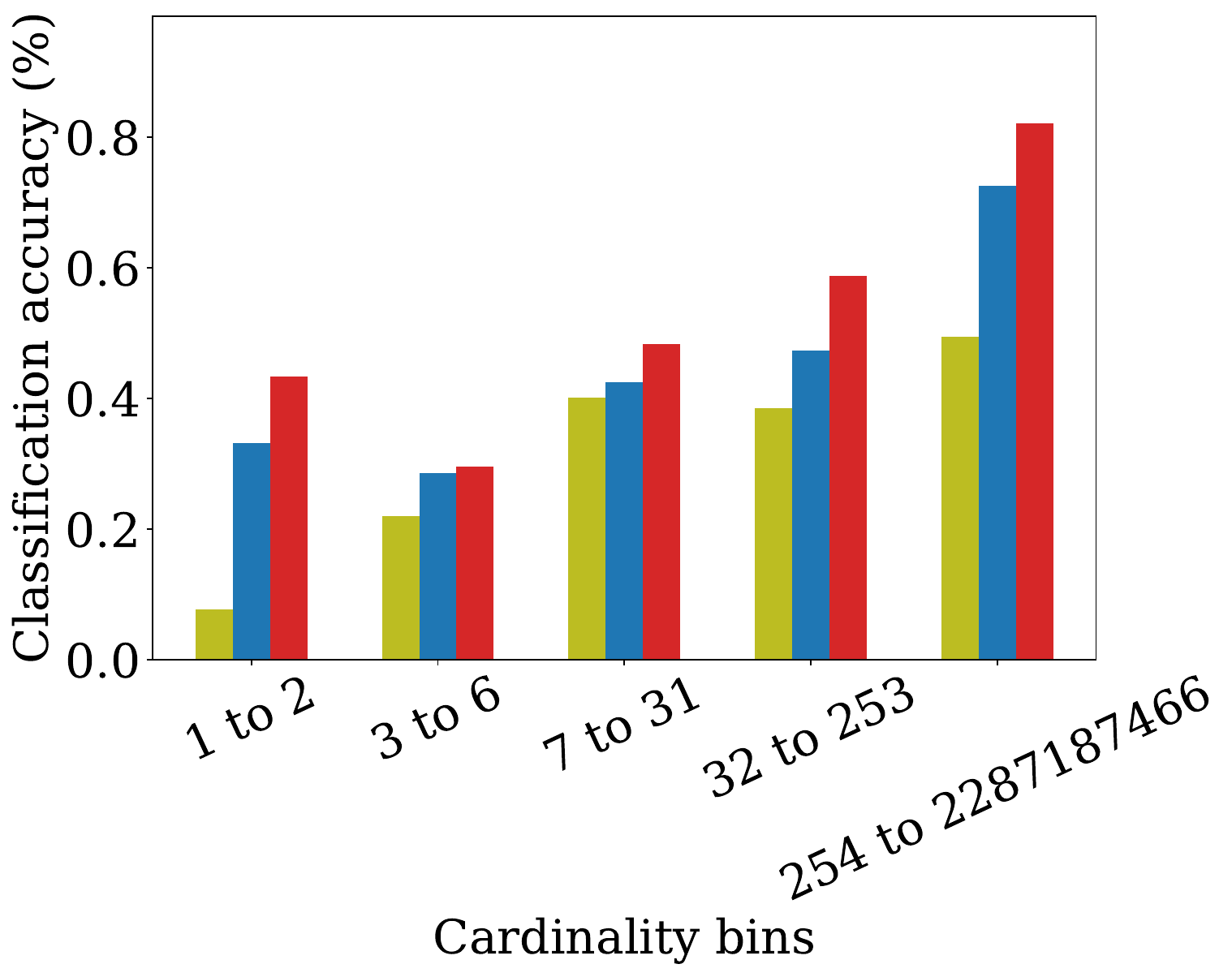}\label{subfig:wiki_star_class}}
\subfloat[Yelp]{\includegraphics[width=0.25\linewidth]{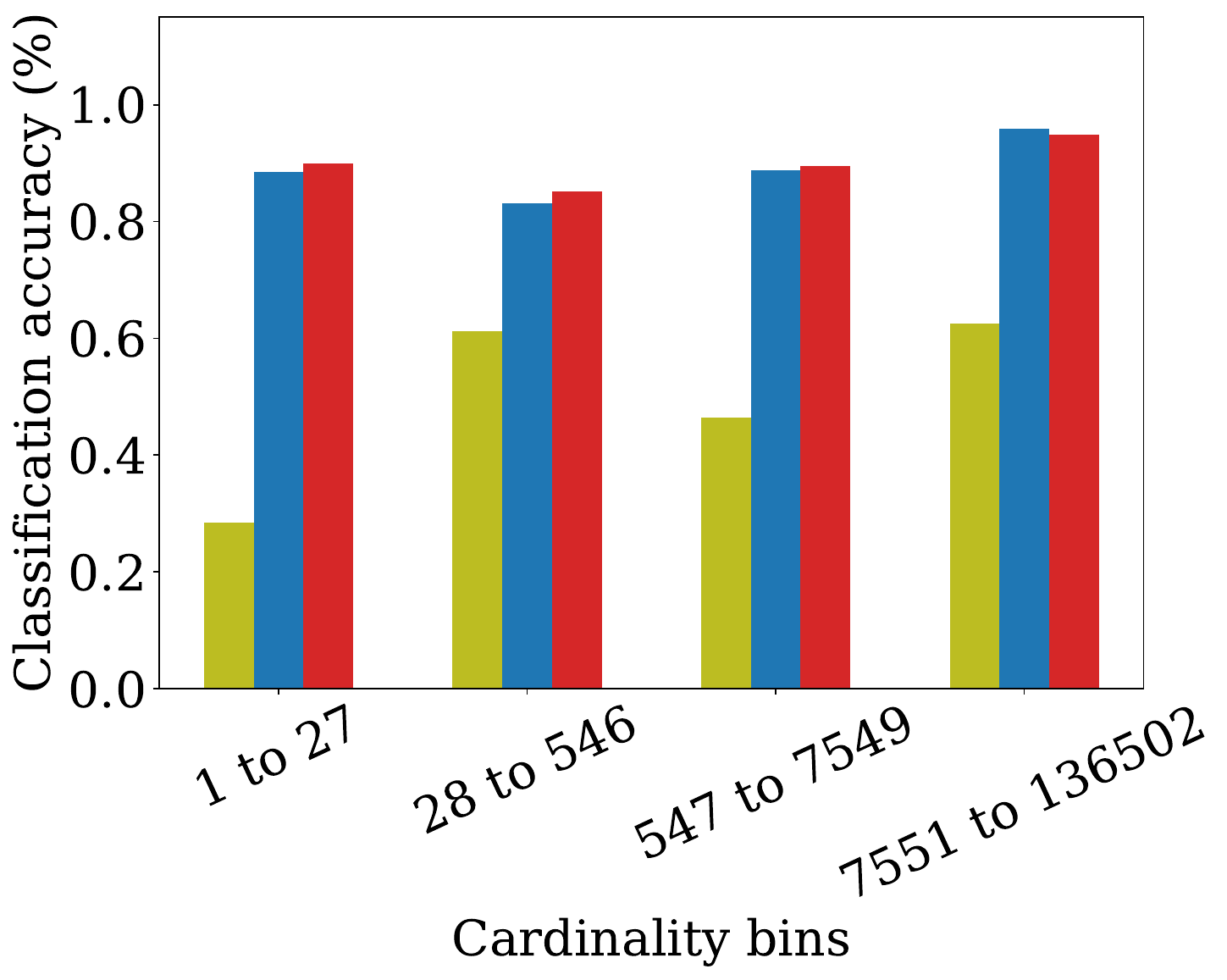}\label{subfig:yelp_star_class}}
\subfloat[SWDF]{\includegraphics[width=0.25\linewidth]{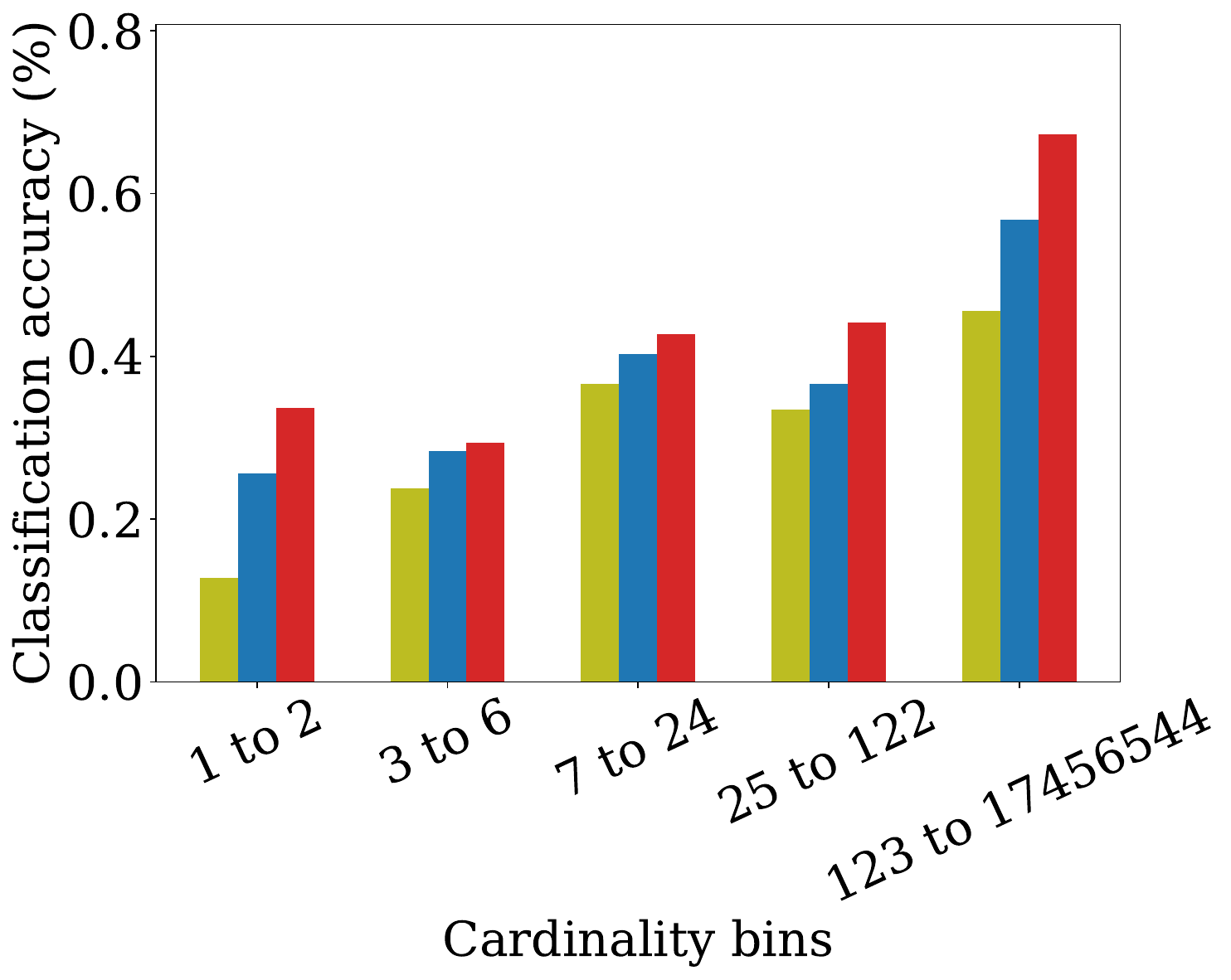}\label{subfig:swdf_star_class}}
\subfloat[LUBM]{\includegraphics[width=0.25\linewidth]{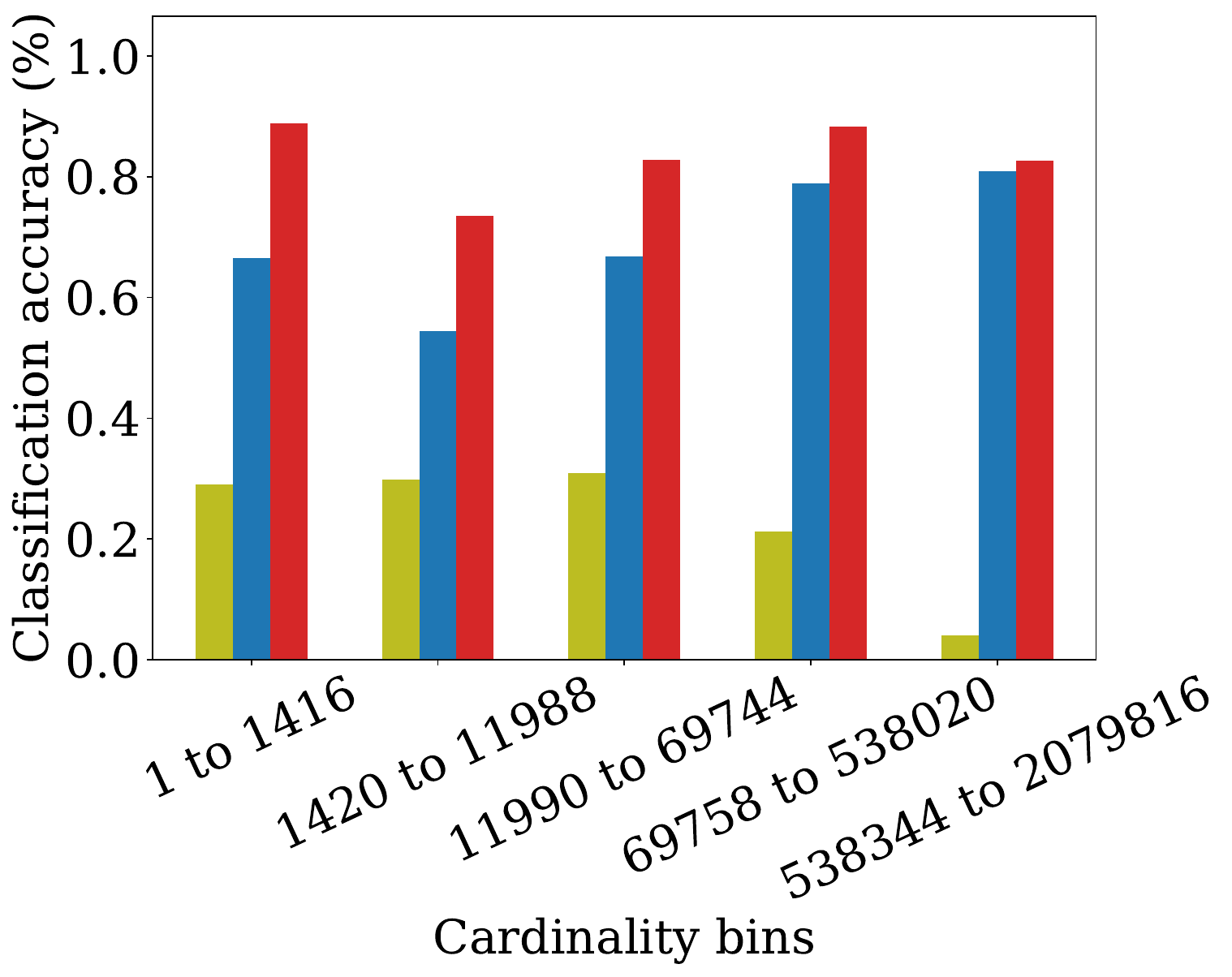}\label{subfig:lubm_star_class}} \\
\vspace{-6pt}
\subfloat[Wikidata]{\includegraphics[width=0.25\linewidth]{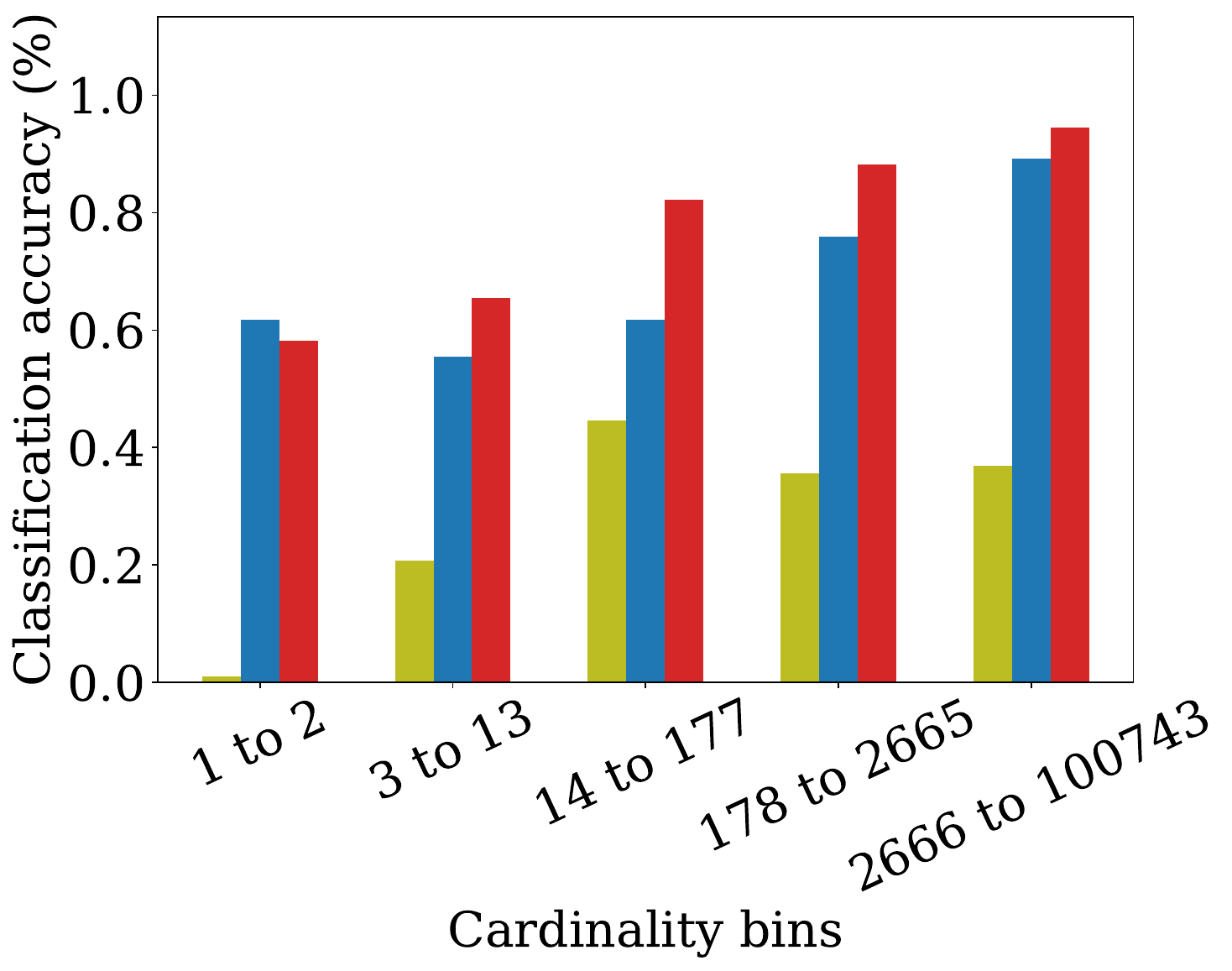}\label{subfig:wiki_chain_class}}
\subfloat[Yelp]{\includegraphics[width=0.25\linewidth]{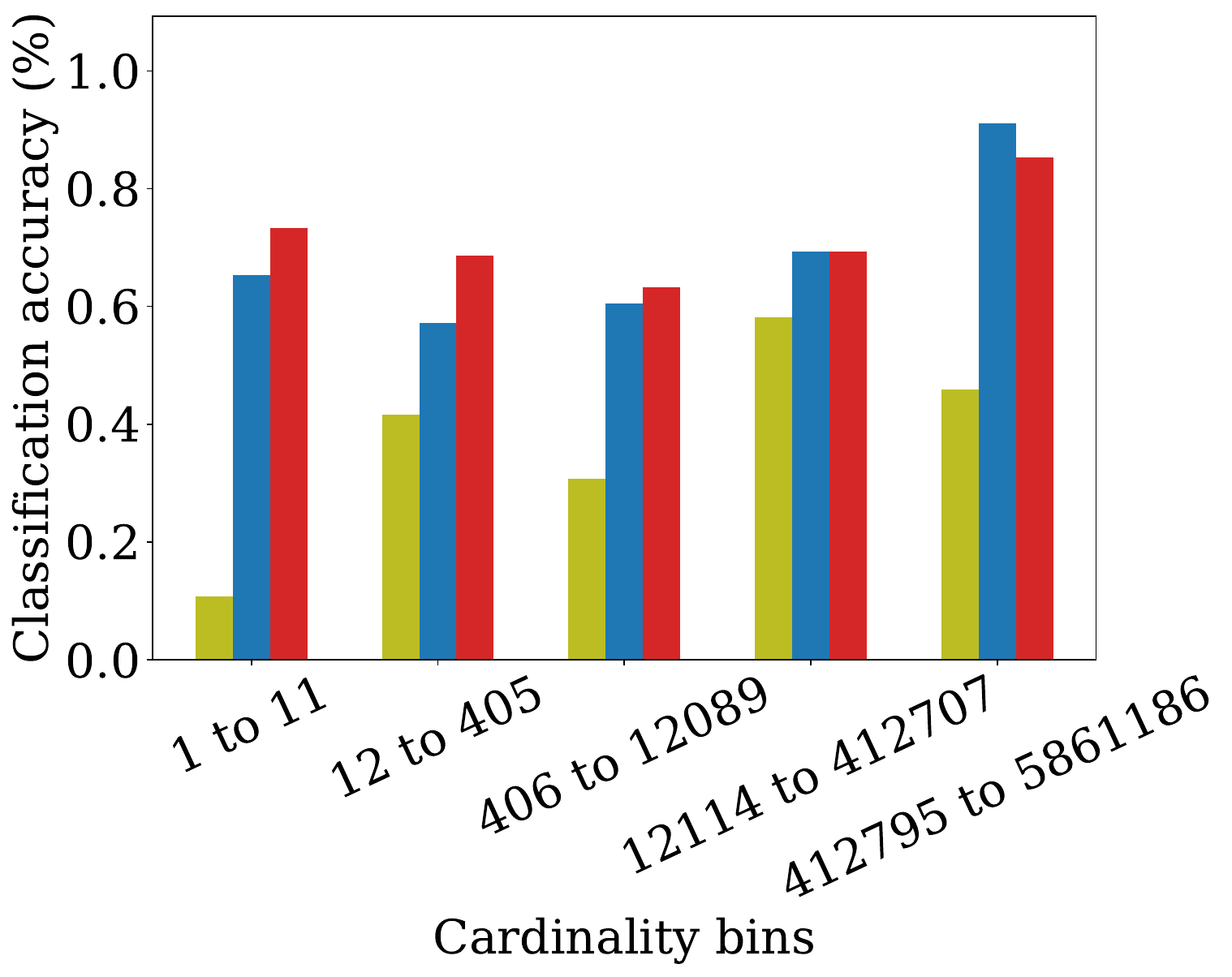}\label{subfig:yelp_chain_class}}
\subfloat[SWDF]{\includegraphics[width=0.25\linewidth]{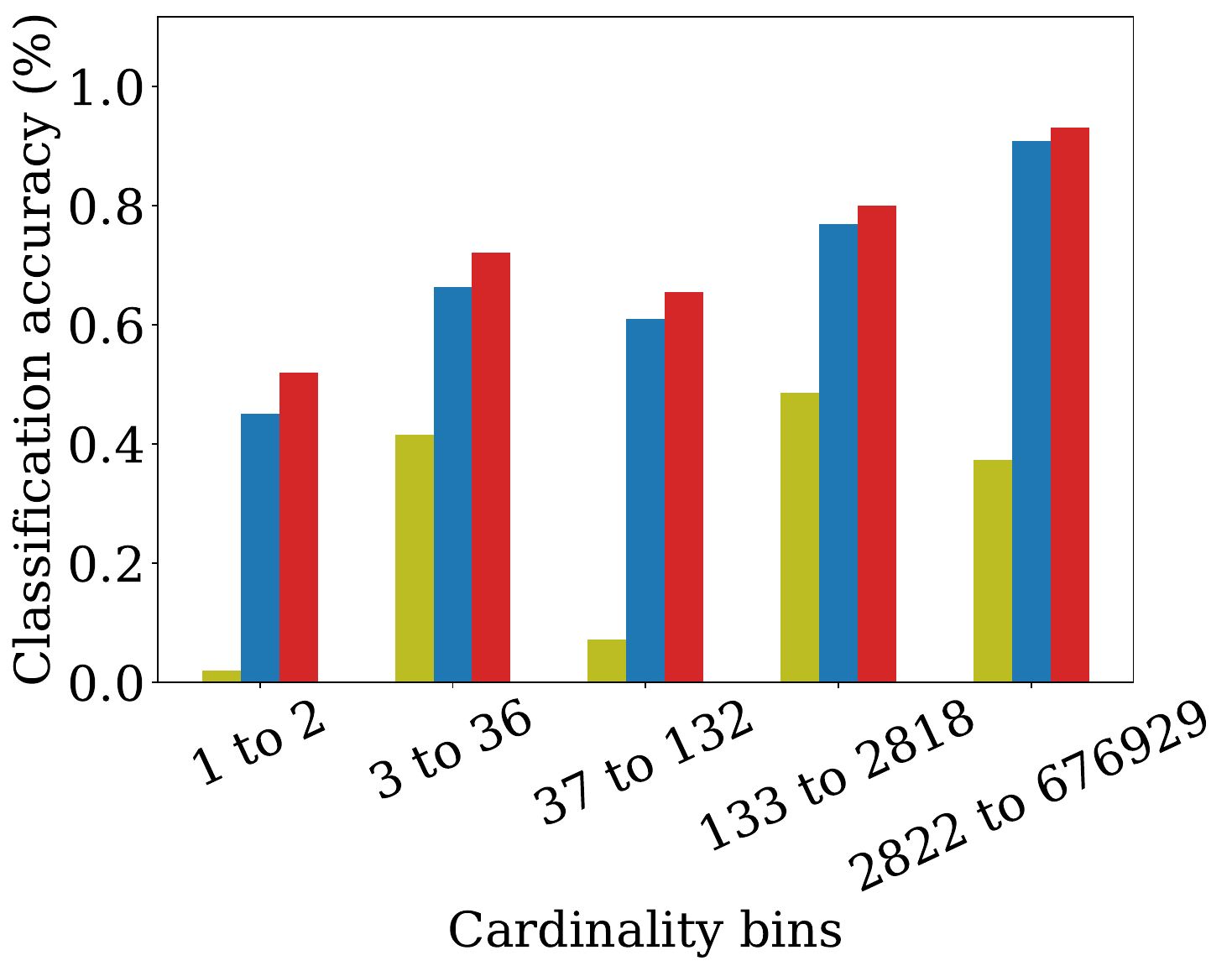}\label{subfig:swdf_chain_class}}
\subfloat[LUBM]{\includegraphics[width=0.25\linewidth]{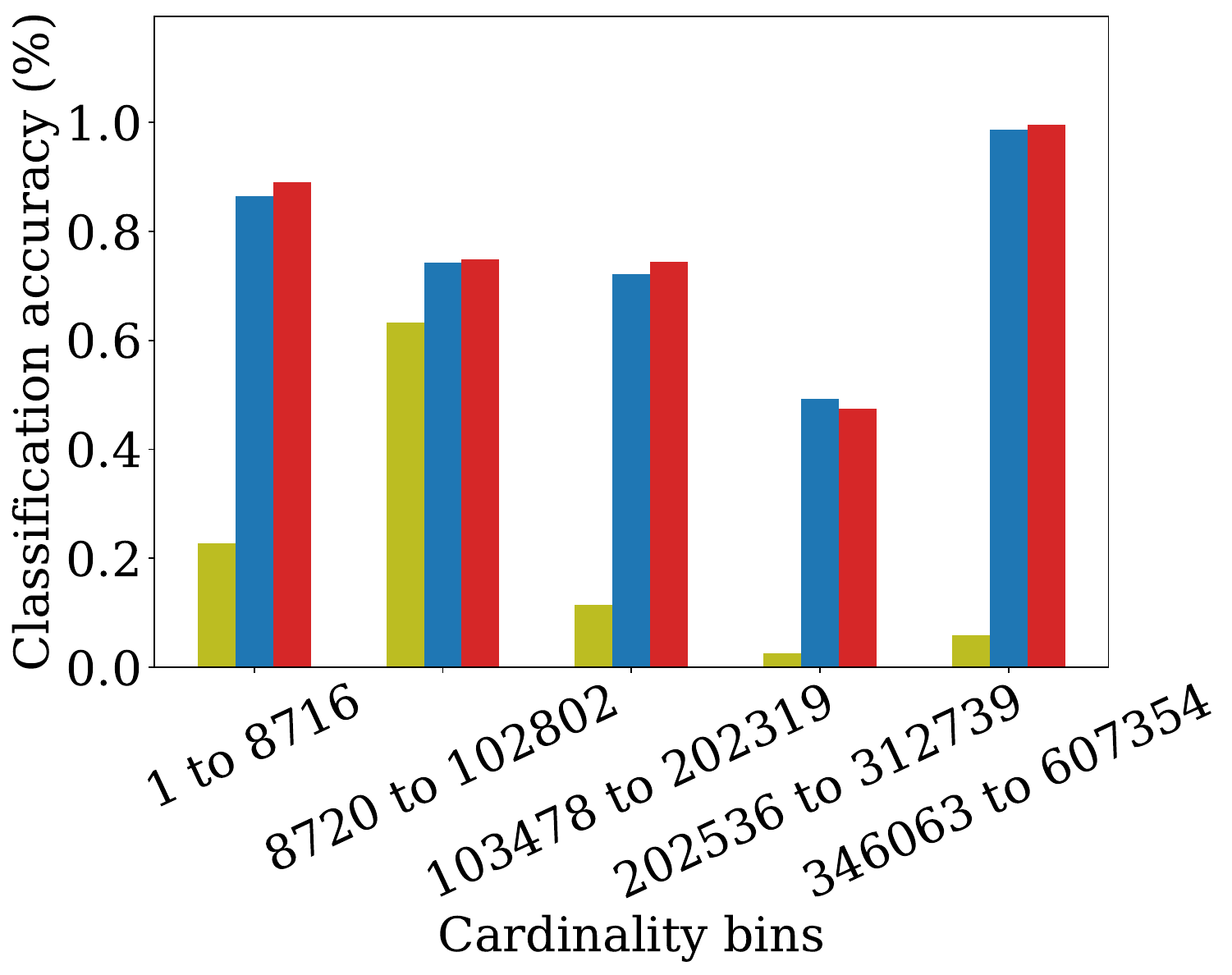}\label{subfig:lubm_chain_class}}
\vspace{-4pt}
\caption{Classification accuracy for star (a-d) and chain (e-h) queries}
\vspace{-6pt}
\label{fig:star_class}
\end{figure*}
}

\begin{figure*}[!ht]
\centering
\subfloat{\includegraphics[width=0.35\linewidth]{Figures/plots/legend.png}}
\\
\vspace{-12pt}
\setcounter{subfigure}{0}
\subfloat[Wikidata]{\includegraphics[width=0.25\linewidth]{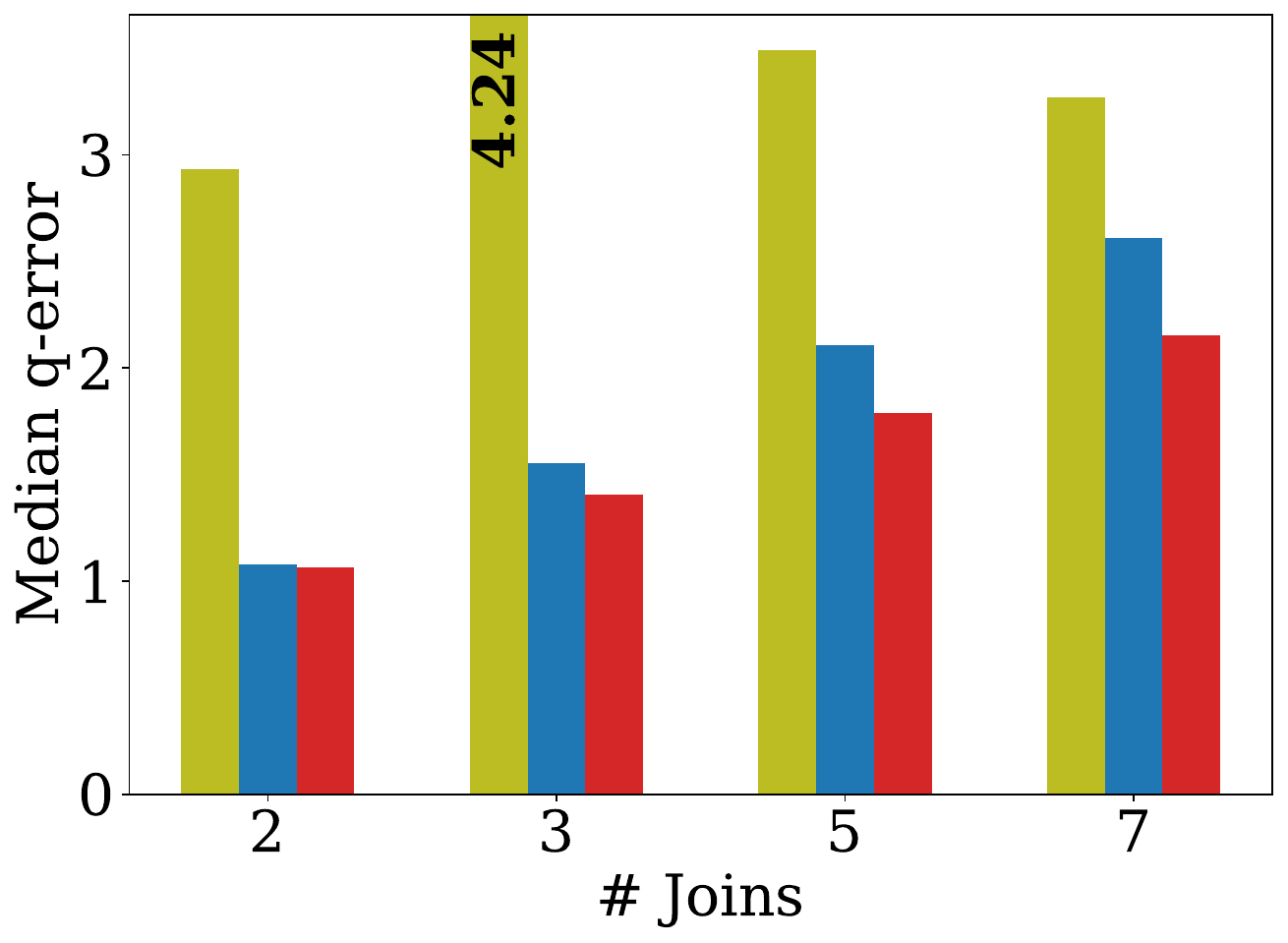}\label{subfig:wiki_joins}}
\subfloat[Yelp]{\includegraphics[width=0.25\linewidth]{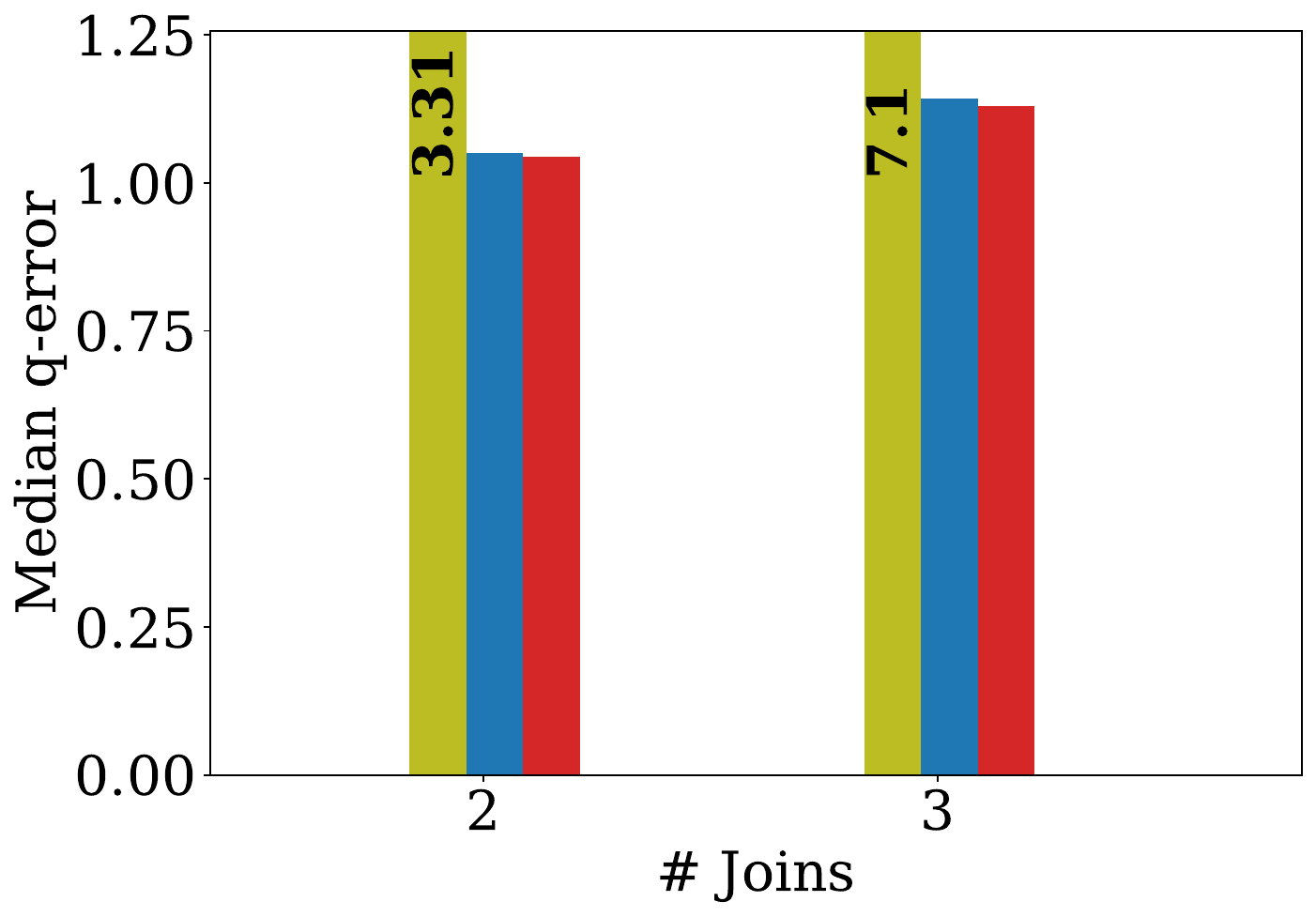}\label{subfig:yelp_joins}}
\subfloat[SWDF]{\includegraphics[width=0.25\linewidth]{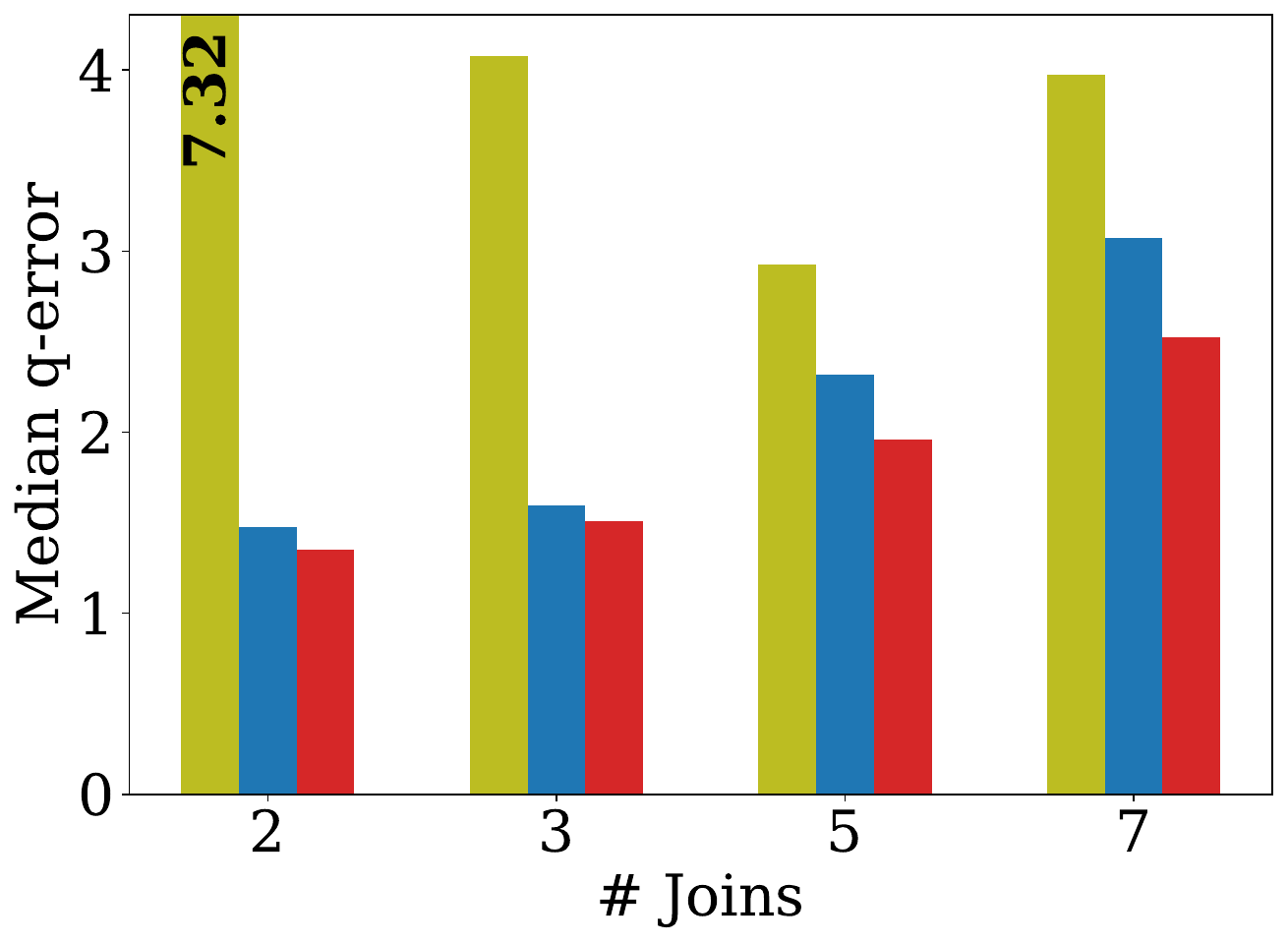}\label{subfig:swdf_joins}}
\subfloat[LUBM]{\includegraphics[width=0.25\linewidth]{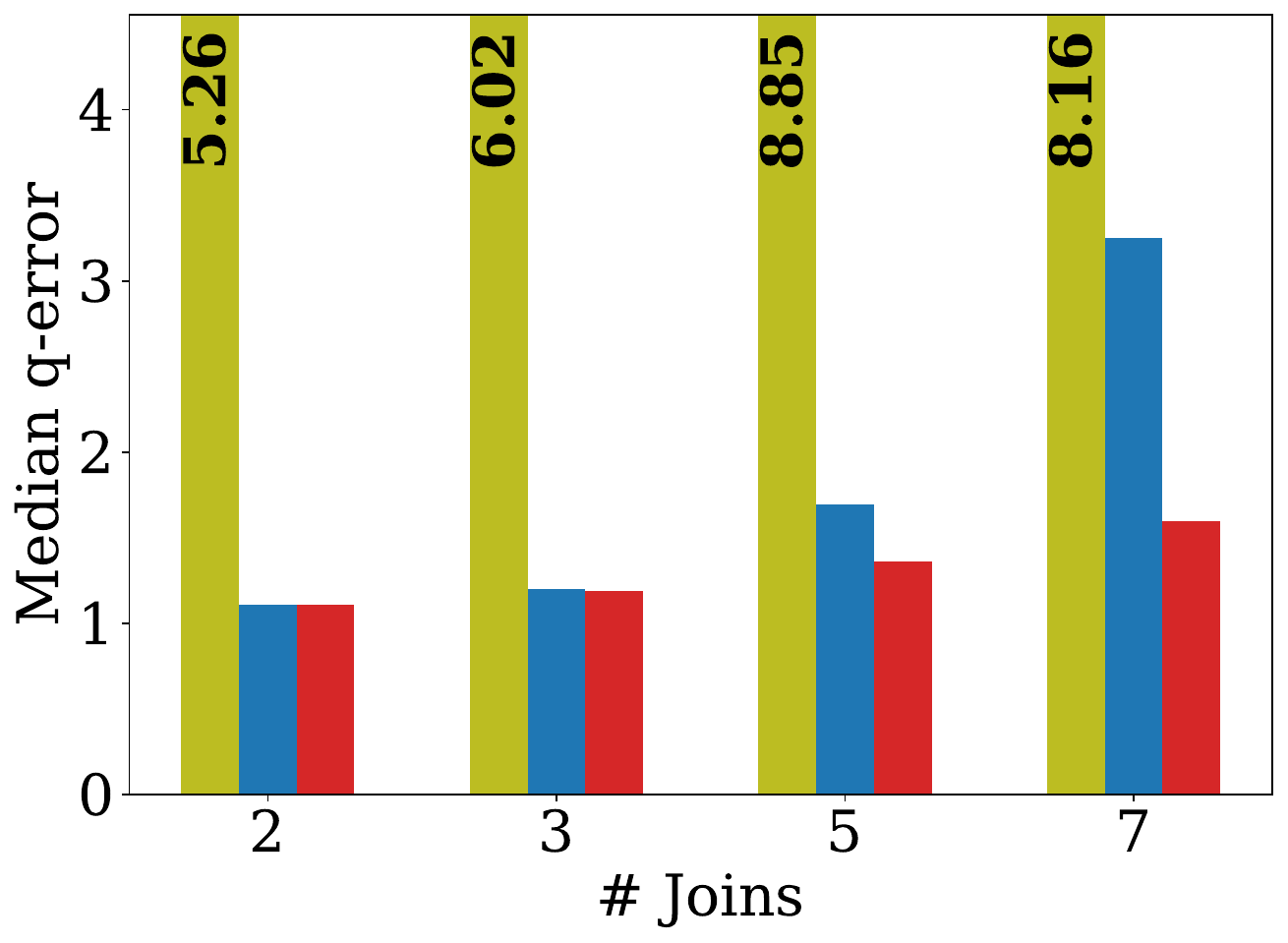}\label{subfig:lubm_joins}}
\vspace{-6pt}
\caption{Median \textit{q}-error with varying number of joins}
\vspace{-6pt}
\label{fig:no_joins}
\end{figure*}

\subsubsection{Varying selectivity}

In Figure~\ref{fig:selectivity}, we evaluate the \textit{q}-error of each approach according to the selectivity of the issued queries, which, as discussed, depends on the given similarity thresholds. Specifically, we grouped the queries based on their similarities and calculated the average similarity per group. Clearly, \MODELb outperforms the rest of the approaches in almost all cases (\mnote{i.e., up to approx. $30\%$ lower \textit{q}-error compared to \MODELa}), except for larger similarities in the Yelp dataset, where the estimation accuracy is similar. \baseline again yields significantly worse performance,  with up to $25$x larger \textit{q}-error. 
For smaller similarity thresholds, \MODELb performs even better, 
due to giving more emphasis on the similarities during training and estimation. 
As the similarities approach $1$, we fall back to the case discussed in Section~\ref{subsubsec:no_sims}.

\subsubsection{Effect of clustering}

Figure~\ref{fig:no_clusters} depicts the \textit{q}-error of \MODELa and \MODELb with the test datasets using the clustering optimization for $1$, $2$, and $3$ clusters.
Note that as clustering is performed on similarities, \baseline is not evaluated in this experiment as it does not support similarity.
For \MODELb, increasing the number of clusters improves the \textit{q}-error. This reveals an interesting trade-off. Increasing the number of clusters tends to improve the quality of the estimation due to narrowing down the --increased due to similarities-- diversity of the training data. But at the same time, the task becomes harder due to decreasing the number of training elements to train in each sub-model. Recall from Section~\ref{subsec:clustering} that a separate model is trained for each cluster.
Interestingly, \MODELa cannot cope well for more than $1$ clusters, presenting up to $50\%$ larger \textit{q}-error. This is attributed to the reduction of training data size and its inherent characteristic of downplaying the provided similarities. This experiment drove our decision to select $1$ cluster as default for the experimental evaluation, to maintain a fair setting for all methods.

\begin{figure*}[!ht]
\centering
\subfloat{\includegraphics[width=0.35\linewidth]{Figures/plots/legend.png}}
\\
\vspace{-12pt}
\setcounter{subfigure}{0}
\subfloat[Wikidata]{\includegraphics[width=0.25\linewidth]{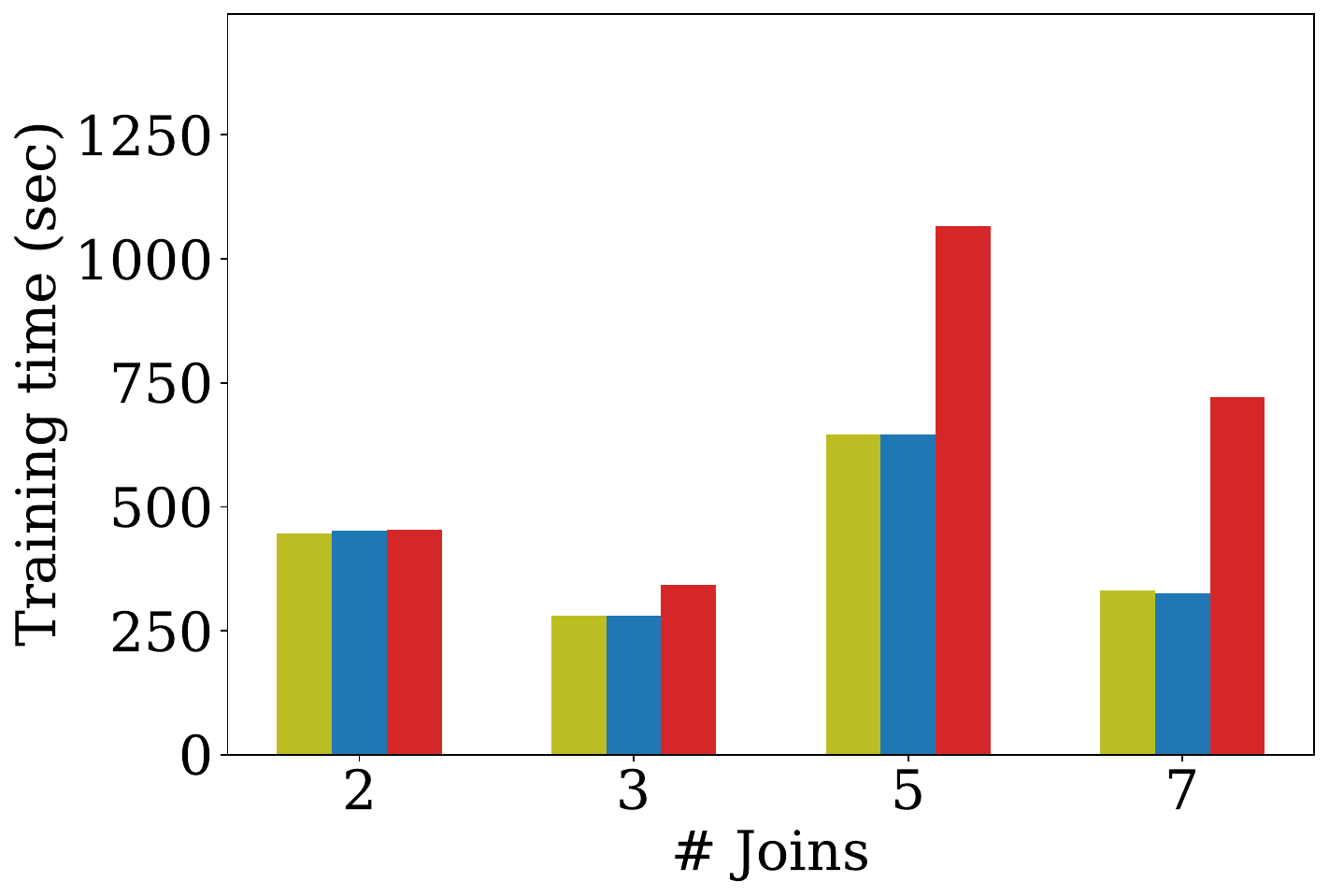}\label{subfig:training_wiki}}
\subfloat[Yelp]{\includegraphics[width=0.25\linewidth]{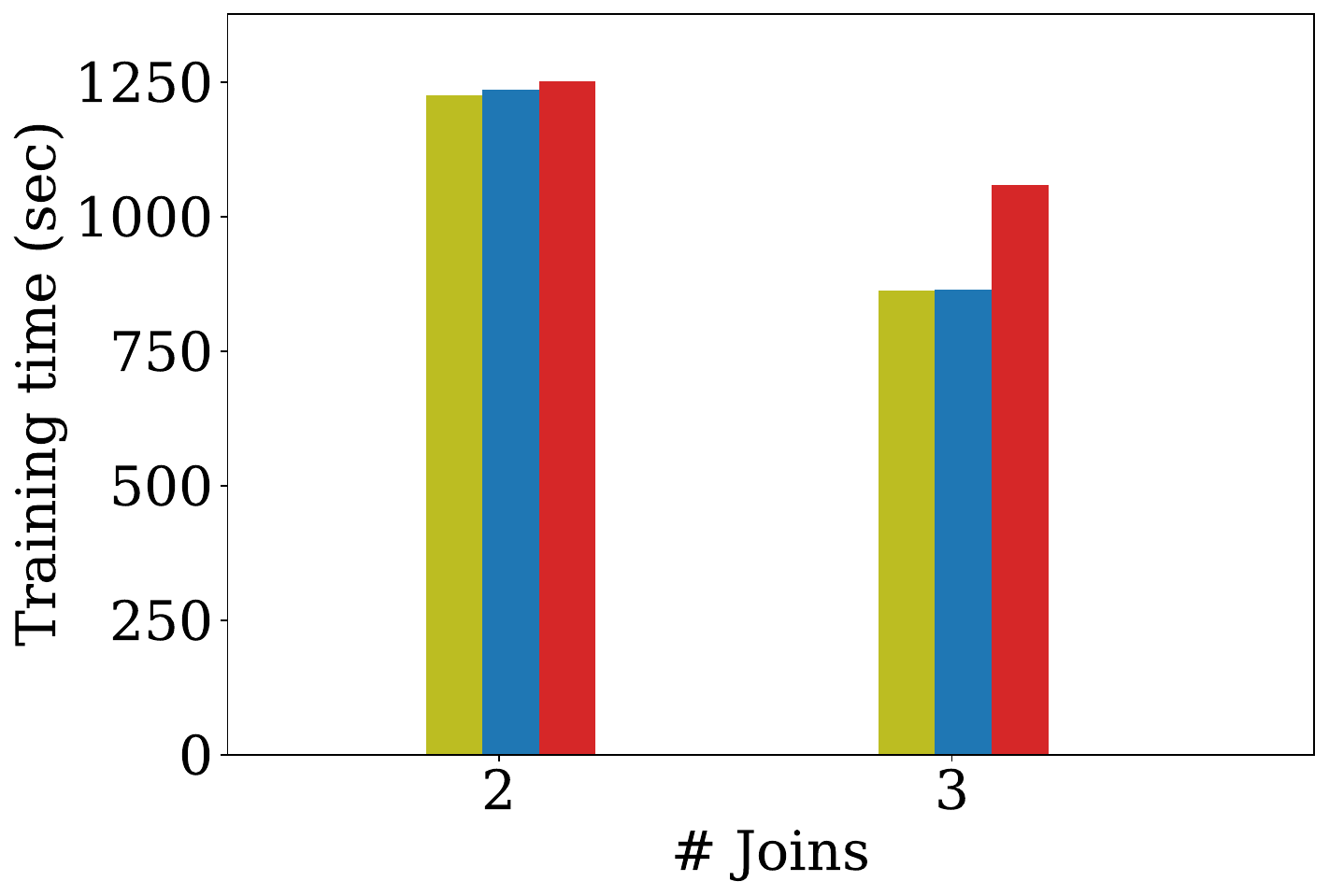}\label{subfig:training_yelp}}
\subfloat[SWDF]{\includegraphics[width=0.25\linewidth]{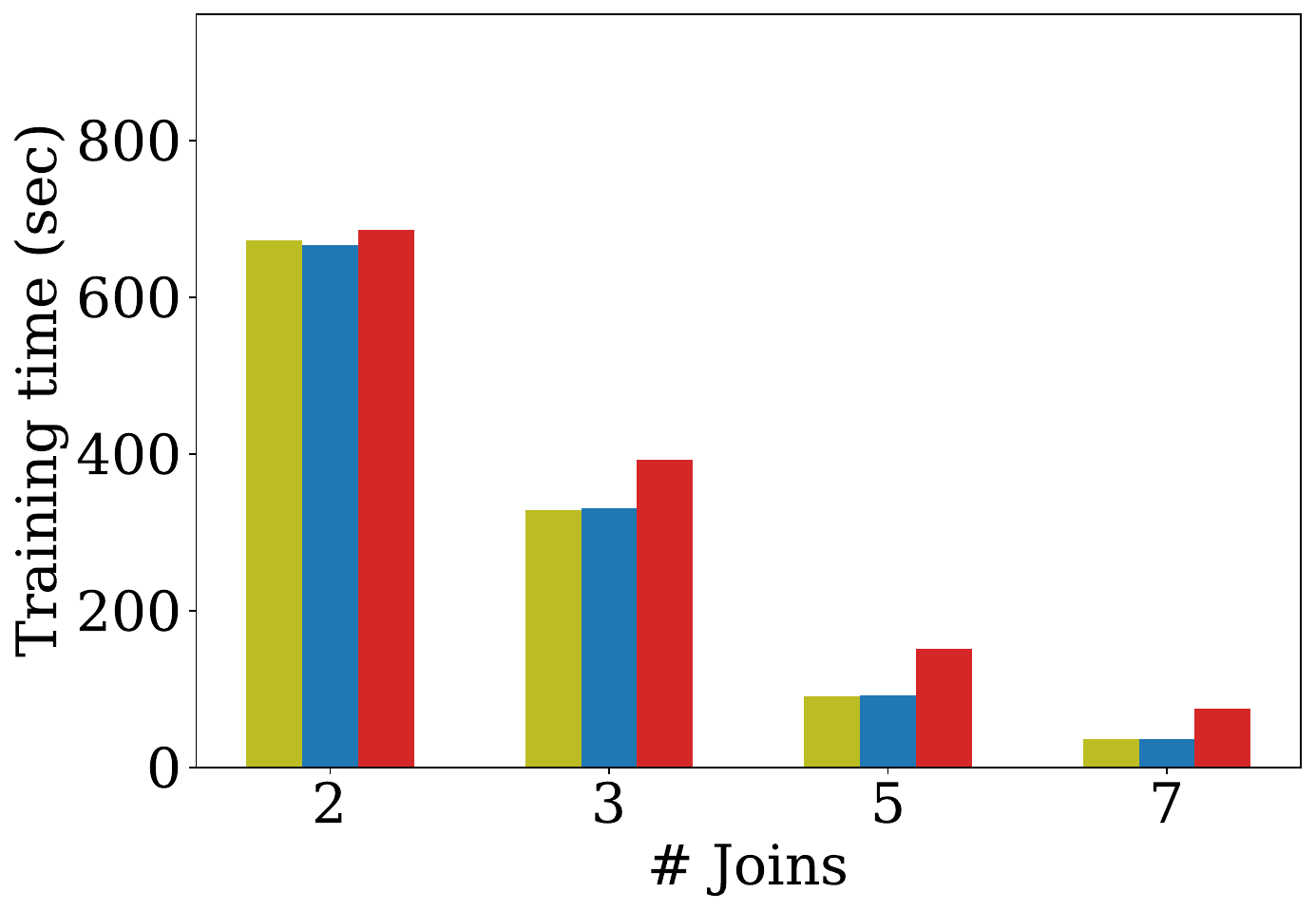}\label{subfig:training_swdf}}
\subfloat[LUBM]{\includegraphics[width=0.25\linewidth]{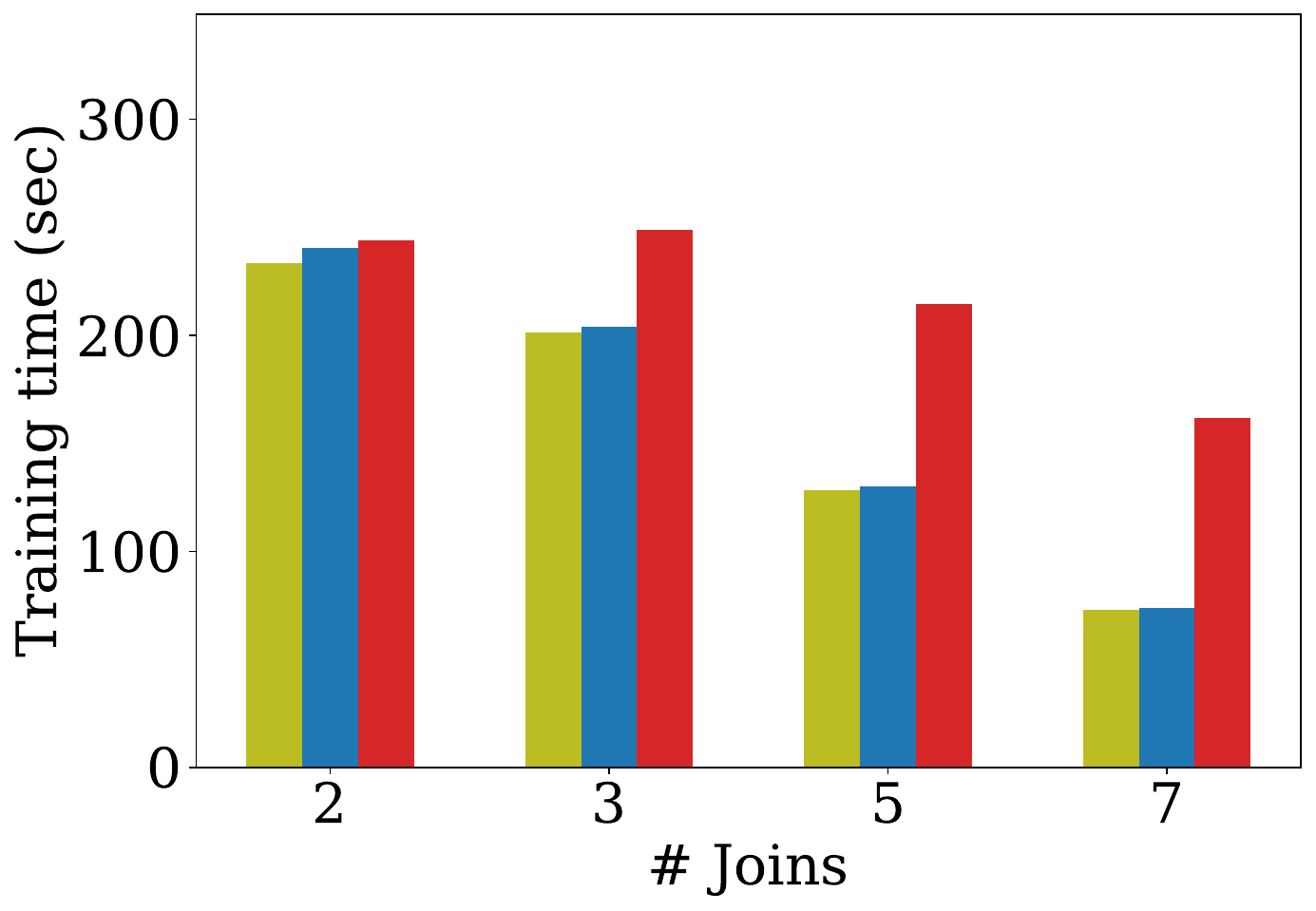}\label{subfig:training_lubm}}
\vspace{-6pt}
\caption{Training times (offline phase) with varying query complexity}
\vspace{-6pt}
\label{fig:train_time}
\end{figure*}

\begin{figure*}[t]
\centering
\subfloat{\includegraphics[width=0.75\linewidth]{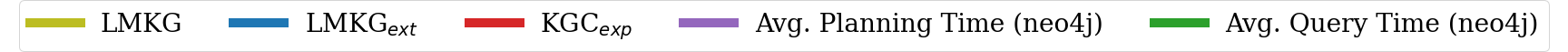}}
\\
\vspace{-8pt}
\setcounter{subfigure}{0}
\subfloat[Wikidata]{\includegraphics[width=0.25\linewidth]{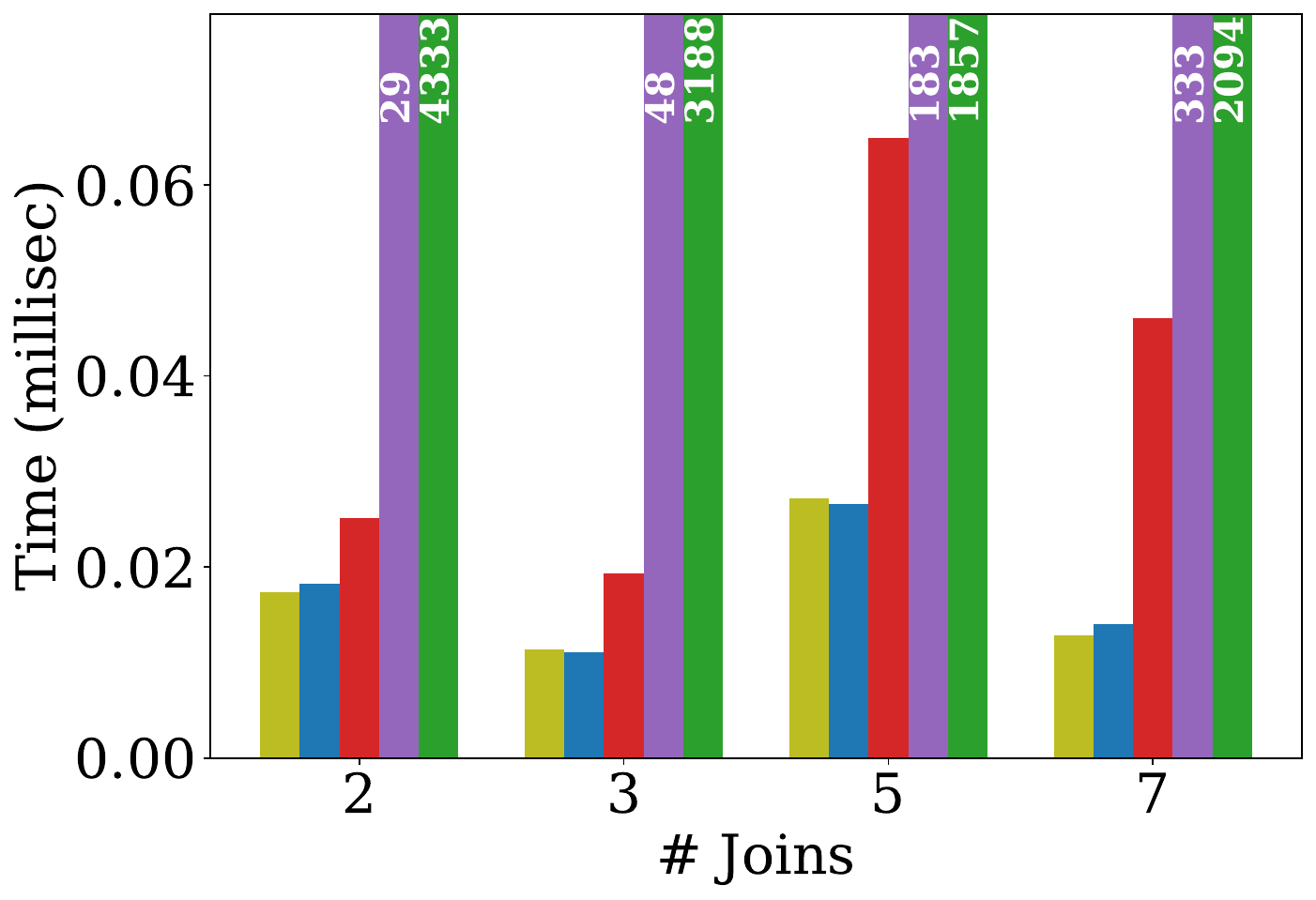}\label{subfig:estimation_wiki}}
\subfloat[Yelp]{\includegraphics[width=0.25\linewidth]{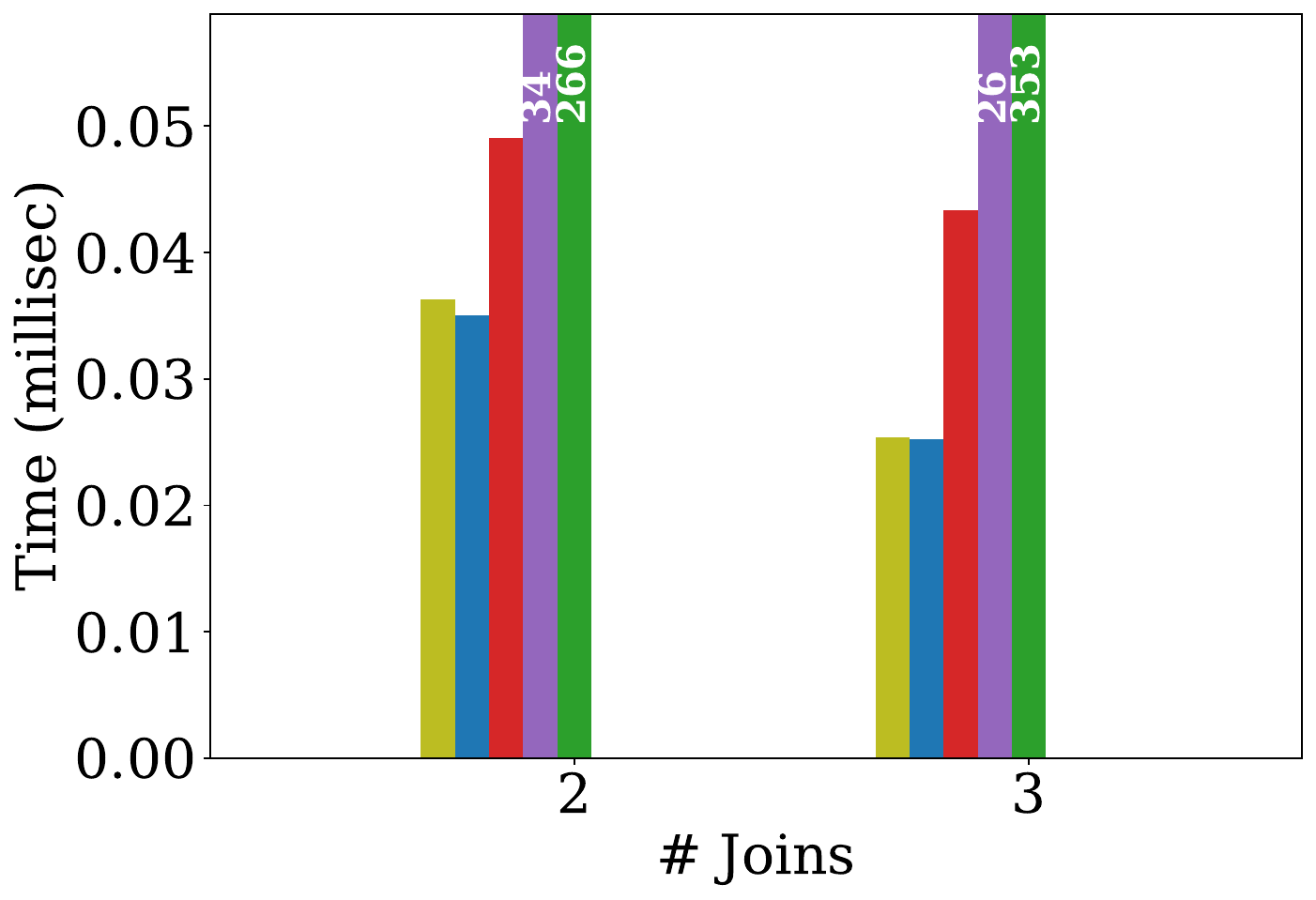}\label{subfig:estimation_yelp}}
\subfloat[SWDF]{\includegraphics[width=0.25\linewidth]{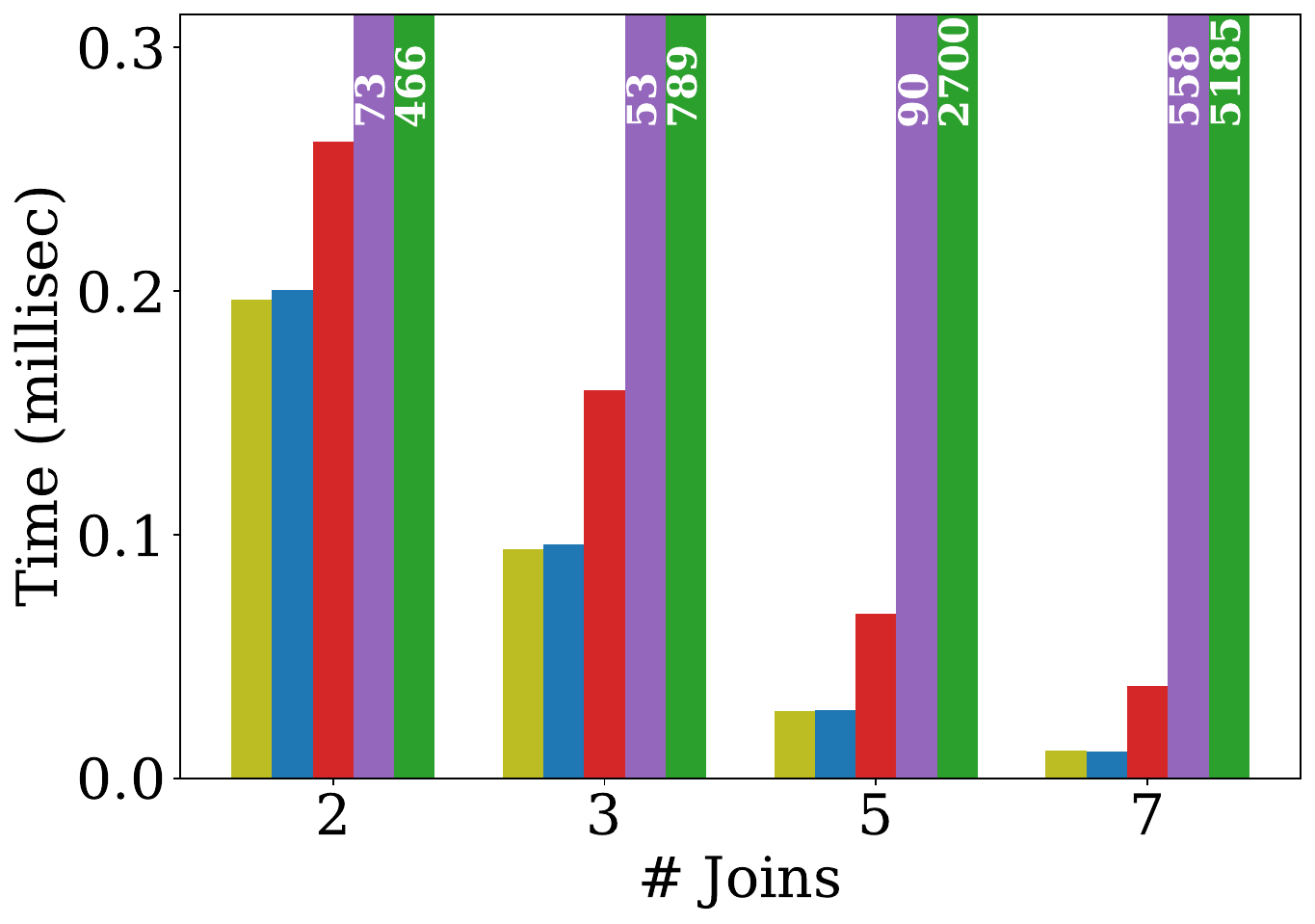}\label{subfig:estimation_swdf}}
\subfloat[LUBM]{\includegraphics[width=0.25\linewidth]{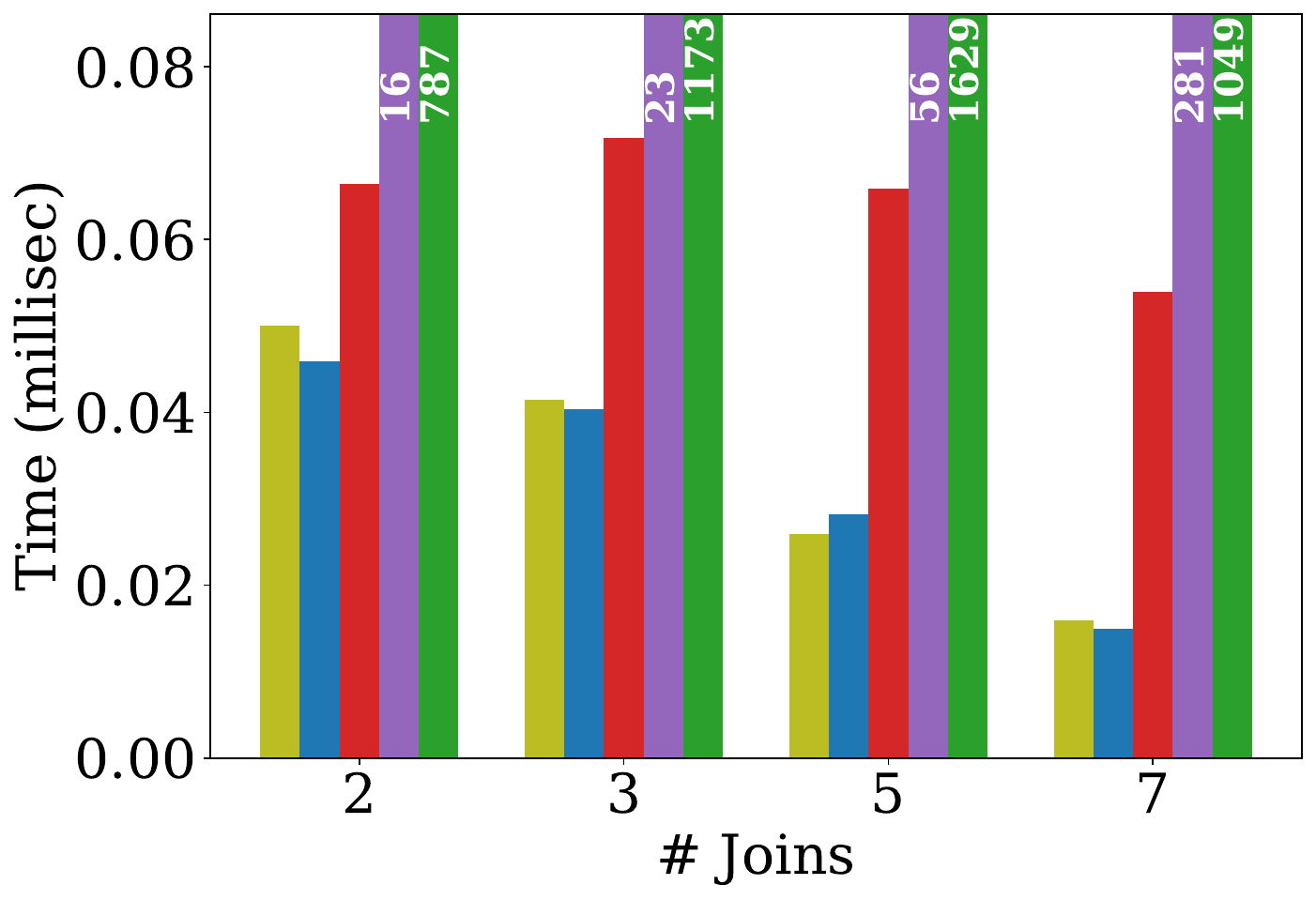}\label{subfig:estimation_lubm}}
\vspace{-6pt}
\caption{Estimation overhead (online phase) vs. avg query planning and execution times with varying query complexity}
\vspace{-6pt}
\label{fig:estimation_time}
\end{figure*}

{\color{red}
\begin{figure*}[t]
\centering
\includegraphics[width=0.95\linewidth]{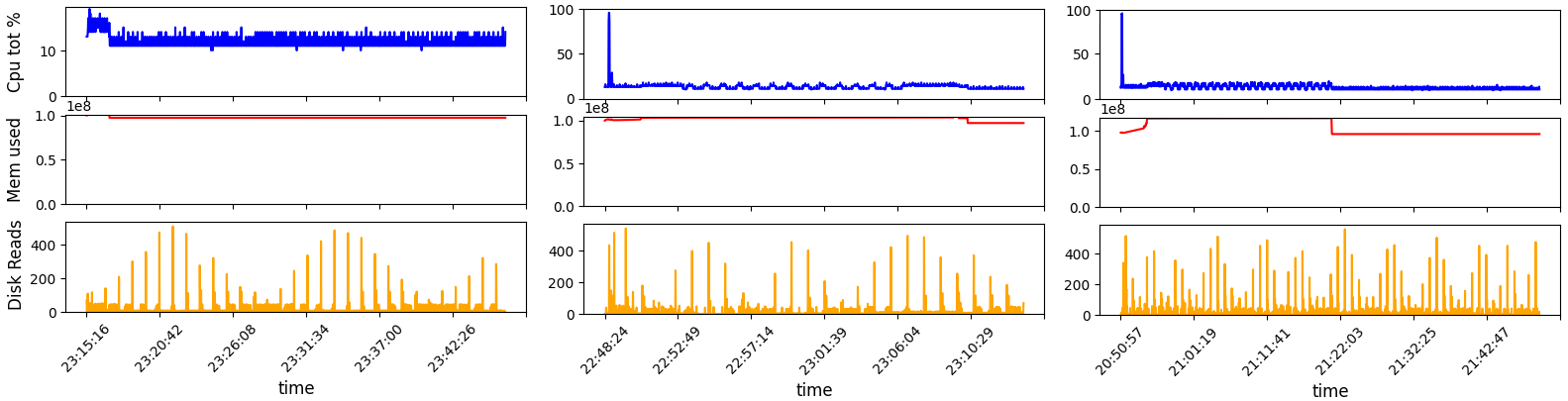}
\vspace{-6pt}
\caption{Resource utilization: cpu, memory, and disk (training on Wikidata with 5 joins). (a)~\baseline, (b)~\MODELa, (c)~\MODELb.}
\vspace{-6pt}
\label{fig:resource-utilization}
\end{figure*}
}

\subsubsection{Query types}

In Figure~\ref{fig:query_type}, we evaluate in more detail the quality of cardinality estimation separately for similarity star (Figures~\ref{subfig:wiki_star}, \ref{subfig:yelp_star}, \ref{subfig:swdf_star} and \ref{subfig:lubm_star}) and similarity chain queries (Figures~\ref{subfig:wiki_chain}, \ref{subfig:yelp_chain}, \ref{subfig:swdf_chain} and \ref{subfig:lubm_chain}). We show for each dataset 
the median, the 75$^{th}$ and 95$^{th}$ percentile of \textit{q}-errors; that is, the maximum error ignoring the $25\%$ and $5\%$ of the largest errors, respectively. 
For the same reasons we explained above, \baseline does not perform well in this scenarios either.
For star similarity queries, 
\MODELb performs better in all cases. 
\mnote{For Yelp, \MODELb has a comparable median \textit{q}-error with \MODELa.}
The 75$^{th}$ and 95$^{th}$ percentiles, however, show an increasing rate of \textit{q}-error differences as a result of \MODELb's ability to better exploit the given similarities. 
For the same reason, the differences are even larger
in the other datasets. For instance, \MODELb has up to $3$x lower \textit{q}-error for the 95$^{th}$ percentile in the LUBM dataset. 
Note that for
chain similarity queries we only have $2$ joins as the datasets under evaluation did not contain large enough paths within the graphs to generate an adequately large training set. Consequently, the differences are smaller here but still apparent.

The results are similar for the classification accuracy of $5$ equi-sized bins, as shown in Figure~\ref{fig:star_class}. Notice that the accuracy is better (up to $20\%$) for star similarity queries on the LUBM dataset (Figures~\ref{subfig:yelp_star_class} and \ref{subfig:lubm_star_class}) since they are less skewed than Wikidata and SWDF. As in the case of \textit{q}-errors, \MODELb outperforms both \MODELa and \baseline in all cases, except for the Yelp dataset, where the accuracy, especially for star similarity queries, is similar.

\subsubsection{Varying query joins}

Increasing the query triples (i.e., joins) is expected to introduce additional complexity to the
model training and hinder the process of cardinality estimation. 
Adding similarities to the mix, the whole process becomes significantly more complex. 
Figure~\ref{fig:no_joins} depicts the results in terms of \textit{q}-error for 
a varying number of joins (\#joins) in the queries. 
Note that for the Yelp dataset (Figure~\ref{subfig:yelp_joins}), the largest possible \#joins is $3$.
\baseline cannot cope well in this scenario either. 
\mnote{As the query becomes more complex (i.e., the \#joins increase), \MODELb clearly outperforms \MODELa}. For $2$ and $3$ joins on Yelp and LUBM (Figure~\ref{subfig:lubm_joins}) datasets, the two approaches are on par. 
However, on the latter dataset the difference increases significantly in favor of \MODELb as the \#joins increase (up to $2$x lower \textit{q}-error). 
This is due to the similarities have significant impact in model training and estimation, as described in Section~\ref{subsec:model2}. The key takeaway from this experiment is that \MODELb scales excellently with increasingly complex queries involving a large \#joins. 

\subsubsection{Offline and online overheads}

Figure~\ref{fig:train_time} depicts the model training times (offline phase) for all approaches. \MODELb employs a more complex model structure as a result of the \SGSPLIT encoding that increases the number of MLPs required in the levels $1$ and $2$ of the model (see Figure~\ref{fig:simcard_kg}). \mnote{This reflects to its training time, which is higher that those of the \MODELa and \baseline models that employ a simpler architecture}. For the same reason, \MODELb has higher average query estimation time (online phase), as shown in Figure~\ref{fig:estimation_time}. 

\mnote{To put things into perspective, we should compare the query estimation time overhead imposed by \MODELa and \MODELb to the query planning and total execution time.}
Typically, query execution involves multiple preparation steps (e.g., parsing, semantic analysis, planning) before the query is dispatched for execution~\cite{neo4j-optimizer}. 
Figure~\ref{fig:estimation_time} shows that the time overhead of all three methods tested are a small fraction of the average planning (purple) and execution time (dark green) of the queries considered in this experiment. In fact, the three methods' overhead is sub-millisecond, while the planning times are in the order of 10s-100s of milliseconds and the execution times in the order of seconds. 
\mnote{At the same time, $95\%$ of the time the \textit{q}-error of \MODELb is significantly smaller than \MODELa and \baseline (see also Figure~\ref{fig:query_type}).}
\mnote{Hence, we anticipate that the more accurate cardinality estimates will result into better query plans, which will significantly compensate for the difference in both the offline and online phase of \MODELb compared to \MODELa and \baseline.}

Our analysis however reveals another interesting observation. A lightweight approach such as MLP consistently offers reasonably satisfactory estimates at a very low cost. Hence, an interesting future direction would be to explore more complex architectures aiming at even more accurate cardinality estimates at the expense of some additional budget.

\subsubsection{Resource utilization}

Finally, we measured the resource utilization of the three methods. In particular, we monitored the training phase of each method on the Wikidata dataset with queries containing $5$ joins.

Figure~\ref{fig:resource-utilization} plots for each method the CPU utilization, the memory used, and the disk reads for the duration of the training phase. We observe that \baseline after a short warm-up phase has a stable CPU utilization around 10\%. \MODELa and \MODELb show a similar behavior with a short peak before they also get stabilized around 10\% as well. The peak in CPU usage accounts for the clustering step (see Section~\ref{subsec:clustering}).

Memory usage seems quite stable with \MODELb requiring somewhat more memory in the initial iterations. We drilled down a bit more on memory consumption for two extreme cases of query types: simple queries with $2$ joins and more complex ones with $7$ joins (Yelp is omitted as it allows up to $3$ joins). Table~\ref{tab:memory} depicts the size of each model (in MBs). The table verifies that \MODELb, due to its more elaborate encoding and architecture, uses memory more aggressively than the other two models for more complex queries.

The disk utilization shows some interesting patterns representing the layered architecture of each model in the duration of training ($50$ epochs). \MODELb demonstrates a busier activity as it processes in parallel many smaller tensors.
One could also see the training times on the x-axis: approximately, 27min (\baseline), 22min (\MODELa), and 52min (\MODELb). \MODELb and \MODELa present a nice trade-off of better accuracy/slower training vs. decent accuracy/faster training.

\begin{table}[t]
  \centering
  \scriptsize
  \caption{Memory consumption}
  \vspace{-2pt}
  \renewcommand{\arraystretch}{0.92}
  \setlength{\tabcolsep}{3pt}
  \begin{tabular}{l|cc|cc|cc}
    \multirow{2}{*}{{\em Dataset}} & \multicolumn{2}{c|}{\MODELb} & \multicolumn{2}{c|}{\MODELa} & \multicolumn{2}{c}{\baseline} \\
    \cline{2-7}
     & 2 joins & 7 joins & 2 joins & 7 joins & 2 joins & 7 joins \\
    \hline
    SWDF     & 4.7MB & 25MB & 2.6MB & 4.3MB & 2.6MB & 4.3MB \\
    Wikidata & 4.8MB & 28MB & 2.8MB & 4.7MB & 2.8MB & 4.7MB \\
    LUBM     & 4.4MB & 25MB & 2.6MB & 4.3MB & 2.6MB & 4.3MB \\
  \end{tabular}
  \label{tab:memory}
\end{table}

\subsection{Augmenting Training Data with an LLM}
\label{subsec:llm_eval}

We compare three training set sources for \MODELb on the LUBM dataset and star queries with $3$ joins: \textit{RW} (random walks), \textit{LLM} (queries generated by an LLM following the prompt of Section~\ref{subsec:llm_training}), and \textit{Hybrid} (a mix of the two). All conditions are evaluated on both LLM-generated test sets of Section~\ref{subsec:exp_setup}. Table~\ref{tab:llm_results} reports the \textit{q}-error for the full training set and a reduced, $30$K-sample one on each test set.

\paragraph{Full training set}
As shown in (Table~\ref{tab:llm_results}(a),(b)), the hybrid approaches outperform the pure RW baseline on every metric and on both test sets. The $50/50$ split has a $12$--$22\%$ lower median \textit{q}-error ($28$--$35\%$ for the $95$th percentile) relative to RW. The pure LLM approach has a lower median \textit{q}-error (it achieves the lowest median on the Gemini test set), however, its max \textit{q}-error is the largest of all cases. This confirms that LLM-generated queries shift the workload distribution differently than random walks. By mixing these two, we take advantage of the tighter center of the LLM data and the bounded tail of the random walk data.

\paragraph{Reduced training set ($30$K samples)}
For the reduced ($30$K) training set (Table~\ref{tab:llm_results}(c),(d)), pure LLM is the clear winner on both test sets, with a median \textit{q}-error $34$--$62\%$ lower ($47$--$74\%$ for the $95$th percentile) relative to RW. The advantage of the LLM-generated data thus grows as the training set size shrinks, indicating that LLM-generated queries are more data-efficient than uniformly sampled random walks.

\begin{table}[t]
    \centering
    \footnotesize
    \caption{\textit{q}-error on the GPT-4o and Gemini test sets; LUBM, $3$-star queries. Best per metric in bold.}
    \vspace{-2pt}
    \label{tab:llm_results}
    \renewcommand{\arraystretch}{0.92}
    \setlength{\tabcolsep}{2pt}

    \begin{minipage}[t]{0.52\linewidth}
    \centering
    \scriptsize
    \textit{(a) Full training set -- GPT-4o}
    \begin{tabular}{l|rrrrr}
    {\em \textbf{Metric}} & RW & LLM & H 50/50 & H 30/70 & H 10/90 \\
    \hline
    median & 1.89   & 1.77    & \textbf{1.66}  & 1.72           & 1.74   \\
    p75    & 3.05   & 2.83    & \textbf{2.39}  & 2.53           & 2.62   \\
    p95    & 7.43   & 8.37    & \textbf{5.32}  & 5.78           & 5.86   \\
    max    & 53.37  & 165.68  & 53.21          & \textbf{34.63} & 37.38  \\
    \end{tabular}
    \end{minipage}\hfill%
    \begin{minipage}[t]{0.44\linewidth}
    \centering
    \scriptsize
    \textit{(b) Full training set -- Gemini}
    \begin{tabular}{rrrrr}
    RW & LLM & H 50/50 & H 30/70 & H 10/90 \\
    \hline
    2.16   & \textbf{1.67} & 1.70          & 1.77   & 1.87  \\
    3.52   & 2.44          & \textbf{2.39} & 2.54   & 2.78  \\
    8.09   & 6.97          & \textbf{5.25} & 5.49   & 5.97  \\
    74.20  & 374.63        & 62.55         & 77.08  & \textbf{53.20} \\
    \end{tabular}
    \end{minipage}

    \vspace{6pt}

    \begin{minipage}[t]{0.52\linewidth}
    \centering
    \scriptsize
    \textit{(c) Reduced ($30$K) training set -- GPT-4o}
    \begin{tabular}{l|rrrrr}
    {\em \textbf{Metric}} & RW & LLM & H 50/50 & H 30/70 & H 10/90 \\
    \hline
    median & 4.51    & \textbf{2.97}   & 4.59    & 4.39    & 5.57    \\
    p75    & 11.77   & \textbf{6.58}   & 10.30   & 10.15   & 13.23   \\
    p95    & 64.05   & \textbf{33.99}  & 38.84   & 35.74   & 53.17   \\
    max    & 1586.49 & 1085.30         & 1343.64 & 1341.02 & \textbf{857.68} \\
    \end{tabular}
    \end{minipage}\hfill%
    \begin{minipage}[t]{0.44\linewidth}
    \centering
    \scriptsize
    \textit{(d) Reduced ($30$K) training set -- Gemini}
    \begin{tabular}{rrrrr}
    RW & LLM & H 50/50 & H 30/70 & H 10/90 \\
    \hline
    6.88    & \textbf{2.61}   & 5.90    & 6.36             & 7.99    \\
    18.45   & \textbf{5.08}   & 12.12   & 13.39            & 18.01   \\
    102.88  & \textbf{26.80}  & 40.03   & 49.39            & 74.29   \\
    2716.19 & 3492.12         & 3207.92 & \textbf{2553.30} & 3166.27 \\
    \end{tabular}
    \end{minipage}
    \vspace{-6pt}
\end{table}

Interestingly, the same ranking holds whether the test queries are created by Gemini or by GPT-4o, even though only Gemini was used to generate the training data; the differences are larger on the Gemini test set, but remain significant on the independent GPT-4o set. Conclusively, an LLM can produce training samples whose distribution complements that of random walks. Combining the two yields the best results when the training set size is large. On the other hand, when training data is scarce, pure LLM data are more effective.

%% file: Conclusions.tex
\section{Conclusions}
\label{sec:conclusions}

We study the problem of supervised similarity-aware cardinality estimation on KGs. We propose a novel approach that explores implicit and explicit means to incorporate similarities into the training process. Specifically, we presented \MODELb, a model that employs a novel encoding scheme for cardinality estimation. Our experiments reveal that \MODELb outperforms state-of-the-art approaches in terms of prediction error, while its online overhead accounts for less than 0.05\% of the average query execution time and less than 1\% of the average query planning time, showing a sub-millisecond latency in all cases. We further demonstrated that an LLM can be leveraged to generate training data resembling realistic user workloads, and that combining such samples with the random walk ones consistently improves estimation accuracy. Promising future directions include extending our approach to other query types (tree, cycle, clique, petal, and flower queries) and investigating unsupervised learning for similarity-aware cardinality estimation on knowledge graphs.